\documentclass[12pt]{iopart}

\newcommand{\beq}{\begin{equation}}
\newcommand{\eneq}{\end{equation}}
\newcommand{\be}{\begin{equation}}
\newcommand{\ee}{\end{equation}}
\newcommand{\bea}{\begin{eqnarray}}
\newcommand{\eea}{\end{eqnarray}}

\usepackage{epsfig}
\usepackage{graphicx}
\begin{document}

\title{Topological Superconductor-Luttinger Liquid Junctions}

\author{Ian Affleck$^{1}$ and Domenico Giuliano$^{2}$  }

\address{$^{1}$ Department of Physics and Astronomy, University of British 
Columbia, Vancouver, B.C., Canada, V6T 1Z1
\\
$^{2}$ Dipartimento di Fisica, Universit\`a della Calabria Arcavacata di Rende I-87036, Cosenza, Italy
and
I.N.F.N., Gruppo collegato di Cosenza, Arcavacata di Rende I-87036, Cosenza, Italy}
\ead{ \\
$^{1}$ iaffleck@phas.ubc.ca \\
$^{2}$ domenico.giuliano@fis.unical.it}
\date{\today}

\begin{abstract}
Experimental evidence was recently obtained for topological superconductivity in spin-orbit coupled nano wires 
in a magnetic field, proximate to an s-wave superconductor.  When only part of the wire contacts the superconductor, 
a localized Majorana mode exists at the junction between superconducting and normal parts of the nanowire. 
We consider here the case of a T-junction between the superconductor and two normal nanowires and also the case 
of a single wire with two (or more) partially filled bands in the normal part.  We find that  coupling  this 
2-channel Luttinger liquid to the single Majorana mode at the junction produces frustration, leading to a  critical point separating 
phases with perfect Andreev scattering in one channel and perfect normal scattering in the other.
\end{abstract}
\maketitle
 
\section{Introduction and Conclusions}
\begin{figure}
  \center
\includegraphics*[width=0.7\linewidth]{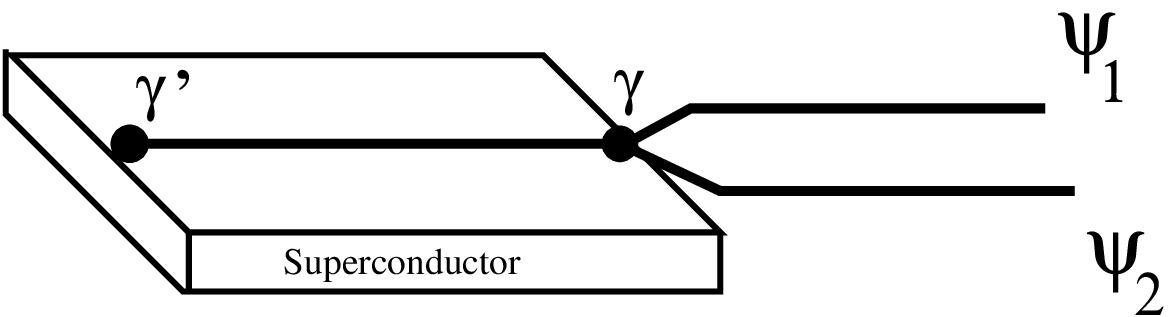}
\caption{Sketch of the SN junction with a topological superconductor, characterized by the localized
Majorana modes $\gamma , \gamma'$ at its boundaries, interfaced to two interacting electronic normal channels: these
may either be regarded as two active channels within the same wire, or as two different quantum wires
interfaced to the topological superconductor.} 
\label{fig_0}
\end{figure}

The existence of  Majorana modes in various topological phases has attracted great theoretical 
and experimental interest due to possible applications to quantum computing. Recently, experimental evidence for such 
a Majorana mode was obtained in indium antimonide quantum wires, where only part of the wire was proximate to an 
s-wave superconductor \cite{Mourik}, following theoretical proposals in \cite{Lutchyn,Oreg}. 
A Majorana mode is expected to be localized at the (SN) junction between superconducting and normal parts of the wire.  
[See figure (\ref{fig_0}).]  
Due to the strong spin-orbit coupling, ideally, the quantum wire might have only one active channel in the normal region. Depending 
on the applied field and other details, two or more channels might instead be active. 
Many theoretical papers have appeared recently on this and related topics, including \cite{Law,Sau,Alicea0,Wimmer,Cook,Klinovaja,Sticlet,Chevalier,Beri0}.
The inspiration for our work was 
the low energy theoretical approach, in both one and two channel cases,  developed in  \cite{lut_majo}, using Luttinger liquid (LL) theory and methods 
introduced in \cite{ACZ} to treat LL SN junctions.
Due to its energy gap, all the electronic degrees of freedom in the superconducting part of the wire may be integrated out, 
except for the Majorana mode at the junction. 
In the single 
channel case, it was found that the system renormalizes, at low energies, to a fixed point characterized by perfect Andreev reflection, 
leading to an enhanced conductance through the  junction. The 2-channel case was also discussed in  \cite{lut_majo}, {\it but only in the simplified 
limit where the Majorana mode  couples to just 1 of the channels}.  It was found that the low energy fixed point has 
perfect Andreev reflection in the channel coupling to the Majorana mode and perfect normal reflection in the other channel. 

Here we extend this analysis to the case where both channels couple to the Majorana mode. The physical situation could correspond 
to a T-junction between a topological superconductor and 2 single-channel quantum wires, or to a single wire containing two active channels in its 
normal part. [See Fig. (\ref{fig_0}).] We find that coupling a single Majorana mode to a two-channel Luttinger liquid leads to an unusual  type of frustration. 
Assuming that the couplings, $t_1$ and $t_2$ of the two channels to the Majorana mode are unequal, we find that the 
larger coupling grows under renormalization and the smaller one shrinks, so that, at low energies, there is perfect Andreev reflection  
in one channel and perfect normal reflection in the other.  This means that the Majorana mode acts as  a switch at the T-junction. 
Even a slight imbalance in tunnel couplings of the two normal wires to the superconductor leads to all the current flowing 
to the more strongly coupled wire, at low energies. This feature might be of use for implementing gate operations in a 
proposed \cite{Alicea} quantum 
computer built from T-junctions.
  By tuning the tunnel couplings
at the junction so $t_1=t_2$, a  critical point can be reached which exhibits non-trivial conductances 
to both channels and associated unusual scaling exponents. The universal conductance and other critical properties 
vary continuously with the Luttinger parameters of the 2-channel Luttinger liquid.  We are able to calculate these 
universal numbers in a certain range of Luttinger parameters using ``$\epsilon$-expansion'' techniques.  We find 
that this critical behaviour is quite robust, surviving when the 2 Luttinger liquids have different velocities and Luttinger parameters and
when they are coupled together in a single quantum wire. Only a single parameter needs to be tuned at the junction 
to reach the critical point. Similar behaviour occurs for $>$2 channels. Analogous universal properties are expected
to appear, for instance, in the equilibrium Josephson current flowing across the normal region contacted
with two topological superconductors at fixed phase difference \cite{giu_af1} which, in the low temperature long junction limit, 
has been shown to depend only on reflection amplitudes at the Fermi level \cite{giu_af2}.

In Sec. II we analyze the phase diagram of the 2-channel model. Sec. III studies its conductance. In Sec. IV we discuss the 
generalization to more than 2 channels. \ref{bosonize} develops bosonization for 2 coupled channels with different velocities. 
 \ref{RG} derives the renormalization group equations in the $\epsilon$ expansion. \ref{spin} presents 
a mapping of the 2-channel Hamiltonian onto an xxz spin chain model containing  unusual impurity couplings. 
\ref{AxN} analyzes the stable fixed point with perfect Andreev scattering in 
one channel and perfect normal scattering in the other.  \ref{NTCP} argues for the stability of the non-trivial critical point 
for general Luttinger parameters.  \ref{single} considers a single channel uniform wire, coupled to 
the topological superconductor far from its endpoints, confirming the proposed phase diagram. \ref{entropy}
calculates the impurity entropy at the various fixed points.  
\section{The  topological superconductor- 2-channel Luttinger liquid junction}
\label{rg_eq}
After integrating out the gapped excitations in the superconductor, we are left with a low energy 
theory describing 2 channels of interacting Dirac fermions coupled to a boundary Majorana mode. 
The left and right moving Dirac fermions contain the wave-vector components near $\pm k_{Fj}$, the Fermi points, for each channel:
\be \psi_j(x)\approx \exp [ik_{Fj}x]\psi_{R  j} (x)+\exp [-ik_{Fj}x]\psi_{Lj} (x).\ee
We write the Hamiltonian:
\be H=H_0+H_{\hbox{int}}+H_b.\label{HT}\ee
Here $H_0$ is the Hamiltonian of 2 channels of non-interacting electrons:
\be H_0\equiv \sum_{j=1}^2 iv_{Fj}\int_0^\infty dx [\psi^\dagger_{Rj}\partial_x\psi_{Rj}-\psi^\dagger_{Lj}\partial_x\psi_{Lj}]
\: , \ee
and $\psi_{L/R,j}(x)$ annihilates left/right moving electrons in channel $j=1,2$. We impose ``open boundary conditions'' at $x=0$:
\be \psi_{Rj}(0)=\psi_{Lj}(0)\label{bc0}\ee
corresponding to disconnected channels before turning on the coupling to the superconductor. $v_{Fj}$ are the Fermi velocities 
in the two channels. The bulk interactions in general consist of intra-channel and inter-channel terms:
\be H_{\hbox{int}}=H_{\hbox{intra}}+H_{\hbox{inter}}.\ee
The most important intra-channel interaction is:
\be H_{\hbox{intra}}=\sum_jV_j\int_0^\infty dx (:\psi^\dagger_{Rj}\psi_{Rj}:+:\psi^\dagger_{Lj}\psi_{Lj}:)^2.\label{intra}\ee
(The $:\ldots :$ denote normal ordering.) 
We assume Umklapp processes can be ignored, true below a critical interaction strength even 
for commensurate electron densities.   The interaction in Eq. (\ref{intra}) changes the 
Luttinger parameter away from its non-interacting value. Other interactions only change the Fermi velocity or have no effect 
on the low energy theory.  In the case of a T-junction of single channel nano wires,  the only important bulk interaction 
is $H_{\hbox{intra}}$. In the case of a 2-channel wire, we assume that the bulk interactions conserve the number 
of electrons in each channel.  We also assume $k_{F1}\neq k_{F2}$, likely to be true in a spin-orbit coupled quantum wire 
in a magnetic field. 
The most important inter-channel interaction is then:
\be H_{\hbox{inter}}=U\int_0^\infty dx (:\psi^\dagger_{L1}\psi_{L1}:+:\psi^\dagger_{R1}\psi_{R1}:)(
:\psi^\dagger_{L2}\psi_{L2}:+:\psi^\dagger_{R2}\psi_{R2}:).\ee
The term $\psi^\dagger_{L1}(x)\psi_{R1}(x)\psi^\dagger_{R2}(x)\psi_{L2}(x)$ comes with an oscillating factor 
$\exp [2i(k_{F1}-k_{F2})x]$ and can be ignored at low energies. [ In the case $k_{F1}=k_{F2}$, this term 
must be taken into account and could produce a gap.]    $U$ mixes the 2 Luttinger liquids. Note that, in the special 
case $v_{F1}=v_{F2}$, $U/2=V_1=V_2$ the model has SU(2) symmetry and it would be natural to interpret $1$ and $2$ as spin indices. 
This symmetry is unlikely to occur in a spin-orbit coupled system in a magnetic field. However, it provides a 
useful consistency check on some of our results, so we will occasionally consider it.  Finally, the most important 
boundary interaction is with the localized Majorana mode, $\gamma$, of the topological superconductor:
\be H_b=\gamma \sum_jt_j[\psi_j(0)-\psi^\dagger_j(0)].\label{Hb}\ee
Here $\psi_j(0)\equiv \psi_{Lj}(0)=\psi_{Rj}(0)$. 
By redefining the phases of $\psi_j$, we will also choose the $t_j$ real and non-negative, $t_j\geq 0$. 
Most of the new results obtained in this paper arise  from considering the general case where $t_1$ and $t_2$ are both non-zero. 
These single electron tunnelling terms from the Luttinger liquid to the superconductor are generally strongly relevant and 
determine the low energy physics. In addition, we could add various other boundary interactions, quadratic 
in $\psi_j(0)$ and $\psi^\dagger_j(0)$, which are generally irrelevant and were discussed in \cite{lut_majo}.

To study the low energy physics we bosonize the Luttinger liquid.  In order to take account of the fermion anti-commutation relations, 
it is very convenient to introduce Klein factors, $\Gamma_i$.  
The usefulness of bosonization Klein factors for studying related Majorana mode models was pointed out in \cite{Beri}. 
Thus we write:
\be \psi_{L/R,j}(x)\propto \Gamma_j \exp \{i\sqrt{\pi}[\phi_j(x)\pm \theta_j(x)]\} \label{bos}\ee
with:
\be \{\Gamma_i,\Gamma_j\} =2\delta_{ij},\ \  \{\Gamma_i,\gamma \}=0,\ \ \gamma^2=1.\label{anticom}\ee
In the T-junction case, with $H_{\hbox{inter}}=0$, the bulk Hamiltonian is diagonal in the $\phi_j$, $\theta_j$ bosons:
\be H_0+H_{\hbox{intra}}={1\over 2}\sum_ju_j\int_0^\infty dx 
\left[ K_j\left( \frac{\partial \phi_j}{\partial x} 
\right)^2 + K_j^{-1}\left( \frac{\partial \theta_j}{\partial x} 
\right)^2\right] .\label{H0j}\ee
Here $K_j$ and $u_j$ are the Luttinger parameters and velocities for the two branches of the T-junction, depending on $V_j$. 
(Our Luttinger parameter $K$ corresponds to $g$ in the notation of \cite{lut_majo}.)
In the non-interacting case, $V_j=0$, $K_j=1$.
When inter-channel interactions, $U$ are included, we can conveniently write $\phi_j$ in terms of bosons $\phi_\rho$, $\phi_\sigma$ which diagonalize 
the bulk Hamiltonian:
\bea \left(\begin{array}{c}\phi_{1}\\ \phi_{2}\end{array}\right)&=&
\left(\begin{array}{cc}r^{-1} \cos \alpha&r^{-1}\sin \alpha \\
-r\sin \alpha & r\cos \alpha \end{array}\right) \left( \begin{array}{c} 
\phi_\sigma \\ \phi_\rho \end{array}\right),\nonumber \\
\left(\begin{array}{c}\theta_{1}\\ \theta_{2}\end{array}\right)&=&
\left(\begin{array}{cc} r\cos \alpha&r\sin \alpha \\
-r^{-1}\sin \alpha & r^{-1}\cos \alpha \end{array}\right) \left( \begin{array}{c} 
\theta_\sigma \\ \theta_\rho \end{array}\right).
\label{phal}
\eea
The parameters $r$ and $\alpha$ are a measure the asymmetry between the two channels.  In the symmetric case, 
$v_{F1}=v_{F2}$, $V_1=V_2$, $\alpha =\pi /4$ and $r=1$.  See \ref{bosonize} for a derivation of these results. 
(\cite{lut_majo}  only considered the symmetric case, $\alpha =\pi /4$, $r=1$.)
The bulk Hamiltonian becomes:
\beq
H_0 = \sum_{\lambda = \rho , \sigma} \: \frac{u_\lambda}{2 } \: 
\int_0^\infty \: d x \: \left[ K_\lambda \left( \frac{\partial \phi_\lambda}{\partial x} 
\right)^2 + K_\lambda^{-1}\left( \frac{\partial \theta_\lambda}{\partial x} 
\right)^2\right].                   
\label{rg2.1}
\eneq
The subscripts $\rho$ and $\sigma$ refer to charge and spin in the Luttinger liquid literature but 
that interpretation is general not appropriate in this case. 
A T-junction with inequivalent wires can be regarded as the special case $\alpha =0$, with $K_1=K_\sigma r^2$ and $K_2=K_\rho /r^2$. 
In the non-interacting limit, $   K_\rho /r^2=K_\sigma  r^2=1$.   
In the SU(2)  symmetric case, $V_1=V_2=U/2$, $v_{F1}=v_{F2}$, $r=1$, $\alpha =\pi /4$ and $K_\sigma$ 
remains fixed at the value $1$.
The normal reflection boundary condition of Eq. (\ref{bc0})
corresponds to:
\be \theta_j(0)=\hbox {constant}\label{bc1}\ee
so the bosonized boundary Hamiltonian is:
\be H_b=i\gamma \sum_jt_j \tau_0^{-1+d_j} \Gamma_j\{\exp [i\sqrt{\pi}\phi_j(0)]+
\exp [-i\sqrt{\pi}\phi_j(0)]\} , \label{Hbb}\ee \noindent
with $\tau_0^{-1}$ being a high-energy cutoff.
For convenience, we  redefine the normalization of the dimensionless tunnelling parameters, $t_j$, so that the boundary operators 
$\exp [\pm i\sqrt{\pi}\phi_j(\tau )]$ are 
unit normalized:
\be{\bf T}  <  \exp [i\sqrt{\pi}\phi_j(\tau )]\exp [-i\sqrt{\pi}\phi_j(0) ]   >={1\over |\tau |^{2d_j}}.\label{norm}
\ee 
Here ${\bf T}$ denotes (imaginary) time-ordering and $\phi_j(\tau )\equiv \phi_j(\tau ,x=0)$.
Taking into account the boundary conditions of Eq. (\ref{bc1}), in the T-junction case:
\be 2d_j={1\over K_j}.\ee
With inter-channel interactions:
\bea 2 d_1&=&r^{-2}\left({\cos^2\alpha \over K_\sigma}+{\sin^2\alpha \over K_\rho}\right) \nonumber \\
2d_2&=&r^2\left( {\sin^2\alpha \over K_\sigma}+{\cos^2\alpha \over K_\rho}\right).\label{di}
\eea
The $d_i$ are the renormalization group scaling dimensions of the boundary interactions, implying the scaling equations:
\be -{d t_i\over d\ln D}=(1-d_i) t_i+\ldots ,\label{beta0}\ee
with $D$ being a running high energy cut-off. 
For non-interacting electrons, $d_i=1/2$ so that the couplings to the Majorana mode are strongly relevant. We expect the $d_i$ to increase 
as repulsive interactions are turned on, but the $t_i$ remain relevant ($d_i<1$) up to quite strong repulsive interactions. 

The low temperature conductance and other low energy properties of the SN junction are determined by the infrared stable 
fixed point of these renormalization group (RG) equations. In the case where $t_2=0$ (or equivalently $t_1=0$) it was argued 
in \cite{lut_majo} that $t_1$ renormalizes to large values and $t_2$ remains at zero.  This fixed point corresponds to a 
conformally invariant boundary condition:
\bea \phi_1(0)&=&0\ \ \hbox{or}\ \ \sqrt{\pi}\nonumber \\
\theta_2(0)&=&0.\label{bc2}
\eea

The two possible boundary conditions \cite{lut_majo} on $\phi_1$ in Eq. (\ref{bc2}) correspond to eigenstates of $i\gamma \Gamma_1$ with 
eigenvalue $-1$ and $+1$ respectively.  To see that such eigenstates exist, note that we may combine $\gamma$ and $\Gamma_1$ 
into a Dirac operator  localized at the junction:
\be \psi_0\equiv {\gamma+i\Gamma_1\over 2}\ee
obeying $\{\psi_0,\psi_0^\dagger \}=1$. 
We then have:
\be i\gamma \Gamma_1=2\left( \psi_0^\dagger\psi_0-1/2\right) \label{locdir}\ee
which clearly has eigenvalues $\pm 1$. These are ``Schroedinger cat states'' in which a single electron has equal amplitude to 
be in the superconductor or in the nanowire (in channel 1). This can be seen by observing that there is actually a second Majorana 
mode, $\gamma '$ localized at the opposite end of the superconductor, far from the SN junction, as sketched 
in Fig. 1. We may construct a different Dirac zero 
mode operator
\be \psi_S\equiv  {\gamma+i\gamma '\over 2}.\label{psiS}\ee
This annihilates an electron which is located entirely inside the superconductor but is highly delocalized, with equal amplitudes 
at both ends.  Denoting the corresponding states as $|0S>$ and $|1S>$, we see that that $\gamma |0S>=|1S>$. Thus the 
eigenstates of $i\gamma \Gamma_1$ are linear superpositions of $|0S>$ and $|1S>$.

As shown in \cite{ACZ,lut_majo}, the boundary condition of Eq. (\ref{bc2}) corresponds to perfect Andreev scattering at the SN junction in channel 1 and 
perfect normal scattering in channel 2 and the corresponding boundary conditions are labelled $A\otimes N$. 

\begin{figure}
 \center
\includegraphics*[width=0.3\linewidth]{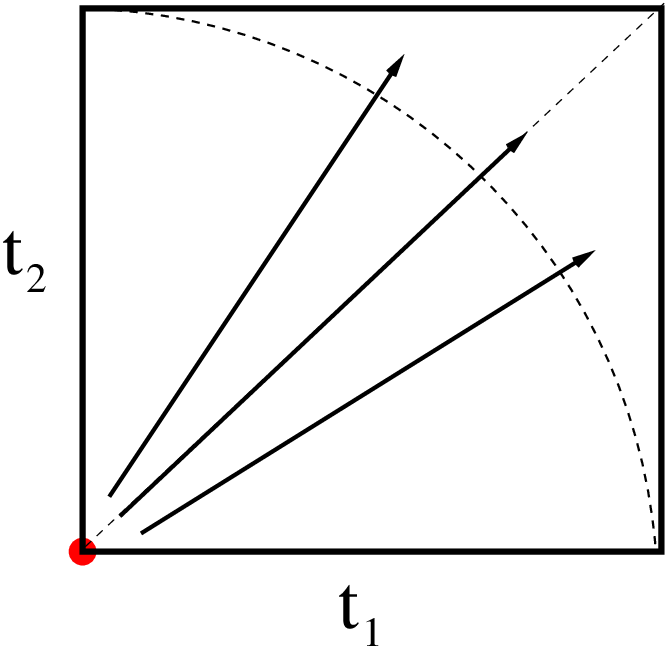}
\caption{ Plot of the flow of the effective tunnelling amplitudes with decreasing energy scale in the $SU(2)$ symmetric case.} 
\label{fig_SU2}
\end{figure}

The main question we wish to address in this paper is the nature of the ground state when $t_1$ and $t_2$ are both non-zero. This 
is readily addressed for the non-interacting model, $V_i=U=0$. Then we can make a change of basis:
\bea \tilde \psi_1&=&{t_1\psi_1+t_2\psi_2\over \sqrt{t_1^2+t_2^2}}\nonumber \\
\tilde \psi_2&=&{-t_2\psi_1+t_1\psi_2\over \sqrt{t_1^2+t_2^2}}.\label{SU2}
\eea
so that
\be H_b\to \gamma \sqrt{t_1^2+t_2^2}[\tilde \psi_1(0)-\tilde \psi_1^\dagger (0)] .\ee
Clearly the ground state corresponds to perfect Andreev scattering in channel $\tilde 1$ and perfect normal scattering in channel $\tilde 2$. An electron from channel $\tilde 1$ is in an entangled state with the superconductor.
We refer to these as rotated $A\otimes N$ boundary conditions.  It is basically the SU(2) symmetry of the free 
electron model which allows us to form this superposition. We might worry that unequal Fermi velocities in the two non-interacting 
channels destroy this SU(2) symmetry. However, since we only need to make the unitary transformation at the boundary, 
and the system is non-interacting, we can 
rescale the $x$-coordinate differently for the two channels and again make this transformation. In general, such a transformation 
cannot be conveniently made in the interacting case. An exception occurs when $v_{F1}=v_{F2}$ and $V_1=V_2=U/2$ so that 
the model has SU(2) symmetry. In this case we can always make the transformation of Eq. (\ref{SU2}) and obtain a
rotated $A\otimes N$ 
boundary condition. Formally, we may make the SU(2) transformation first, and then bosonize. The RG flow diagram in 
this case corresponds to Fig. (\ref{fig_SU2}). 

It is important to note that the rotated $A \otimes N$ boundary condition is very different from an $A\otimes A$ boundary condition
 despite the fact that the tunnelling amplitudes, $t_i$, to both channels, renormalized to large values. 
In an $A\otimes N$ state, the Majorana mode is strongly entangled with {\it one linear combination} of fermion fields and the orthogonal 
linear combination decouples. This corresponds to a stable fixed point of the system as we  show later in this section 
and in \ref{AxN}. On the other hand, 
in an $A\otimes A$ state, the Majorana mode is strongly entangled with {\it both} fermion fields.  This turns out to be an unstable fixed point, 
as we will show in \ref{NTCP}. 

It is far from obvious at this stage what the low energy behavior is for the general non-SU(2) invariant interacting case, 
with $t_1$, $t_2>0$.  To gain some insight into this question, we calculate the terms of next order in $t_i$ 
in the $\beta$-functions 
of Eq. (\ref{beta0}).  This calculation is of interest primarily in the case where $0<1-d_i\ll 1$, since then 
the $\beta$-functions may have a zero (corresponding to a renormalization group fixed point) at small $t_i$
where stopping the expansion at next order can be justified. 
It is then convenient to define:
\be \epsilon_i\equiv 1-d_i.\ee
To calculate the next order terms in the $\beta$-functions 
we thus set $\epsilon_1=\epsilon_2=0$, corresponding to marginal interactions. 
 Thus the four parameters labelling the bulk interactions are reduced to 2 independent parameters by the conditions, following from Eq. (\ref{di}):
 \be r^{-2}\left({\cos^2\alpha\over K_\sigma}+{\sin^2\alpha\over K_\rho}\right)=r^2\left({\cos^2\alpha\over K_\rho}+{\sin^2\alpha\over K_\sigma}\right) =2.\label{marg}\ee
The detailed calculation of these $\beta$-functions in given in \ref{RG}. 
The quadratic terms  can easily be seen to vanish and the cubic terms are given in terms of a single function of 
the interactions parameters, 
\be \nu\equiv {\sin 2\alpha \over 2}\left({1\over K_\rho}-{1\over K_\sigma}\right).\label{nu}\ee
These have the form:
\begin{eqnarray}
 \frac{d t_1}{d l } &=& \epsilon_1  t_1 - {\cal F} (  \nu )  t_1(  t_2)^2 \nonumber \\
  \frac{d t_2}{d l } &=& \epsilon_2  t_2 - {\cal F} (  \nu )  t_2(  t_1)^2.
\label{rg2.19}
\end{eqnarray}
Here $l\equiv -\ln D$. 
The function $ {\cal F} (  \nu )$ is given as a 1-dimensional integral in Eq. (\ref{fnu}) and plotted in Fig. (\ref{fig_nu}). 
It is monotone decreasing, passing through $0$ at $\nu =1$. This latter value provides an important check on our 
calculations. The bulk theory has $SU(2)$ invariant when $r=1$, $\alpha =\pi /4$ and $K_\sigma =1$. The condition Eq. (\ref{marg}) 
for the interactions be  marginal then determines $K_\rho=1/3$ and hence, from Eq. (\ref{nu}), $\nu =1$.  Thus, ${\cal F}=0$ 
for the $SU(2)$ invariant model. This is consistent with our observation that the RG flows are along rays in the $t_1-t_2$ plane, flowing
towards strong coupling in that case, as shown in Fig. (\ref{fig_SU2}). 
Note that the T-junction corresponds to $\alpha =0$ and then Eq. (\ref{marg}) gives $K_\rho=K_\sigma =1/2$ and thus 
$\nu =0$ and ${\cal F}=4$. 

 \begin{figure}
 \center
\includegraphics*[width=0.7\linewidth]{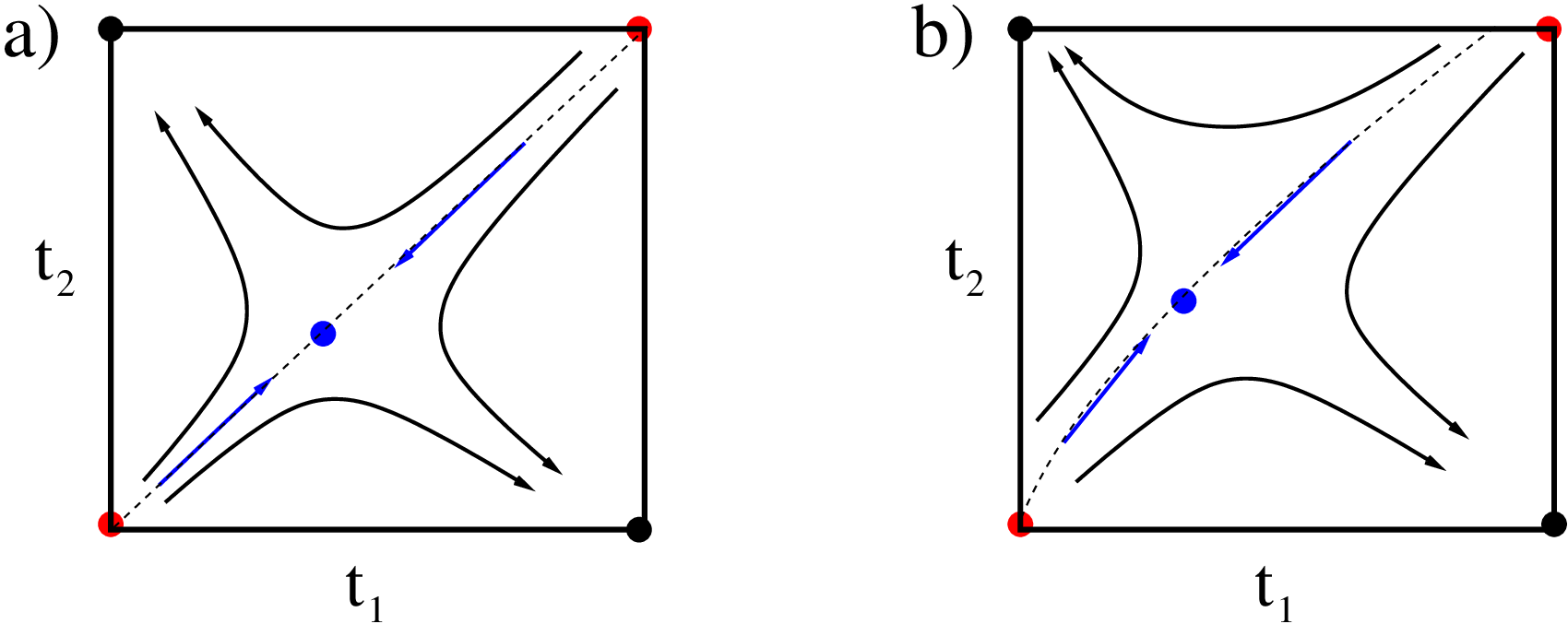}
\caption{Flow of the effective tunnelling amplitudes with decreasing energy scale.  a) The symmetric case $\epsilon_1=\epsilon_2$. 
The red dots represent the completely unstable $N\otimes N$ and $A\otimes A$ fixed points, the blue dot represents 
the non-trivial critical point and the black dots represent the stable $A\otimes N$ and $N\otimes A$ critical points.  The unstable 
$A\otimes A$ fixed is discussed in \ref{NTCP}. 
b) An example of the general case.} 
\label{phd_1}
\end{figure}

\begin{figure}
 \center
\includegraphics*[width=0.3\linewidth]{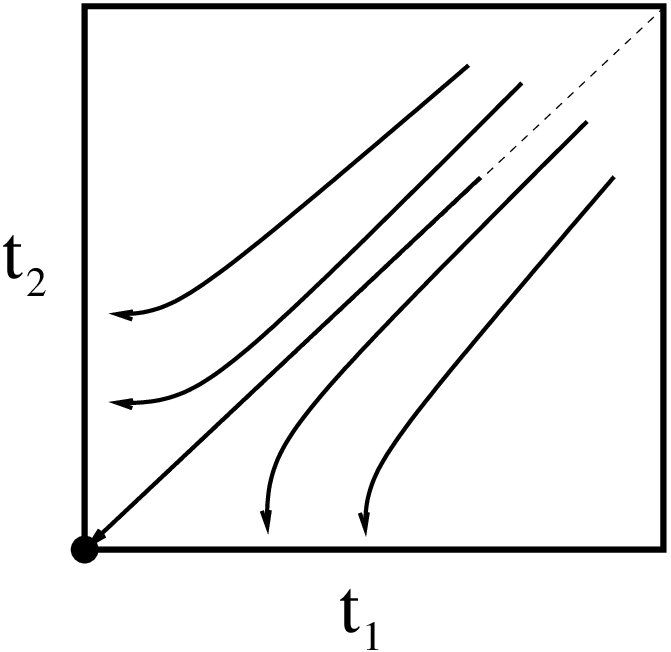}
\caption{Flow of the effective tunnelling amplitudes with decreasing energy scale. for the case of marginal 
boundary couplings, $\epsilon_i=0$.} 
\label{fig_e=0}
\end{figure}

In general, for $\nu <1$, where ${\cal F}(\nu ) >  0$, the RG equations have a fixed point at:
\bea t_{1c}&=& \sqrt{\epsilon_2\over {\cal F}(\nu )}\nonumber \\
t_{2c}&=&\sqrt{\epsilon_1\over {\cal F}(\nu )}.\label{tfp}\eea
Taking $\epsilon_i\to 0$ for any fixed $\nu$, this fixed point occurs at weak coupling where higher order terms in the 
$\beta$-function can be ignored. The corresponding RG flow diagram is sketched in Fig. (\ref{phd_1}). This nontrivial 
critical point (NTCP) has one unstable direction, and one stable direction in the $t_1-t_2$ plane. In the symmetric case $\epsilon_1=\epsilon_2$ , 
the unstable direction corresponds to making the $t_i$'s unequal:
\bea t_1&=&t_c+\delta t\nonumber \\
t_2&=&t_{c}-\delta t.\eea
Eqs. (\ref{rg2.19}) then imply:
\be {d(\delta t)\over dl}=2\epsilon \ \delta t +\ldots \ee
On the other hand, the NTCP is stable along the symmetry line where:
\bea t_1&=&t_c+\delta t\nonumber \\
t_2&=&t_{c}+\delta t,\eea
and thus
\be {d(\delta t)\over dl}=-2\epsilon \ \delta t +\ldots \ee
 Note that when the $\epsilon_i$ are 
strictly zero, the RG equations, Eq. (\ref{rg2.19}) imply lines of stable fixed points along the $t_1$ and $t_2$ axes 
as shown in Fig. (\ref{fig_e=0}).  This type of flow diagram was discussed in \cite{affl_1,affl_2} as a model 
for a 2-level system interacting with 2 independent heat baths, where it corresponded to  ``frustration of decoherence''. 
In \ref{spin}, we show that a special case with $\epsilon_i=0$ corresponds 
to an unusual anisotropic version of the 2-channel Kondo model.   The model with decoupled equivalent chains, $U=0$, $v_{F1}=v_{F2}$, 
$V_1=V_2$ arbitrary,  turns out to be identical to a formulation of the dissipative Hofstadter 
model \cite{Callan}, studied in \cite{Novais}.  
This can be seen from the spin chain representation of our model, introduced in \ref{spin}; precisely the same model 
was studied in \cite{Novais}, in this special case.  
The $\beta$ functions of Eq. (\ref{rg2.19}) were already obtained, and their fixed point analyzed, for this  case. 
This model and $\beta$-function are also closely related to ones studied earlier in the context of 
the dissipative Hofstadter model and of the Bose-Fermi Kondo model \cite{Zarand}.  
For $\epsilon_1 = \epsilon_2 = \nu = 0$ the $\beta$ function in Eq.(\ref{rg2.19})
also describes the ``paperclip model'' at topological angle $\vartheta = \pi$ studied in
\cite{lukyanov}.
However  for the general case, with $V_1\neq V_2$, $U\neq 0$ and arbitrary $\alpha$, 
these NTCP's are {\it not} equivalent to any other previously discovered ones that we are aware of. 
Note that, moving away from the NTCP along the unstable direction, the flow is towards $(t_1,t_2)=(\infty ,0)$ 
or $(0,\infty)$ corresponding to $A\otimes N$ or $N\otimes A$ fixed points. These fixed points 
are stable \cite{lut_majo} and thus it is plausible that the RG flow goes to them in all cases. 

So far, we have only established this phase diagram for small $\epsilon_i$. However, we expect that it 
 remains qualitatively similar for a finite range of  bulk interaction parameters ($K_\rho$, $K_\sigma$, $r$ and $\alpha$). 
Arguments are presented for this in a series of Appendices. In \ref{AxN} we show that the $A\otimes N$ and $N\otimes A$
fixed points are stable over a large range of bulk interaction parameters.   In \ref{NTCP}, 
we give arguments for the existence of the NTCP. In particular,  
we discuss the unstable $A\otimes A$ fixed point, shown in Fig. (\ref{phd_1}) at $t_1=t_2=\infty$. 
We also mention a possible alternative to the phase diagram of Fig.  (\ref{phd_1})  which 
might conceivably occur for a range of Luttinger parameters but requires additional nontrivial critical points. 
In \ref{single} we consider a model of a uniform single channel quantum wire coupled far from its ends to the topological superconductor. 
For $2-\sqrt{3}\approx .268<K<1$, we find that the RG flow is again to $A\otimes N$ or $N\otimes A$ or else to the NTCP
if an appropriate symmetry is respected. This strongly supports the universality of our 
proposed phase diagram, Fig. (\ref{phd_1}). In \ref{entropy} we calculate the universal impurity entropy at the various fixed points 
and show that our proposed phase diagram is consistent with the $g$-theorem  \cite{g,aflud,Friedan}. 

The phase diagram of Fig. (\ref{phd_1}) is perhaps especially interesting in the $T$-junction case. If such a junction could be tuned 
close to the critical point by adjusting the $t_i$'s with gate voltages, its behavior would become extremely 
sensitive to small changes in these gates, which would drive it away from the NTCP to $A\otimes N$ or $N\otimes A$.
As discussed in the next section, this behaviour could be measured from the low energy conductance through the junction. 
A slight detuning of the tunnelling amplitudes would produce zero conductance to one wire and perfect 
Andreev conductance to the other, at low energies.

\section{Conductance}
Conductance measurements might provide experimental observation of the NTCP found in the previous section. 
 Labelling the 3 arms of the $T$-junction by $1$ and $2$ for the two Luttinger liquid channels and $0$ 
for the superconducting quantum wire, we consider voltages $V_k$ applied to channel $k$ and currents 
$I_j$ flowing in arm $j$, towards the junction. The linear conductance tensor, $G_{jk}$ is then defined by:
\be I_j=\sum_{k=0}^2G_{jk}V_k.\ee
Conservation of charge and the condition that no current flows when all voltages are equal imply:
\be \sum_{j}G_{jk}=0=\sum_{k}G_{jk}=0.\label{sums}\ee

Let us 
first consider the case where the tunnelling amplitudes have been fine-tuned so that the system flows to the 
NTCP at low energies. Within our $\epsilon$-expansion, we may calculate the conductance at zero temperature, 
zero frequency and zero source-drain voltage, $V$, in second order 
perturbation theory in the tunnelling amplitudes, $t_i$, using their fixed point values $t_{ic}$. 
A closely related  calculation is done in \cite{KF}, Appendix A \cite{typos}. In that paper the bosonized tunnelling 
term was written $H_b=t\cos \sqrt{\pi}\phi$. Apart from a factor of 2 in the definition of $t$, the main 
difference in our case is the presence of the Majorana mode and Klein factor $\gamma \Gamma_i$. The perturbative calculation 
just requires calculating the 2-point Green's function of the tunnelling operator:
\be {\cal T}<\gamma (\tau )\Gamma_j(\tau)e^{i\sqrt{\pi}\phi_j(\tau )}
\gamma (0 )\Gamma_j(0)e^{-i\sqrt{\pi}\phi_j(0)}>.\ee
Using:
\be {\cal T}<\gamma (\tau )\gamma (0)>= {\cal T}<\Gamma_j (\tau )\Gamma_j (0)>=\epsilon (\tau ),\ee
the anti-symmetric step function, we see that the product of Majorana mode and Klein factor Green's function is unity leading to
\be {\cal T}<\gamma (\tau )\Gamma_j(\tau)e^{i\sqrt{\pi}\phi_j(\tau )}
\gamma (0 )\Gamma_j(0)e^{-i\sqrt{\pi}\phi_j(0)}>= - {1\over \tau^2}.\ee
Here we have set $\epsilon_j=0$, valid to lowest order. 
This corresponds to $n^2g=1$, $\lambda =2$, in the notation of \cite{KF}, Appendix A. 
Noting that the functions of \cite{KF} Appendix A take values $f_1(2)=f_3(2)=\pi$,  the current 
flowing to channel $j$ in response to a voltage 
difference $V_j$ is $I=G_{jc}V_j$, with:
\be G_{jc}=(e^2/h)( \pi t_{jc})^2.\label{Gjc}\ee
From Eq. (\ref{tfp}) this gives:
\bea G_{1c}&=&{e^2\pi^2\over h} {\epsilon_2\over {\cal F}(\nu )}\nonumber \\
G_{2c}&=&{e^2\pi^2\over h}{\epsilon_1\over {\cal F}(\nu )}.\label{GNT}\eea
It can readily be seen that, to $O(\epsilon_i)$, no current flows from channel 1 to channel 2 in linear response 
to a voltage difference. Following the method of  \cite{KF} and \cite{oca} for the $T$-junction case, this 
holds because the perturbative expansion of the partition function  does 
not contain a term of $O(t_1t_2)$. It then follows, using Eq. (\ref{sums}) that the full conductance tensor
at the non-trivial critical point, to $O(\epsilon_i)$ is:
\be G=\left(\begin{array}{ccc}
G_{1c}+G_{2c}&-G_{1c}&-G_{2c}\\
-G_{1c}&G_{1c}&0\\
-G_{2c}&0&G_{2c}
\end{array}\right).\label{GNTCP}\ee

So far we have failed to take into account the fact that the Luttinger liquid quantum wire (or wires) will be of finite length and may 
be contacted by Fermi liquid leads far from the SN junction. As discussed in \cite{lut_majo}, in the case of 
adiabatic contacts of the interacting quantum wires  to Fermi liquid leads, this may be 
modelled by  frequency dependent Luttinger parameters which have the values ($K_\rho$, $K_\sigma$) at 
higher frequencies but eventually cross over to the free Fermion values (1,1) at low frequencies, of order 
$\omega_{\ell}\equiv v_F/\ell$ where $\ell$ is the length of the wires.  We might then expect the results of Eq. (\ref{GNT}) to hold 
at frequencies above $\omega_\ell$ but the result for a Majorana SN junction to free fermion channels to hold 
below $\omega_\ell$.  This would correspond to the rotated $A\otimes N$ fixed point discussed above with 
total conductance, summed over both channels, of $G_{11}+G_{22}=2e^2/h$. 

\begin{figure}
  \center
\includegraphics*[width=0.7\linewidth]{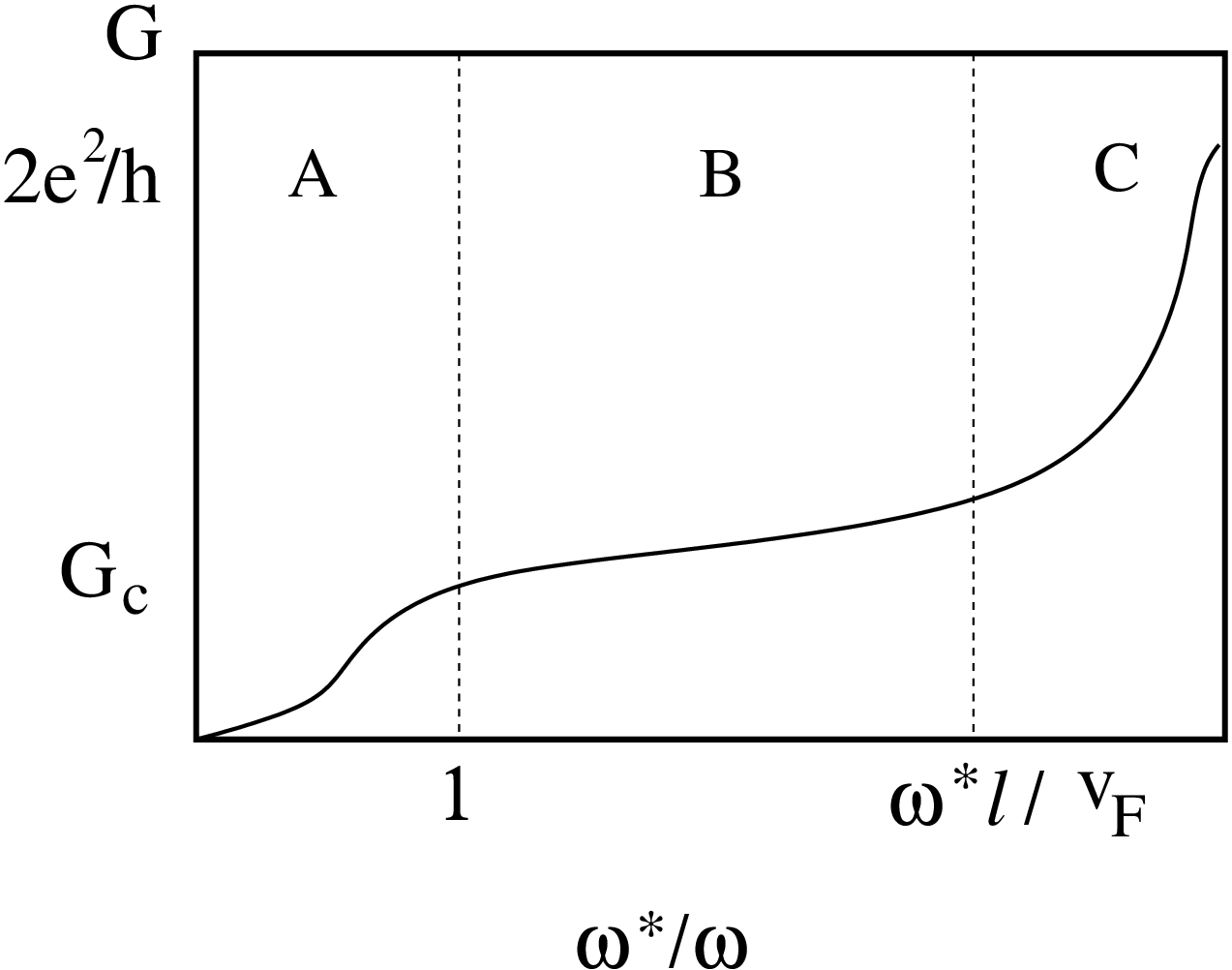}
\caption{Sketch of  the conductance $G$ as a function of $\omega^* / \omega$ across
the NTCP for  a junction made with wires of finite length $\ell$.
For $\omega^* / \omega \ll 1$, the conductance gets small and eventually
goes to 0 for   $\omega^* / \omega \to 0$, which is the appropriate 
behavior in the neighborhood of the $N \otimes N$-fixed point (region A of the plot).
For $1 \leq \omega^*  / \omega \leq \omega^* \ell / v_F$, the conductance 
takes a value of the order of $G_c$, showing that, in this region, the transport
properties of the junction are controlled by the NTCP (region B of the plot). Finally, 
for $\omega^* / \omega > \omega^* \ell / v_F$ the conductance flows towards
$2e^2 / h$, corresponding to the onset of the rotated $A \otimes N$-fixed point, due
to coupling of wires of finite length $\ell$ to Fermi liquid reservoirs. } 
\label{fig_Gcross}
\end{figure}

At frequencies above $\omega_\ell$, the conductance should exhibit a universal scaling behaviour as the NTCP is approached. 
For small $\epsilon$, and choosing, for simplicity, $\epsilon_1=\epsilon_2=\epsilon$
\be G_{jj}(\omega )\to G_{jc} \left\{ 1 -\left(\frac{\omega }{\omega^*} \right)^{2\epsilon}
\right\} \: , \label{corrw}\ee
where $\omega^*$ is a cross-over energy scale. [This is assuming weak bare tunnelling, $t$. For stronger tunnelling, larger 
than the critical value, 
the sign would be positive for the second term in Eq. (\ref{corrw}).] Using our exact solution for the RG flow 
at small $\epsilon$, in the limit of weak bare tunnelling, the full crossover is given by:
\be G_{jj}={e^2\pi^2\over h} {\epsilon \over {\cal F}(\nu )}{1\over 1+(\omega /\omega^*)^{2\epsilon}}\ee
approaching zero at $\omega$ much greater than the crossover scale, $\omega^*$ where the $N\otimes N$
 fixed point is approached. This behaviour is sketched in Fig. (\ref{fig_Gcross}) showing the crossover with 
 increasing frequency between the 3 different fixed points. 
 
 If the ratio of tunnelling amplitudes, $t_1/t_2$ is not perfectly tuned to 1 we expect more complicated crossover 
 behaviour with increasing $\omega$. As $\omega$ is raised the system goes from rotated $A\otimes N$ to 
 unrotated $A\otimes N$ to non-trivial to $N\otimes N$ fixed points.  Assuming, for simplicity, $\epsilon_1=\epsilon_2$, 
 and $t_1$ slightly bigger than $t_2$, the flow away from the NTCP is described by:
 \bea G_{11}&=&{e^2\pi^2\over h} {\epsilon \over {\cal F}(\nu )}[1+\delta t(\tilde{\omega}^*/\omega )^{2\epsilon}]\nonumber \\
 G_{22}&=&{e^2\pi^2\over h} {\epsilon \over {\cal F}(\nu )}[1-\delta t(\tilde{\omega}^*/\omega )^{2\epsilon}]
 \eea
 where $\delta t\equiv t_1-t_2$ and $\tilde{\omega}^*$ is another crossover scale. 

At the $A\otimes N$ fixed point the conductance tensor is given by \cite{lut_majo}
\be  G={K_12e^2\over h}\left(\begin{array}{rrr}
1&-1&0\\
-1&1&0\\
0&0&0
\end{array}\right).\label{CAN}\ee
Including adiabatic connections to Fermi liquid leads, the factor of $K_1$ in Eq. (\ref{CAN}) should be set to 1. 
At low but non-zero temperature the other components of the conductance are non-zero and are controlled 
by the leading irrelevant operators at the $A \otimes N$ fixed point
which can transport electrons between the superconductor and wire 2 or 
between wire 1 and wire 2. These are given in Eq. (\ref{SW}) and (\ref{p1p2}) respectively, leading to:
\bea G_{02}&\propto& T^{4/K_2-2}\nonumber \\
G_{12}&\propto& T^{K_1+1/K_2-2}.\eea

\section{More channels}
\begin{figure}
 \center
\includegraphics*[width=0.7\linewidth]{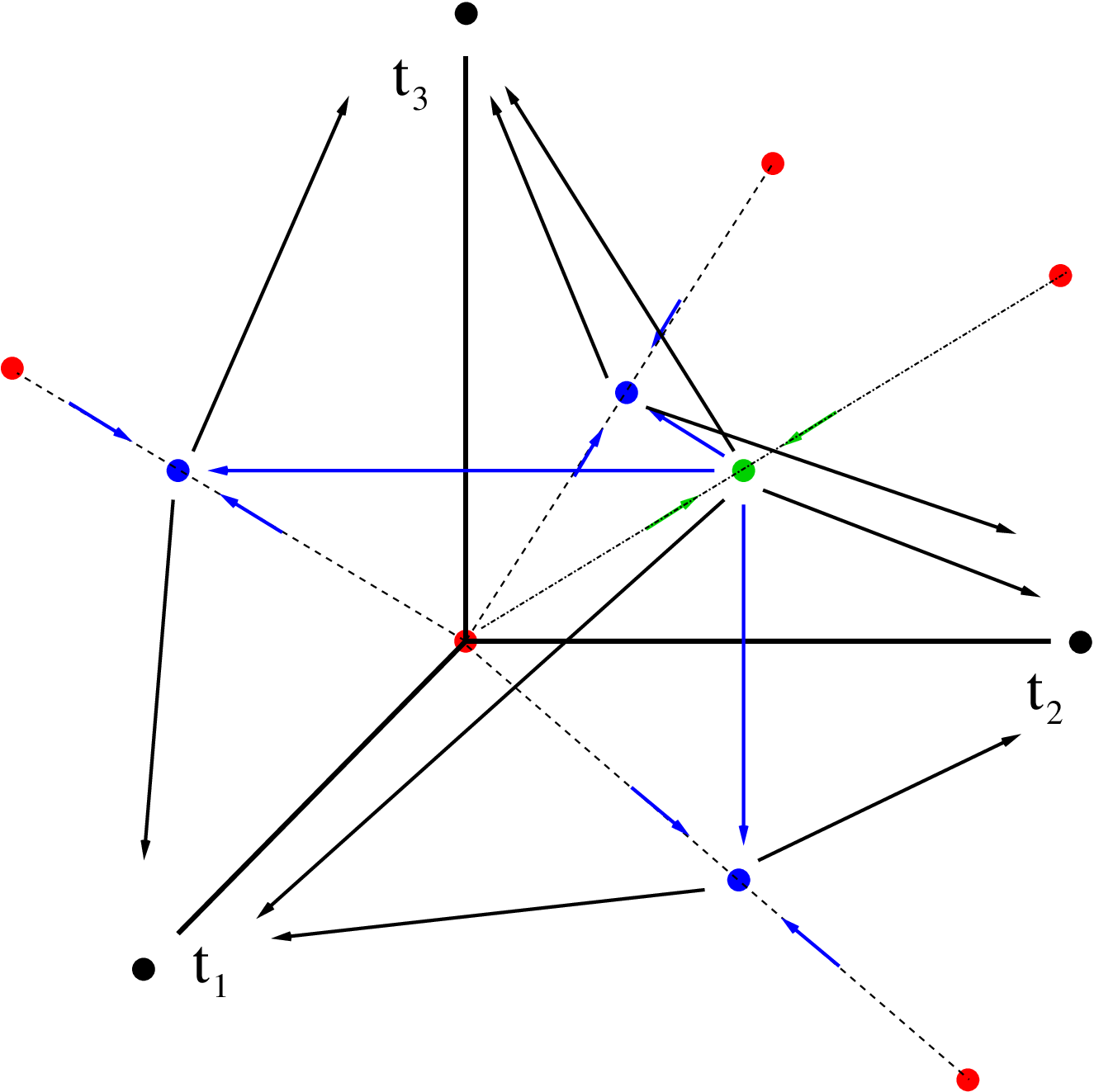}
\caption{Plot of the flows of the effective tunnelling amplitudes with decreasing energy scale, corresponding 
to Eq. (\ref{RG3}). There is a flow from completely unstable $N\otimes N\otimes N$ and $A\otimes A\otimes A$ 
critical points to 
the highly unstable critical point on the main diagonal (green dot) to more 
stable ones on the 3 planes (blue dots), to totally stable ones at $\infty$ along the axes (black dots).} 
\label{plot_3d}
\end{figure}

The  analysis of Sec. II may be straightforwardly generalized to the case of a single superconducting wire interacting 
with $k$ normal channels, corresponding, for example, to a T-junction with several channels in each normal branch. 
The crucial interaction between the single Majorana mode and the normal channels is again given by Eq. (\ref{Hb})  
with the sum now running over $k$ channels. We may again bosonize, introducing $k$ Klein factors. For a wide range of 
Luttinger parameters we expect the only stable RG fixed points to correspond to perfect Andreev reflection in 
one channel and perfect normal reflection in all the rest. This can be confirmed from the RG equations in 
the case where all tunnelling parameters, $t_i$, have RG scaling dimensions $d_i=1-\epsilon_i$, with $0<\epsilon_i\ll 1$.  Due to the Klein factors, 
it is easily seen that the RG equations, to cubic order, generally  take the form:
\be {dt_i\over dl}=\epsilon_it_i-t_i\sum_{j\neq i}{\cal F}_{ij}t_j^2\label{RGk}\ee
where the parameters $\epsilon_i$ and ${\cal F}_{ij}$ depend on the various intra-channel and inter-channel bulk interactions in the 
normal wires.  Again we see that these equations allow one tunnelling parameter, $t_i$,  to flow to strong coupling 
while the rest flow to zero, corresponding to the above fixed points. In addition, there will be various less 
stable non-trivial critical points. Consider, for example, the simplest case with 3 channels, all $\epsilon_i$ 
equal to a common value $\epsilon$ and all ${\cal F}_{ij}$ having 
the common value ${\cal F}$. Then Eq. (\ref{RGk}) reduces to:
\bea {dt_1\over dl}&=&\epsilon t_1-{\cal F}t_1[(t_2)^2+(t_3)^2]\nonumber \\
{dt_2\over dl}&=&\epsilon t_2-{\cal F}t_2[(t_1)^2+(t_3)^2]\nonumber \\
{dt_3\over dl}&=&\epsilon t_3-{\cal F}t_3[(t_1)^2+(t_2)^2].\label{RG3}
\eea
Without loss of generality, we may again assume all $t_i\geq 0$. Clearly there is an unstable critical point along the main diagonal at
\be t_1=t_2=t_3=\sqrt{\epsilon /2{\cal F}},\ee
In addition, there is an unstable critical point in the $t_1$-$t_2$ plane at
\be t_1=t_2=\sqrt{\epsilon  /{\cal F}},\ \  t_3=0, \ee
as well as 2 other equivalent critical points in the $t_2$-$t_3$ and $t_1$-$t_3$ planes. 
The RG flows are sketched in Fig. (\ref{plot_3d}). 2 parameters need to be fine-tuned, 
$t_1=t_2=t_3$, to 
stabilize the critical point on the main diagonal but only 1 parameter needs to 
adjusted, $t_1=t_2$, to stabilize the critical point in the $t_1$-$t_2$ plane. The only stable fixed points 
are along the axes at $t_1=\infty$, $t_2=t_3=0$, etc. A similar hierarchy of non-trivial critical points 
occurs in the general case. 
Thus we see that the Majorana mode  acts as a switch at low energies in a multi-channel T-junction, or in a 
junction with an arbitrary number of normal arms. 
Furthermore  we might expect a non-trivial critical point,  which can be stabilized by tuning a single tunnelling amplitude, 
 to exist in a multi-channel T-junction. 
 
 \vspace{0.2cm}

We would like to thank Hamed Karimi, Yashar Komijani, Eduardo Novais, Zheng Shi 
and Arturo Tagliacozzo for useful comments 
about this work; DG would like to thank the Department of Physics and Astronomy of the
University of British Columbia for the kind hospitality during the completion
of this work.  This research was supported by NSERC and CIfAR.

\appendix

\section{Bosonization of 2 channel model}
\label{bosonize}
The most general parity invariant quadratic Hamiltonian is:
\be {\cal H}={1\over 2}\sum_ju_j \left[K_j(\partial_x\phi_j)^2+K_j^{-1}(\partial_x\theta_j)^2\right] 
+U_\phi\partial_x\phi_1\partial_x \phi_2+U_
\theta \partial_x \theta_1\partial_x \theta_2.\label{H}\ee
Of primary interest is the case $U_\phi =0$, $U_\theta =U/\pi$ corresponding to a pure density-density interaction. 
This follows from
\be \rho_j(x)\equiv :\psi_{Rj}^\dagger \psi_{Rj}:+
:\psi_{Lj}^\dagger \psi_{Lj}:=-{1\over \sqrt{\pi}}\partial_x\theta_j.\ee
The continuum model:
\be {\cal H}=\sum_j[iv_{Fj}(\psi_{Rj}^\dagger \partial_x\psi_{Rj}-\psi_{Lj}^\dagger \partial_x\psi_{Lj})+V_j\rho_j^2]+U\rho_1\rho_2
\ee
bosonizes into Eq. (\ref{H}) with:
\be u_jK_j=v_{Fj},\ \  u_jK_j^{-1}=v_{Fj}+V_j/\pi ,\ \  U_\theta =U/\pi ,\ \  U_\phi =0.\label{uKV}\ee

The most general transformation we can make on $\vec \phi$ and $\vec \theta$ which is canonical, preserving the 
commutation relations:
\be [\phi_j(x),\theta_k(y)]=-{i\over 2}\epsilon (x-y),\ee
is:
\be \vec \phi = M\vec \phi ',\ \  \vec \theta =(M^{-1})^T\vec \theta '\label{transf}\ee
for an arbitrary real matrix $M$. 
A sufficiently general choice of $M$ to diagonalize the Hamiltonian when $U_\phi =0$ is:
\be M=\left(\begin{array}{cc}
r^{-1}\cos \alpha &r^{-1}\sin \alpha\\
-r\sin \alpha &r\cos \alpha 
\end{array} \right),\ \ (M^{-1})^T=\left(\begin{array}{cc}
r\cos \alpha &r\sin \alpha\\
-r^{-1}\sin \alpha &r^{-1}\cos \alpha 
\end{array} \right)\label{M}\ee
for an angle $\alpha$ and a real positive number $r$. 
The other two free parameters in $M$ correspond to rescaling $\phi_j'$s and $\theta_j'$s, equivalently rescaling the $K_j'$'s.
Requiring the off-diagonal terms in {\cal H} to vanish gives:
\be r^4={u_1K_1\over u_2K_2}\ee
and 
\be {\cos^2\alpha -\sin^2\alpha \over \cos \alpha \sin \alpha}={r^{-2}u_2K_2^{-1}-r^2u_1K_1^{-1}\over U/\pi}
\ee
implying
\be \tan 2\alpha =2{\sqrt{u_1u_2K_1K_2}U\over \pi (u_2^2-u_1^2)}.\label{alpha}\ee
The velocities and Luttinger parameters of the transformed fields are then determined by:
\bea u_1'K_1'&=&u_1K_1r^{-2}\cos^2\alpha +u_2K_2r^{2}\sin^2\alpha \\
u_1'(K_1')^{-1}&=&u_1K_1^{-1}r^{2}\cos^2\alpha+u_2K_2^{-1}r^{-2}\sin^2\alpha-2U\cos\alpha \sin \alpha /\pi\\
u_2'K_2'&=&u_1K_1r^{-2}\sin^2\alpha +u_2K_2r^{2}\cos^2\alpha \\
u_2'(K_2')^{-1}&=&u_1K_1^{-1}r^{2}\sin^2\alpha+u_2K_2^{-1}r^{-2}\cos^2\alpha +2U\cos\alpha \sin \alpha /\pi .\label{u'K'}
\eea
Defining:
\be \tilde U\equiv U\sqrt{K_1K_2}/\pi ,\ee
we have
\bea \cos 2\alpha &=& {u_2^2-u_1^2\over \sqrt{(u_1^2-u_2^2)^2+4u_1u_2\tilde U^2}}\nonumber \\
\sin 2\alpha &=& {2\sqrt{u_1u_2}\tilde U\over  \sqrt{(u_1^2-u_2^2)^2+4u_1u_2\tilde U^2}}.
\eea
Thus:
\bea u_{1,2}'(K_{1,2}')^{-1}&=&{1\over 2\sqrt{u_1u_2K_1K_2}}\left[u_1^2+u_2^2\mp \sqrt{(u_2^2-u_1^2)^2+4u_1u_2\tilde U^2}\right]\nonumber \\
u_{1,2}'K_{1,2}&=&\sqrt{u_1u_2K_1K_2}.\label{u'K'2}
\eea
One application of this transformation is to spinful fermions in a magnetic field.  The 
field makes $k_{F1}\neq k_{F2}$ thus eliminating relevant inter-channel backscattering interactions. It also makes 
$v_{F1}\neq v_{F2}$, so $u_1\neq u_2$ giving a 2-component Luttinger liquid exhibiting 
a generalization of spin-charge separation.

 From Eq. (\ref{u'K'2}), we see that $u_1'(K_1')^{-1}$  vanishes when
\be (u_2^2+u_1^2)^2=(u_2^2-u_1^2)^2+4u_1u_2\tilde U^2,\ee
implying 
\be \tilde U=\sqrt{u_1u_2}.\ee
Thus the 2 component Luttinger liquid phase is stable for
\be U\leq U_c\equiv \pi\sqrt{u_1u_2/(K_1K_2)}.\ee
This condition could actually be deduced directly from the Hamiltonian, Eq. (\ref{H}) (for $U_\theta =0$, $U_\phi =U$.) 
The terms depending on the $\theta_j$ are:
\be H_\theta = {1\over 2}(\partial_x \theta_1, \partial_x \theta_2)\left(\begin{array}{cc}
u_1K_1^{-1}&U/\pi \\
U/\pi &u_2K_2^{-1}
\end{array}\right)
\left( \begin{array}{c}
\partial_x \theta_1\\
\partial_x \theta_2
\end{array}\right).\ee
The condition for this to be positive definite is:
\be \hbox{det}\left(\begin{array}{cc}
u_1K_1^{-1}&U/\pi \\
U/\pi &u_2K_2^{-1}
\end{array}\right)=u_1u_2/(K_1K_2)-(U/\pi )^2>0.\ee

If we use the values for $u_j$, $K_j$ given in Eq. (\ref{uKV}), obtained from intra-chain interactions, Eq. (\ref{u'K'}) becomes:
\bea (u_{1',2'})^{2}&=&{1\over 2}\biggl[ v_{F1}(v_{F1}+V_1/\pi )+v_{F2}(v_{F2}+V_2/\pi )\nonumber \\
&&\mp \sqrt{[v_{F2}(v_{F2}+V_2/\pi )-v_{F1}(v_{F1}+V_1/\pi )]^2+4v_{F1}v_{F2} U^2/\pi}\biggr]\nonumber \\
(K_{1',2'})^{-2}&=&{1\over 2v_{F1}v_{F2}}\biggl[v_{F1}(v_{F1}+V_1/\pi )+v_{F2}(v_{F2}+V_2/\pi )\nonumber \\
&&\mp \sqrt{[v_{F2}(v_{F2}+V_2/\pi )-v_{F1}(v_{F1}+V_1/\pi )]^2+4v_{F1}v_{F2} U^2/\pi}\biggr].\nonumber \\ &&
\label{u'K'3}\eea
In the $SU(2)$ symmetric case, $v_{F1}=v_{F2}$, $V_1=V_2=U/2=V$, $r=1$, $\alpha =\pi /4$, Eq. (\ref{u'K'3}) reduces to:
\bea (u_{1',2'})^{2}&=&\left[v_{F}(v_{F}+V/\pi )
\mp v_FV/ \pi \right]\nonumber \\
(K_{1',2'})^{-2}&=&{1\over v_{F}}\left[v_{F}(v_{F}+V/\pi )
\mp  v_FV/ \pi  \right].\nonumber \\ &&
\label{u'K'4}\eea
In this case, we may identify $\theta_1'$ with the spin boson $\theta_\sigma$ and $\theta_2'$ with the charge boson, $\theta_\rho$. 
We see that the Luttinger parameter of the spin boson takes the free fermion value, $K_1'=K_\sigma =1$ in the 
$SU(2)$ symmetric case and the spin velocity is also unaffected by interactions. On the other hand the 
Luttinger parameter decreases, and the velocity increases for the charge boson. 

To understand what may happen when $U>U_c$, note that from Eqs. (\ref{u'K'2}),
$u_1'/K_1'<0$ in this case. The terms in the Hamiltonian involving the $\theta_j$ fields may be written in terms of the densities:
\bea \rho_1'&\equiv&r^{-1}\cos \alpha \rho_1-r\sin \alpha \rho_2\label{rho'} \\
\rho_2'&\equiv&r^{-1}\sin \alpha \rho_1+r\cos \alpha \rho_2\eea
as
\be H_\theta = {\pi\over 2} \sum_ju_j'(K_j')^{-1}(\rho_j')^2.\ee
When $u_1'(K_1')^{-1}<0$, it  
becomes necessary to include  terms of higher order in $\rho_i$ in the Hamiltonian for stability. 
The vanishing of the quadratic term may be associated with a phase transition.  
A mean field Landau theory analysis suggests, 
for equivalent channels corresponding to spinful fermions in zero magnetic field, 
 a continuous transition to a ferromagnetic state at $U=U_c$. For inequivalent channels corresponding to 
spinful fermions in a finite field, Landau theory predicts a first order transition,
  corresponding to a jump in the magnetization and density, at a value of $U$ less than the value $U_c$ where 
$u_1'(K_1')^{-1}$ vanishes.

\section{Renormalization group equations}
\label{RG}
In this Appendix we derive the cubic term in the RG equations of Eq. (\ref{rg2.19}), setting $\epsilon_i=0$.  Our technique for deriving RG equations is based on 
the Operator Product Expansion (OPE).  While derivation of quadratic terms in $\beta$-functions using the OPE is quite 
standard \cite{Cardy} 
we require a generalization of this technique involving unusual 3-point OPE's \cite{AL}. In addition, anti-commutation relations of Klein factors and the Majorana 
mode, Eq. (\ref{anticom}), play a crucial role. We begin with the boundary interaction, $H_b$,  of Eq. (\ref{Hbb}) with the bosonic ``vertex operators'' normalized 
as in Eq. (\ref{norm}).  Perturbation theory in the $t_i$  is ultra-violet divergent so a cut-off is necessary. 
Following \cite{Cardy}  the 
cut-off is defined by a ``hard core repulsion'' in perturbation theory. The $n^{\hbox{th}}$ order term in perturbation theory for the partition function has the form:
\be Z_n/Z_0={(-1)^n\over n!}\int_{-\beta /2}^{\beta /2} d\tau_1 d\tau_2\ldots d\tau_n{\cal T}<H_b(\tau_1)H_b(\tau_2)\ldots H_b(\tau_n)>.\label{nth}\ee
Here we work in imaginary time and ${\cal T}$ represents time-ordering while $\beta$ is the inverse temperature. 
We will eventually take $\beta \to \infty$; it just acts as an infrared regulator in the calculation. Below we use 
the zero temperature Green's functions with that limit in mind. 
We use the fact that the interaction term in the imaginary time action is simply $S_{\hbox{int}}=\int d\tau H_b(\tau )$. 
We may think of $<\ldots >$ as representing a Feynman path integral 
over the bosonic fields.  On the other hand, it is more convenient to take the elementary traces over the Majorana mode and Klein factors directly. 
Ultra-violet divergences occur when two or more of the $\tau_i$ approach each other.  Our ultra-violet cut-off is to restrict the integration in Eq. (\ref{nth})  
by requiring $|\tau_i-\tau_j|\geq \tau_0$ for all $i\neq j$.  Our basic RG step is to increase the short-time cut-off:
\be \tau_0\to \tau_0+\delta \tau \ee
corresponding to reducing a cut-off in energy domain. We study how the renormalized parameters $t_i$ change under this increase of $\tau_0$. This is 
done using the 3-point OPE of boundary operators:
\be {\cal T}\left[ H_b(\tau_1)H_b(\tau_2)H_b(\tau_3)\right] \to \sum_{n=1}^\infty f_n(\tau_2-\tau_1,\tau_3-\tau_1)O_n(\tau_1).\label{OPE}\ee
Here the limit is taken where $\tau_1$, $\tau_2$ and $\tau_3$ approach each other. The 
operators ${\cal O}_n$ on the right hand side of Eq. (\ref{OPE}) are a complete set of boundary operators and 
the $f_n$'s are a set of functions, cubic in the $t_j$'s, which generally become singular when any two $\tau_i$'s become equal.  
We insert this expansion into the cubic term in  Eq. (\ref{nth}), obtaining a correction to the effective action. 
We examine how this correction to the action changes under a small change in $\tau_0$,  
extract the term linear in $\delta \tau$  and from this obtain the renormalization of the coupling constants $t_i$ 
which is cubic in the $t_i$'s and linear in $\delta \tau$.  In doing this, we must be careful to subtract  a 
term which is simply $H_b$ times a perturbation to the 
free energy of quadratic order in $H_b$.  Thus we may write, in cubic order:
\begin{eqnarray} \delta S &=&{1\over 3!}
\int d\tau_1d\tau_2d\tau_3 \sum_{n=1}^\infty f_n (\tau_2-\tau_1,\tau_3-\tau_1)O_n(\tau_1) \nonumber \\
&-&\int d\tau_1H_b(\tau_1){1\over 2!}\int d\tau_2d\tau_3{\cal T}<H_b(\tau_2)H_b(\tau_3)>.\label{dSE}
\end{eqnarray}
\noindent
The last term in Eq. (\ref{dSE}) cancels many of the contributions from the first one. 

The OPE in Eq. (\ref{OPE}) factorizes into contributions from the Majorana modes, Klein factors and bosonic operators. 
Using the defining anti-commutation relations of Eq. (\ref{anticom}), we obtain for the Majorana modes:
\be {\cal T}\left[ \gamma (\tau_1)\gamma (\tau_2)\gamma (\tau_3)\right]=\epsilon (\tau_1,\tau_2,\tau_3)\gamma \label{gOPE}\ee
where $\epsilon$ is the anti-symmetric step function of 3 arguments:
\bea \epsilon (\tau_1,\tau_2,\tau_3)&=&1,\  \  (\tau_1>\tau_2>\tau_3)\nonumber \\
&=&-1\ \  (\tau_2>\tau_1>\tau_3)\label{Tgamma}\eea
et cetera. Note that the operator $\gamma$ doesn't actually have any time dependence in perturbation theory 
since it doesn't appear in the unperturbed Hamiltonian. The time-dependence in Eq. (\ref{Tgamma}) arises from 
the time-ordering due to the anti-commutation relations. Similarly:
\be {\cal T}\left[ \Gamma_1 (\tau_1)\Gamma_1 (\tau_2)\Gamma_1 (\tau_3)\right] =\epsilon (\tau_1,\tau_2,\tau_3)\Gamma_1.
\label{111} \ee
By contrast:
\be {\cal T}\left[ \Gamma_1 (\tau_1)\Gamma_1 (\tau_2)\Gamma_2 (\tau_3)\right] =\epsilon (\tau_1-\tau_2)\Gamma_2
\label{112} \ee
where $\epsilon (\tau )$ is the standard anti-symmetric step function.  Of course, the analogous equations hold 
with the indices $1$ and $2$ interchanged. To calculate the 3-point OPE of bosonic operators, we first consider 
the 4-point function for a boson field $\phi (\tau)\equiv \phi (\tau ,x=0)$, with the Hamiltonian 
\be  H=\frac{u}{2 } \: 
\int_0^\infty \: d x \: \left[ K \left( \frac{\partial \phi}{\partial x} 
\right)^2 + K^{-1}\left( \frac{\partial \theta}{\partial x} 
\right)^2\right].\ee
Using standard Gaussian path integration techniques and the normalization of Eq. (\ref{norm}), this is:
\be {\cal T}<e^{i\sqrt{\pi}\phi (\tau_1)}e^{i\sqrt{\pi}\phi (\tau_2)}e^{-i\sqrt{\pi}\phi (\tau_3)}
e^{-i\sqrt{\pi}\phi (\tau_4)}>=\left|{\tau_{12}\tau_{34}\over \tau_{13}\tau_{14}\tau_{23}\tau_{24}}\right|^{1\over K}\label{4pt}
\ee
where we have defined, for convenience,
\be \tau_{ij}\equiv \tau_i-\tau_j.\ee
We now consider the limit $\tau_4\to \infty$ to extract the 3-point OPE. In this 
limit, Eq. (\ref{4pt}) becomes:
\bea & {\cal T} & <e^{i\sqrt{\pi}\phi (\tau_1)}e^{i\sqrt{\pi}\phi (\tau_2)}e^{-i\sqrt{\pi}\phi (\tau_3)}
e^{-i\sqrt{\pi}\phi (\tau_4)}> \to  
\left|{\tau_{12}\over \tau_{13}\tau_{23}}\right|^{1\over K}\left|{1\over \tau_4}\right|^{1\over K}\nonumber \\
&=&
\left|{\tau_{12}\over \tau_{13}\tau_{23}}\right|^{1\over K}<e^{i\sqrt{\pi}\phi ( \tau_1 )}e^{-i\sqrt{\pi}\phi (\tau_4)}>
.
\eea
Thus we deduce the basic OPE:
\be{\cal T}\left[ e^{i\sqrt{\pi}\phi (\tau_1)}e^{i\sqrt{\pi}\phi (\tau_2)}e^{-i\sqrt{\pi}\phi (\tau_3)}\right] \to 
\left|{\tau_{12}\over \tau_{13}\tau_{23}}\right|^{1\over K}e^{i\sqrt{\pi}\phi (\tau_1)}.\label{bOPE}\ee
This result holds up to higher dimensions operators which are also generated in the effective action, but are not of 
interest to us.  Note that whether the argument of $\phi$ on the right hand side is chosen to be $\tau_1$, $\tau_2$ or $\tau_3$
is immaterial, neglecting these higher dimension operators. 
Eq. (\ref{bOPE}) implies the useful result:
\bea &&{\cal T}\{\cos [\sqrt{\pi}\phi (\tau_1)]\cos [\sqrt{\pi}\phi (\tau_2)]\cos [\sqrt{\pi}\phi (\tau_3)]\}
\nonumber \\ &&
\to {1\over 4}\cos [\sqrt{\pi}\phi (\tau_1)]\left[\left|{\tau_{12}\over \tau_{13}\tau_{23}}\right|^{1\over K}
+\left|{\tau_{13}\over \tau_{12}\tau_{23}}\right|^{1\over K}+\left|{\tau_{23}\over \tau_{12}\tau_{13}}\right|^{1\over K}
\right].\label{O111}\eea
For two independent bosons, $\phi_1$ and $\phi_2$ with the bulk Hamiltonian of Eq. (\ref{H0j}) we use the trivial result:
\be {\cal T}\left\{ \cos [\sqrt{\pi}\phi_1 (\tau_1)]\cos [\sqrt{\pi}\phi_2 (\tau_2)]\cos [\sqrt{\pi}\phi_2 (\tau_3)]\right\} \to
{1\over 2}\left|{1\over \tau_{23}}\right|^{1\over K_2}\cos [\sqrt{\pi}\phi_1 (\tau_1)]\label{O122}\ee
up to higher dimension operators.  We can extend these results immediately to the general case discussed in Sec. II 
with inter-channel coupling. Using the fact that $\phi_\rho$ and $\phi_\sigma$ commute, we simply factorize 
all exponentials of sums:
\be \exp [i(a\phi_\rho +b\phi_\sigma )]=\exp [ia\phi_\rho ]\exp [ib\phi_\sigma ]\ee
and use Eqs. (\ref{O111}) and (\ref{O122}) with $\phi_1$ and $\phi_2$ replaced by $\phi_\rho$ and $\phi_\sigma$
and $K_1$ and $K_2$ replaced by $K_\rho$ and $K_\sigma$. 
Note that the bosonic OPE's all give positive functions of the $\tau_i$'s whereas the fermonic 
ones give factors of $\pm 1$, contributing crucial minus signs. 

Let us begin with the T-junction case, of decoupled channels. In this case, since we are taking the limit $\epsilon_i=0$ 
for this calculation, we have $1/K_1=1/K_2=2$. 
It  can be seen that the cubic $\beta$-functions 
contain no terms proportional to $t_1^3$ or $t_2^3$. To obtain this result, first note that  Eqs. (\ref{Tgamma}) and (\ref{111})  imply 
the time-independent result:
\be 
{\cal T}\left[ \gamma (\tau_1)\Gamma_1 (\tau_1)\gamma (\tau_2)\Gamma_1 (\tau_2)
\gamma (\tau_3)\Gamma_1 (\tau_3)\right]=-\gamma \Gamma_1
.\ee
Inserting Eq. (\ref{O111}) into Eq. (\ref{dSE}) then gives the term   in $\delta S$ which is cubic in $t_1$:
\begin{eqnarray}  \delta S^{111}&=& -it_1^3\int d\tau_1\gamma \Gamma_1\cos [\sqrt{\pi}\phi_1(\tau_1)]
\int d\tau_2d\tau_3\biggl\{ {1\over 3!}2\biggl[ \left|{\tau_{12}\over \tau_{13}\tau_{23}}\right|^2 \nonumber \\
&+&\left|{\tau_{13}\over \tau_{12}\tau_{23}}\right|^2+\left|{\tau_{23}\over \tau_{12}\tau_{13}}\right|^2\biggr]
-2\left|{1\over \tau_{23}}\right|^2\biggr\}
\end{eqnarray}
\noindent
Now we use the remarkable identity
\be \left({\tau_{12}\over \tau_{13}\tau_{23}}\right)^2+\left({\tau_{13}\over \tau_{12}\tau_{23}}\right)^2
+\left({\tau_{23}\over \tau_{12}\tau_{23}}\right)^2=2\left[\left({1\over \tau_{12}}\right)^2
+\left({1\over \tau_{13}}\right)^2+\left({1\over \tau_{23}}\right)^2\right].\label{id}
\ee
This can be verified by inspection but can be understood in a deeper way explained in \ref{spin}. 
Using Eq. (\ref{id}) we may write:
\begin{eqnarray} 
\delta S^{111} &=&-{2it_1^3\over 3}\int d\tau_1\gamma \Gamma_1\cos [\sqrt{\pi}\phi_1(\tau_1)]
\int d\tau_2d\tau_3\biggl[\left({1\over \tau_{12}}\right)^2
+\left({1\over \tau_{13}}\right)^2\nonumber \\
&-&2\left({1\over \tau_{23}}\right)^2\biggr]\theta (|\tau_{12}|-\tau_0)
\theta (|\tau_{23}|-\tau_0)\theta (|\tau_{13}|-\tau_0)
\end{eqnarray}
\noindent
where in the last line we have inserted explicitly the ultra-violet cut-off which was previously not 
explicitly written.  [$\theta (\tau )$ is the Heaviside step function.] The change in $t_1$ under a small change in the cut-off, $\tau_0$ is proportional to:
\be {d\over d\tau_0}\int d\tau_2d\tau_3\left[\left({1\over \tau_{12}}\right)^2
+\left({1\over \tau_{13}}\right)^2-2\left({1\over \tau_{23}}\right)^2\right]\theta (|\tau_{12}|-\tau_0)
\theta (|\tau_{23}|-\tau_0)\theta (|\tau_{13}|-\tau_0).
\ee
Using:
\be {d\over d\tau_0}\theta (|\tau |-\tau_0)=-\sum_{\pm}\delta (\tau_0\pm \tau )\ee
this derivative is a sum of 3 terms, from differentiating each of the step functions. For example, the contribution form 
differentating $\theta (|\tau_{12}|-\tau_0)$ is:
\begin{eqnarray} &-&\sum_{\pm}\int d\tau_3\left[\left({1\over \tau_0}\right)^2+\left({1\over \tau_{13}}\right)^2-
2\left({1\over  \tau_{13}\pm \tau_0}\right)^2\right]\theta (|\tau_{13}\pm \tau_0|-\tau_0)\theta (|\tau_{13}|-\tau_0)
\nonumber \\
&\approx& -{2\beta \over \tau_0^2}+{3\over \tau_0}\end{eqnarray} \noindent
 up to negligible terms of O($\tau_0/\beta $).
Adding together the contributions from differentiating the 3 step functions, we obtain zero (up to negligible terms).

Now let us consider the term in $\delta S$ proportional to $t_1t_2^2$. The bosonic part of the calculation is 
much simpler, using Eq. (\ref{O122}). There is an essential complication in the fermion part however:
\begin{eqnarray} {\cal T}\left[\gamma (\tau_1)\Gamma_1(\tau_1)\gamma (\tau_2)\Gamma_2(\tau_2)\gamma 
(\tau_3)\Gamma_2(\tau_3)\right]
&=&-\epsilon (\tau_1,\tau_2,\tau_3)\epsilon (\tau_2-\tau_3)\gamma \Gamma_1 \nonumber \\
&=&-\epsilon (\tau_1-\tau_2)\epsilon (\tau_1-\tau_3)
\gamma \Gamma_1.
\end{eqnarray}\noindent
Thus we obtain
\be \delta S^{122}=-i{4t_1t_2^2\over 2!}\int d\tau_1\gamma \Gamma_1\cos [\sqrt{\pi}\phi_1(\tau_1)]
\int d\tau_2d\tau_3\left({1\over \tau_{23}}\right)^2[\epsilon (\tau_1-\tau_2)\epsilon (\tau_1-\tau_3)-1].\label{dS112}\ee
Note this is non-zero only due to the sign functions coming from the Majorana mode and Klein factors. We may rewrite this as:
\be \delta S^{122}=i8t_1t_2^2\int d\tau_1\gamma \Gamma_1\cos [\sqrt{\pi}\phi_1(\tau_1)]
\int_{-\beta /2}^{\tau_1-\tau_0}d\tau_2\int_{\tau_1+\tau_0}^{\beta /2} d\tau_3\left({1\over \tau_{23}}\right)^2.\ee
Up to terms which are negligible at $\beta \to \infty$, $\tau_0\to 0$, this becomes:
\be \delta S^{122}=i8t_1t_2^2\ln (\beta /\tau_0)\int d\tau_1\gamma \Gamma_1\cos [\sqrt{\pi}\phi_1(\tau_1)].\ee
Thus:
\be {d\over d\tau_0}\delta S^{122}=-i{8t_1t_2^2\over \tau_0}\int d\tau_1\gamma \Gamma_1\cos [\sqrt{\pi}\phi_1(\tau_1)].\ee
This determines the cubic term in the $\beta$-function:
\be {dt_1\over d\ln \tau_0}=-4t_1t_2^2.\ee
Referring to Eqs. (\ref{nu},\ref{rg2.19}), we see that we have proven ${\cal F}(0)=4$. 

\begin{figure}
 \center
\includegraphics*[width=0.7\linewidth]{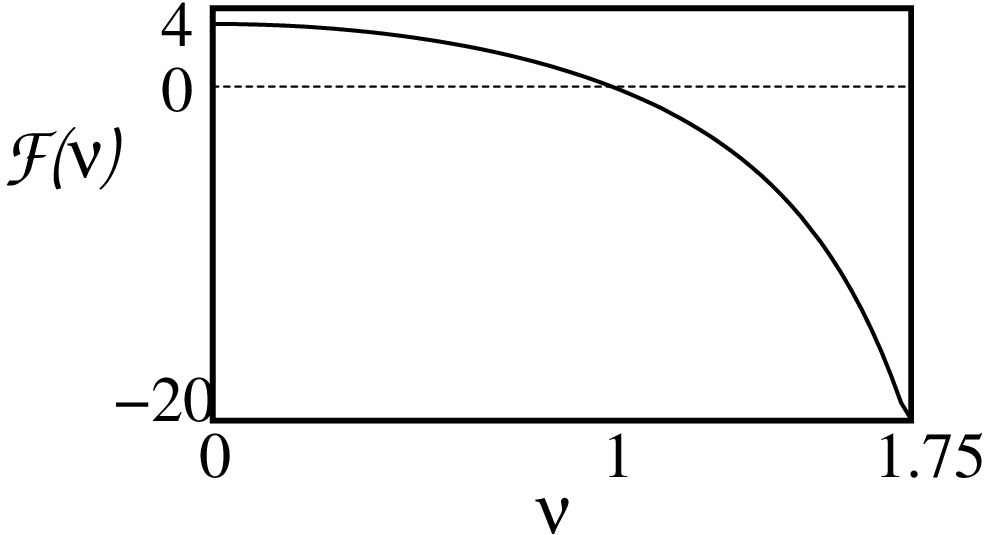}
\caption{Plot of the function $F(\nu )$ appearing in the $\beta$ function of Eq. (\ref{rg2.19}).} 
\label{fig_nu}
\end{figure}

We now consider the general case, discussed in Sec. II. Using Eq. (\ref{phal}), we have:
\bea&& {\cal T}\left[ 8\cos [\sqrt{\pi}\phi_1(\tau_1)]\cos [\sqrt{\pi}\phi_2(\tau_2)]\cos [\sqrt{\pi}\phi_2(\tau_3)]
\right]\nonumber \\ &&
=\sum_{s_i=\pm 1}{\cal T} \exp \left\{  i\sqrt{\pi}\left[  s_1r^{-1}\cos \alpha \phi_\sigma (\tau_1)-ir\sin \alpha [s_2\phi_\sigma (\tau_2)
+s_3\phi_\sigma (\tau_3)]\right] \right\} \nonumber \\ &&
\times {\cal T}\exp \left\{  i\sqrt{\pi}\left[ s_1r^{-1}\sin \alpha \phi_\rho (\tau_1)+ir\cos \alpha [s_2
\phi_\rho (\tau_2)+s_3\phi_\rho (\tau_3)]\right] \right\}  .
\eea
Only the terms with $s_3=-s_2$ will renormalize the original couplings. Then we use the OPE's, derived as above:
\bea &{\cal T}& \exp \left\{  i\sqrt{\pi}\left[  s_1r^{-1}\cos \alpha \phi_\sigma (\tau_1)-ir\sin \alpha [s_2\phi_\sigma (\tau_2)
+s_3\phi_\sigma (\tau_3)]\right] \right\} \nonumber \\
&\to& \exp [is_1r^{-1}\cos \alpha \sqrt{\pi}\phi_\sigma (\tau_1)]
\left|{\tau_{13}\over \tau_{12}}\right|^{s_1s_2\cos \alpha \sin \alpha /K_\sigma}\left|{1\over 
\tau_{23}}\right|^{r^2\sin^2\alpha /K\sigma}
\eea
and
\bea
&{\cal T}&\exp \left\{  i\sqrt{\pi}\left[ s_1r^{-1}\sin \alpha \phi_\rho (\tau_1)+ir\cos \alpha [s_2
\phi_\rho (\tau_2)+s_3\phi_\rho (\tau_3)]\right] \right\} \nonumber \\&\to&
\exp [is_1r^{-1}\sin \alpha \sqrt{\pi}\phi_\rho (\tau_1)]\left|{\tau_{12}\over \tau_{13}}\right|^{s_1s_2\cos \alpha 
\sin \alpha /K_\rho}\left|{1\over \tau_{23}}\right|^{r^2\cos^2\alpha /K\rho}.
\eea
Using Eqs. (\ref{marg}), we obtain:
\bea &{\cal T}& \left[ 8\cos [\sqrt{\pi}\phi_1(\tau_1)]\cos [\sqrt{\pi}\phi_2(\tau_2)]\cos [\sqrt{\pi}\phi_2(\tau_3)]\right]
\nonumber \\ 
&\to& 2\cos  [\sqrt{\pi}\phi_1(\tau_1)]\left({1\over \tau_{23}}\right)^2
\left[\left|{\tau_{12}\over \tau_{13}}\right|^\nu +\left|{\tau_{13}\over \tau_{12}}\right|^\nu \right]\label{dS122}
\eea
where $\nu$ is defined in Eq. (\ref{nu}). Therefore Eq. (\ref{dS112}) is modified to:
\bea \delta S^{122}&=&-i{2t_1t_2^2\over 2!}\int d\tau_1\gamma \Gamma_1\cos [\sqrt{\pi}\phi_1(\tau_1)]
\int d\tau_2d\tau_3\left({1\over \tau_{23}}\right)^2 \nonumber \\
&\times& \biggl[\biggl(\left|{\tau_{12}\over \tau_{13}}\right|^\nu 
+ \left|{\tau_{13}\over \tau_{12}}\right|^\nu \biggr)\epsilon (\tau_1-\tau_2)\epsilon (\tau_1-\tau_3)-2\biggr].\eea
This gives the $\beta$-functions of Eq. (\ref{rg2.19}) with
\bea {\cal F}(\nu )&=&{\tau_0 \over 2}{d\over d\tau_0}
\int d\tau_2d\tau_3\left({1\over \tau_{23}}\right)^2\left[\left(\left|{\tau_{12}\over \tau_{13}}\right|^\nu 
+\left|{\tau_{13}\over \tau_{12}}\right|^\nu \right)\epsilon (\tau_1-\tau_2)\epsilon (\tau_1-\tau_3)-2\right] \nonumber \\
&\times&
\prod_{i<j}\theta (|\tau_{ij}|-\tau_0).\label{int}\label{F}\eea
(Note that, by time-translation invariance, this double integral is independent of $\tau_1$. 
We will show that this object becomes independent of $\tau_0$ and $\beta$ in the limit $\tau_0/\beta \to 0$, that we are considering.)
Before evaluating $F(\nu )$ in general, it is interesting to consider the SU(2) invariant case, $\nu =1$. 
Then we can use:
\be \left(\left|{\tau_{12}\over \tau_{13}}\right|
+\left|{\tau_{13}\over \tau_{12}}\right| \right)\epsilon (\tau_1-\tau_2)\epsilon (\tau_1-\tau_3)-2=
{\tau_{12}\over \tau_{13}}+{\tau_{13}\over \tau_{12}}-2={\tau_{23}^2\over \tau_{12}\tau_{13}}
\ee
to reduce Eq. (\ref{int}) to:
\be {\cal F}(1)={\tau_0\over 2}{d\over d\tau_0}
\int d\tau_2d\tau_3 {1\over \tau_{12}\tau_{13}}.\ee
This is a product of 2 ultraviolet finite principal value integrals and is consequently independent of the cut-off 
up to corrections that vanish when $\tau_0\ll \beta$. Therefore, ${\cal F}(1)=0$, a result that must hold 
due to SU(2) symmetry as discussed in Sec. II.  To simplify Eq. (\ref{F}) in general, we first rewrite it as:
\bea {\cal F}(\nu )&=&\tau_0 {d\over d\tau_0}\biggl\{ \int_{\tau_0}^{\beta /2}{d\tau_2d\tau_3\over \tau_{23}^2}
\left[\left({\tau_2\over \tau_3}\right)^\nu+\left({\tau_3\over \tau_2}\right)^\nu-2\right]\theta (|\tau_{23}|-\tau_0)
\nonumber \\
&-&\int_{\tau_0}^{\beta /2}{d\tau_2d\tau_3\over (\tau_{2}+\tau_3)^2}
\left[\left({\tau_2\over \tau_3}\right)^\nu+\left({\tau_3\over \tau_2}\right)^\nu+2\right]\biggr\}.\label{F2}
\eea
Differentiating the explicit step function in Eq. (\ref{F2}) gives:
\bea &-& {2\over \tau_0}\int_{\tau_0}^{\beta /2}d\tau_2\left[\left({\tau_2+\tau_0\over \tau_2}\right)^\nu 
+\left({\tau_2\over \tau_2+\tau_0}\right)^\nu -2\right] \nonumber \\
&\to& -2\int_1^\infty dx\left[
\left({x+1\over x}\right)^\nu +\left({x\over x+1}\right)^\nu -2\right] \eea
where we have rescaled the integration variable $\tau =\tau_0x$ and taken $\beta \to \infty$ 
in the last expression. Differentiating the lower limits of integration in Eq. (\ref{F2}) gives:
\bea &-&2\tau_0\int_{2\tau_0}^{\beta /2}{d\tau_2\over (\tau_2-\tau_0)^2}\left[ 
\left({\tau_2\over \tau_0}\right)^\nu + \left({\tau_0\over \tau_2}\right)^\nu 
-2\right] \nonumber \\
&+& 2\tau_0\int_{\tau_0}^{\beta /2}{d\tau_2\over (\tau_2+\tau_0)^2}\left[ 
\left({\tau_2\over \tau_0}\right)^\nu + \left({\tau_0\over \tau_2}\right)^\nu 
+2 \right]
\eea
Doing the simple  integrals exactly, again rescaling the integration variable, taking $\beta \to \infty$,  and collecting terms gives
an expression for the needed function $F(\nu )$ in terms of a convergent dimensionless integral:
\bea {\cal F}(\nu )&=& 6-2\int_1^\infty dx\biggl[{(x+1)^\nu +(x+1)^{-\nu}\over x^2}-{x^\nu +x^{-\nu}\over (x+1)^2}
\nonumber \\ &+& \left({x+1\over x}\right)^\nu +\left({x+1\over x}\right)^{-\nu}-2
\biggr].
\label{fnu}
\eea
We did the integral numerically 
and ${\cal F}(\nu )$ is plotted in Fig. (\ref{fig_nu}). Note that it is monotone decreasing, with ${\cal F}(0)=4$ and ${\cal F}(1)=0$ 
as calculated analytically above. 

\section{Spin chain representation}
\label{spin}
A mapping of a tight binding version of our model onto a spin chain Hamiltonian provides intuition about the phase diagram 
as well as a useful representation for possible future numerical work. This mapping generalizes 
an approach introduced in  	\cite{lut_majo} for the single channel case.  We first represent each channel by a semi-infinite spinless fermion tight-binding chain and represent the 
two Majorana modes of the superconductor, $\psi_S$ of Eq. (\ref{psiS}), by the electron operator at the origin, $c_0$. For 
the T-junction (no inter-channel interactions) the various terms in the Hamiltonian of Eq. (\ref{HT}) become:
\bea H_0&=&-(J_1/2)\sum_{j=-\infty}^{-2}c^\dagger_jc_{j+1}-(J_2/2)\sum_{j=1}^{\infty}c^\dagger_jc_{j+1} + {\rm h.c.} \nonumber \\
H_{\hbox{int}}&=&V_1\sum_{j=-\infty}^{-1}(n_j-1/2)(n_{j-1}-1/2)+V_2\sum_{j=1}^{\infty}(n_j-1/2)(n_{j+1}-1/2)\nonumber \\
H_b&=&-t_1(c^\dagger_0+c_0)(c_{-1}^\dagger -c_{-1})/2-t_2(c^\dagger_0+c_0)(c_{1}^\dagger -c_{1})/2
\label{bione}
.\eea
We now make an ``inverse Jordan-Wigner transformation'' to a set of S=1/2 variables on each lattice site:
\beq c^\dagger_jc_j = S^z_j+1/2 \;\; , \;
c_j =\left[ \prod_{l<j}(-2S^z_l)\right] S^-_j
\:\:\:\: . 
\label{remib.3}
\eneq
The resulting Hamiltonian in terms of spin-1/2 variables, after the transformation $S^\pm_j\to (-1)^jS^\pm_j$ is:
\bea H_0+H_{\hbox{int}}&=&
\sum_{j=-\infty}^{-2}[J_1(S^x_jS^x_{j+1}+S^y_jS^y_{j+1})+V_1S^z_jS^z_{j+1}] \nonumber \\
&+&\sum_{j=1}^{\infty}[J_2(S^x_jS^x_{j+1}+S^y_jS^y_{j+1})+V_2S^z_jS^z_{j+1}]
\nonumber \\
H_b&=&-2t_{1}S_0^xS_{-1}^x-2t_2S_0^yS_{1}^y.\label{Htb}
\eea
We see that, when $t_1=t_2=0$, we have 2 independent xxz model defined on semi-infinite lines, $j>0$ and $j<0$
with the impurity spin at the origin decoupled. Each spin chain obeys ``open boundary conditions'' at the origin. 
Turning on $t_1$, the chain at $j>0$ remains 
decoupled and $[S^x_0,H]=0$. As discussed in 	\cite{lut_majo}, there are 2 equivalent ground states 
in which $S^x_0=\pm 1/2$.  The spin chain defined at $j<0$ then experiences a boundary magnetic field pointing 
in the $\pm x$ direction:
\be  B_x=\pm t_1.\ee
This transverse boundary field model was analyzed in \cite{Ian_1}. 
Along the  xxz critical line, $0<V_1/J_1<1$, this is a relevant boundary    interaction. To see this we may bosonize the two spin chains, introducing bosons $(\phi_j,\theta_j)$ with the
 bulk Hamiltonian of Eq. (\ref{H0j}) and boundary terms:
\bea H_b &=& 2t_1 \tau_0^{-1+d_1} S^x_0\left\{\exp [i\sqrt{\pi}\phi_1(0)]+\exp [-i\sqrt{\pi}\phi_1(0)]\right\} 
\nonumber \\
&+&
2t_2 \tau_0^{-1+d_2}  S^y_0\left\{ \exp [i\sqrt{\pi}\phi_2(0)]+
\exp [-i\sqrt{\pi}\phi_2(0)]\right\} .\label{Hbbs}
\eea
This Hamiltonian could have been obtained directly from Eq. (\ref{Hbb}) by the substition:
\bea i\gamma \Gamma_1&\to& 2S^x_0\nonumber \\
i\gamma \Gamma_2&\to& 2S^y_0.\eea
The conclusion of \cite{Ian_1} was that, for $t_2=0$, the relevant interaction $t_1$ renormalizes 
to large values, pinning $\phi_1(0)$.  $<S^x_j>$ then becomes non-zero, with power-law decay away from the boundary. 
This boundary field fixed point of the spin chain corresponds to perfect Andreev reflection in the fermion 
system an analogy that was discussed  in \cite{ACZ}. When $t_1$ and $t_2$ are both non-zero, 
there is a competition between $S^x_0$ and $S^y_0$ being pinned. This frustration leads to the 
non-trivial critical point discussed in Sec. II. 

The case of marginal boundary Hamiltonian, $d_i=1$, $K_j=1/2$, corresponds to Heisenberg antiferromagnetic chains, $V_i=J_i$.
This leads to new insights into the cubic terms in the $\beta$-functions for decoupled chains, $\nu =0$. In this 
case  the SU(2) symmetry of each spin chain [not to be confused with the SU(2) symmetry discussed 
earlier in the fermion basis], may be used to rewrite the boundary Hamiltonian as:
\be H_b=2t_1S_0^xS_{-1}^z+2t_{2}S_0^yS_{1}^z.\ee
Upon bosonizing we obtain:
\be H_b=t_1'S_0^x\partial_x\phi_1(0)+t_2'S_0^y\partial_x\phi_2(0)\label{Hbz}\ee
for some rescaled boundary couplings $t_i'$. 
The equivalence of the operators $\partial_x\phi $ and $\cos \sqrt{\pi}\phi$ in the case $K=1/2$ follows from 
the hidden SU(2) symmetry of the bosonized Heisenberg model.  It is  this hidden symmetry 
which explains the remarkable identity in Eq. (\ref{id}). 
Eq. (\ref{Hbz}) was precisely the interaction studied in 
\cite{affl_1},\cite{affl_2} as a model of a 2 level system with non-commuting interactions 
with 2 independent heat baths. The cubic RG equations derived in \ref{RG} of this paper are identical, 
for $\nu =0$, to the ones derived by a different method in \cite{affl_1},\cite{affl_2}.  
Precisely the Hamiltonian of Eq. (\ref{Htb}), for the case $J_1=J_2$, $V_1=V_2$ arbitrary,  was studied 
in \cite{Novais} as a version of the dissipative Hofstadter model, and the RG equations were analyzed there, 
following \cite{Callan} and \cite{Zarand}.

The model of Eq. (\ref{Htb}) also has an interesting connection with the 2-channel Kondo model
in the marginal case of Heisenberg chains, $V_1=J_1=V_2=J_2$. The 2-channel Kondo model  
involves 2 channels of conduction electrons interacting with an impurity spin. Upon bosonizing the 
low energy effective Hamiltonian, only 
the spin degrees of freedom of each fermion channel participates in the Kondo interaction. The model 
then becomes equivalent to two semi-infinite Heisenberg chains with a weak 
coupling to a central spin. This is again the Hamiltonian of Eq. (\ref{Htb})  but with $H_b$ replaced by 
an isotropic version:
\be H_b=2\vec S_0\cdot [t_1\vec S_{1}+t_2\vec S_{-1}].\ee
After an appropriate rescaling of the $t_i$'s this model has a quadratic plus cubic $\beta$-function:
\be {dt_j\over dl}=2t_j^2-8t_j\sum_{k=1}^2t_k^2
.\ee
(See \cite{AL} for a derivation of this  $\beta$-function using OPE methods.)  Now consider 
an anisotropic 2-channel Kondo model with:
\be H_b=2\sum_{a=x,y,z}S_0^a\cdot [t_1^aS_{1}^a+t_2^aS_{-1}^a].\ee
By a simple extension of the methods in \cite{AL} it can be seen that the $\beta$-function becomes:
\be {dt_j^a\over dl}=\sum_{b,c}|\epsilon^{abc}|t_j^bt_j^c-4t_j^a\sum_{b\neq a}\sum_{k=1}^2t_k^bt_k^b
. \ee
We see that the Majorana interaction of Eq. (\ref{Htb}) corresponds to $t_1^x=-t_1$, $ t_2^y=-t_2$ all other 
$t_i^a$'s zero. Then we see that the quadratic terms in the Kondo $\beta$-function vanish and the 
cubic term reduces to the one in our Majorana model. 

\begin{figure}
\center
\includegraphics*[width=0.7\linewidth]{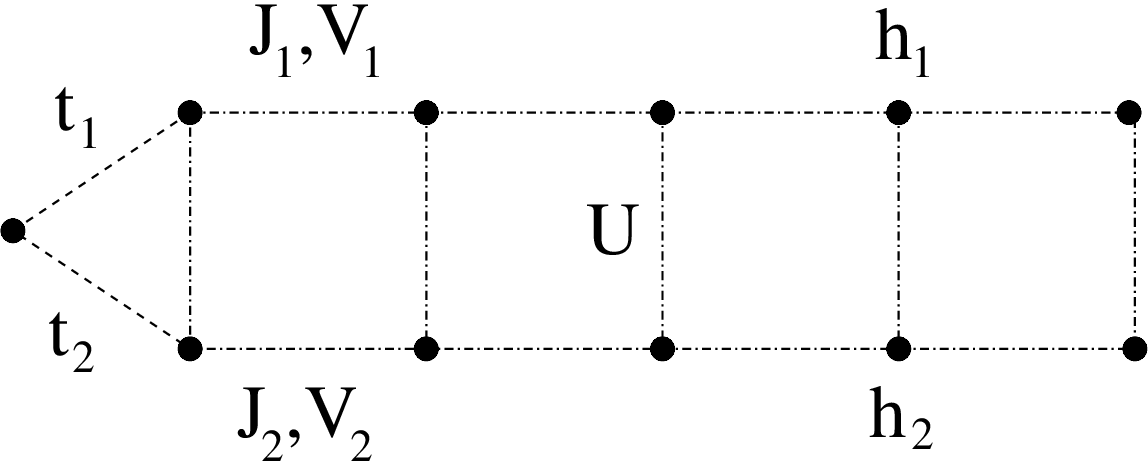}
\caption{The 2 leg spin ladder coupled to an impurity corresponding to the coupled 2 channel fermion model.} 
\label{fig:2legspin}
\end{figure}

So far, we have discussed the T-junction case of decoupled channels coupled to a Majorana mode, showing that 
it maps onto 2 semi-infinite spin chains coupled to an impurity spin. We may extend this mapping to the 
case of coupled channels.  In this case we get an impurity spin end-coupled to a 2-leg spin ladder
 as sketched in Fig. (\ref{fig:2legspin}). Labelling 
the spins $\vec S_{i,j}$ where $i=1,2,3,\ldots \infty$ measures distance along the chain and $j=1$, $2$ 
indexes the 2 legs, 
the Hamiltonian, after the inverse Jordan-Wigner transformation,  becomes:
\bea H_0&=&\sum_{i=1}^{\infty}\sum_{j=1}^2[J_j(S^x_{i,j}S^x_{i+1,j}+S^y_{i,j}S^y_{i+1,j})-h_jS^z_{i,j}]
\nonumber \\
H_{\hbox{intra}}&=&\sum_{i=1}^{\infty}\sum_{j=1}^2V_jS^z_{i,j}S^z_{i+1,j}\nonumber \\
H_{\hbox{inter}}&=&U\sum_{i=1}^{\infty}S^z_{i,1}S^z_{i,2}\nonumber \\
H_b&=&-2t_{1}S_0^xS_{1,1}^x-2t_2S_0^yS_{1,2}^y.
\eea
We have added unequal magnetic fields , $h_1\neq h_2$, acting on the 2 legs, in order to avoid developing a gap, 
as discussed in Sec. II.

These one dimensional spin chain models could be readily studied using either Density Matrix Renormalization Group 
or Quantum Monte Carlo. (The spin chain version developed in this Appendix makes it clear there is no fermion sign problem.) This would be 
very useful for checking our conjectured phase diagram and, in particular, the existence and properties of the non-trivial 
critical point.

\section{$A\otimes N$ fixed point}
\label{AxN}
As discussed in Sec. II, our RG equations, for $0<\epsilon_i\ll 1$, indicate that the only stable fixed points 
are $A\otimes N$ (and $N\otimes A$). In this Appendix we investigate the stability and other properties of 
this fixed point in more detail. These are characterized by the conformally invariant boundary conditions:
\bea \phi_1(0)&=&0 \: {\rm or} \: \sqrt{\pi} \nonumber \\
\theta_2(0)&=&0 \: {\rm or} \: \sqrt{\pi}.\label{bcAA}
\eea
In the T-junction case, where channels 1 and 2 are decoupled, the properties of these fixed points are well-known. 
As for the normal reflection boundary condition, the starting point of our analysis  in Sec. II, 
the channel-two fermion fields  at the junction are bosonized as:
\be \psi_2(0)\propto \exp [i\sqrt{\pi}\phi_2(0)],\ \ \psi_2^\dagger (0)\propto \exp [-i\sqrt{\pi}\phi_2(0)],\ee
giving operators of dimension $d_{2}=1/(2K_2)$. 
On the other hand, we see from Eq. (\ref{bos}) and (\ref{bcAA}) that the channel-one fermion fields at the junction 
are bosonized as:
\be \psi_1(0)=\psi_1^\dagger (0)\propto \cos [\sqrt{\pi}\theta_1(0)]\ee
of dimension $d_{A1}=K_1/2$.
From these results we can work out the dimensions of the various boundary operators at this fixed point. Normal 
reflection in channel 1 corresponds to 
\be \psi_1^\dagger (0)\psi_1(0)\propto \cos[ 2\sqrt{\pi }\theta_1(0)]\ee
of dimension $2K_1$. This is $1/d_1$ where $d_1$ is the dimension of the coupling to the Majorana mode 
analyzed in Sec. II. Thus, whenever that Majorana coupling is relevant, and the system flows to the $A\otimes N$ 
fixed point, this normal scattering is irrelevant. 

We now consider Andreev scattering in channel 2 at the $A\otimes N$ fixed point, corresponding to the 
perturbation analysed in Sec. II
\be H_{b2}\equiv 2it_2 \tau_0^{-1+d_2} \gamma \Gamma_2\cos [\sqrt{\pi}\phi_2(0)].\ee
In analyzing the effect of this operator it is important to take into account the ``Schroedinger cat'' nature of the 
$A\otimes N$ ground states, as discussed in Sec. II. Ignoring channel 2, there are 2 ground states in which 
$i\gamma \Gamma_1=\pm 1$ and correspondingly $<\cos \sqrt{\pi} \phi_1(0)>$ is negative or positive.  The 
localized Dirac state constructed from $\gamma$ and $\Gamma_1$ is  occupied or empty in these 2 states respectively. 
The operator $\gamma \Gamma_2$ acts off-diagonally in the local mode space since 
\be \gamma =\psi_0+\psi_0^\dagger \ee
where $\psi_0$, defined in Eq. (\ref{locdir}),  annihilates the local mode state. Therefore acting with $H_{b2}$ on one of these states produces a 
high energy state where the local mode occupancy is switched without changing the sign of  $<\cos \sqrt{\pi} \phi_1(0)>$ 
(i.e. without changing the electron number parity in channel 1). 
We thus should treat $H_{b2}$ using a ``Schrieffer-Wolff'' type procedure \cite{SW}, familiar from the mapping of the Anderson 
impurity model into the Kondo model.  The perturbation then corresponds to
\be \delta H\propto {t_2^2\over t_1}[\psi_2^\dagger \partial_x\psi_2^\dagger +h.c.]
\propto {t_2^2\over t_1}\cos [2\sqrt{\pi}\phi_2(0)].\label{SW}\ee
Low energy processes can only transfer pairs of electrons between channel 2 and the superconductor, 
to avoid changing the occupancy 
of the local mode. This simply corresponds to regular Andreev scattering processes that could take place 
without the presence of the Majorana mode. Once the Majorana mode is strongly entangled with channel 1, 
it cannot enable Andreev scattering in channel 2.   The perturbation in Eq. (\ref{SW}) has dimension $2/K_2=4d_2$ 
and is thus strongly irrelevant.  

\begin{figure}
\center
\includegraphics*[width=0.85\linewidth]{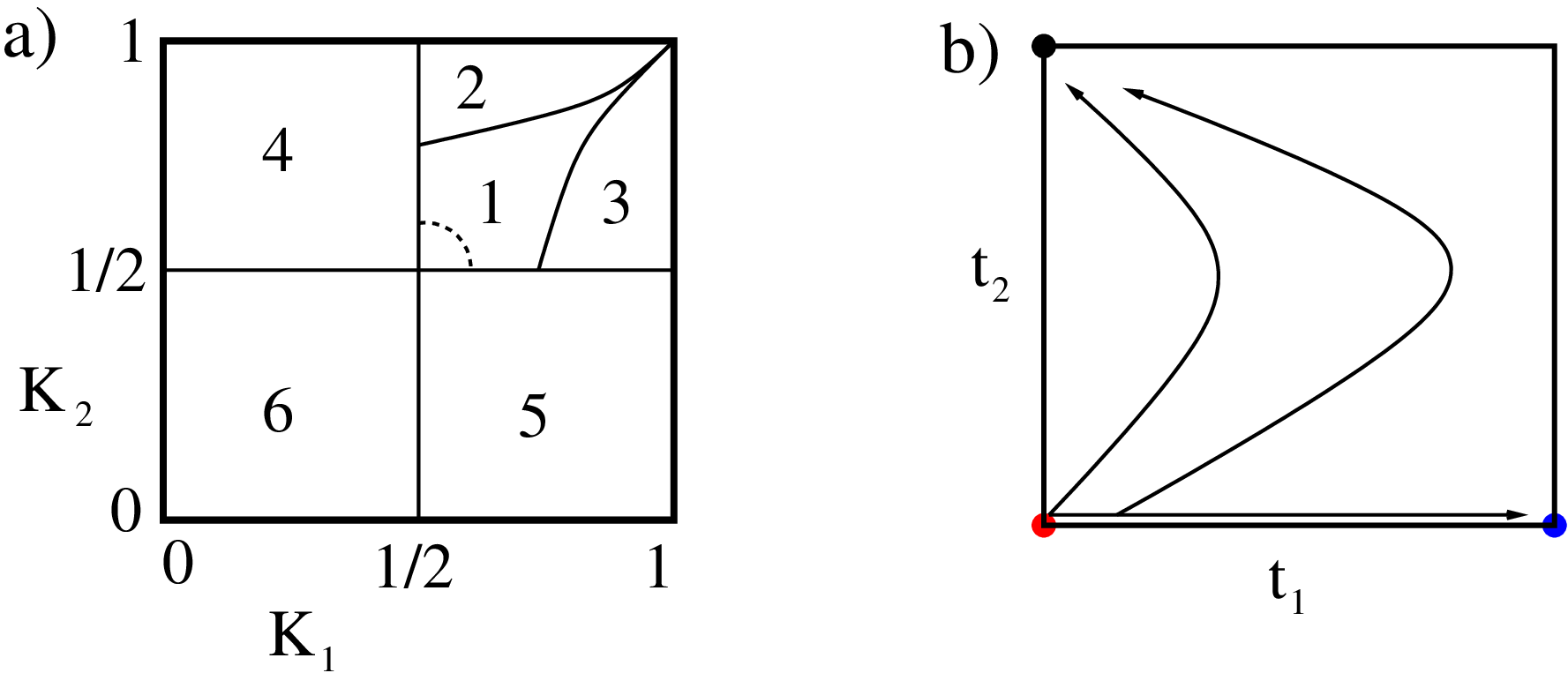}
\caption{{\bf a)}:  Sketch of various regions in the $K_1$-$K_2$ plane which have different phase diagrams and RG 
flows in the $t_1$-$t_2$ plane, for the $T$-junction (decoupled channels). The dashed quarter-circle indicates qualitatively the parameter region where the $\epsilon$-expansion
 is valid and predicts a NTCP.\\
{\bf b)}: Qualitative sketch of the RG flow in region 2.} 
\label{fig_asym}
\end{figure}

Finally, we must consider processes that change the number of electrons 
in each channel by $\pm 1$. These can either correspond to  Andreev tunnelling of a pair of electrons, one drawn 
from each channel,  into (or out of) the superconductor or to transfer of a single electron between the 2 channels. 
In this case, the Schrieffer-Wolff transformation is not necessary and a representative operator is:
\be H_{12} \propto \psi_1(0)\psi_2(0)\propto \cos [\sqrt{\pi}\theta_1(0)]\exp [i\sqrt{\pi}\phi_2(0)]\label{p1p2}\ee
of dimension:
\be d=d_{A1}+d_2={1\over 2}\left(K_1+{1\over K_2}\right).\label{d12d}\ee
This is irrelevant, $d>1$, when $K_1=K_2<1$ corresponding to equal repulsive interactions in both channels. However, 
it can be relevant when $K_1<K_2<1$, to the left of the line $K_1=2-1/K_2$. This is physically reasonable. If channel 
1 has stronger repulsive interactions than channel 2 then the fixed point with perfect Andreev scattering in channel 
1 and perfect normal scattering in channel 2 can become unstable.  Taking into account the stability 
conditions at $A\otimes N$, $N\otimes A$  and $N\otimes N$ fixed points, 
 we can identify 6 different regions in the $K_1$-$K_2$ plane, with $K_i<1$, which have different 
phase diagrams and RG flows in the $t_1$-$t_2$ plane.  These regions are numbered in Fig. (\ref{fig_asym}a). 
In Table (\ref{table_s}) the fixed points are indicated which are either stable or semi-stable (meaning 
stable under moving in precisely one direction in the $t_1$-$t_2$ plane) in each of these 6 regions.   In Fig. (\ref{fig_asym}b), 
a qualitative sketch of the RG flow diagram is given for a point in region 2.

\begin{table}
\centering
\begin{tabular}{c|c|c}
Region&Stable&Semi-stable\\
\hline \hline \\
1&$A\otimes N$, $N\otimes A$ & NTCP\\
\hline \\
2&$N\otimes A$&$A\otimes N$\\
\hline \\
3&$A\otimes N$& $N\otimes A$\\
\hline \\
4&$N\otimes A$&$N\otimes N$\\
\hline \\
5&$A\otimes N$&$N\otimes N$\\
\hline \\
6&$N\otimes N$&--
\end{tabular}
\caption{Stable and semi-stable fixed points in different regions of $K_1$-$K_2$ plane for the $T$-junction (decoupled channels). 
The six regions are labelled as in Fig. (\ref{fig_asym}).}
\label{table_s}
\end{table}

We now consider the case with inter-channel interactions, which is a rather novel fixed point. It's properties were 
discussed in \cite{lut_majo} 
using the technique of integrating out the boson fields everywhere except at the origin. Here we wish to 
discuss this fixed point using boundary conformal field theory (BCFT) techniques where a conformally invariant 
boundary condition is applied to the bulk conformal field theory of 2 free bosons. We will corroborate some of the 
conclusions of \cite{lut_majo} as well as gaining new insight. The BCFT approach is generally more powerful 
since it lends itself to calculating Green's functions at arbitrary spatial locations as well as the impurity entropy, 
which we discuss in \ref{entropy}. 

When both couplings to the Majorana mode are turned off, $t_1=t_2=0$, the boson fields obey the boundary 
conditions $\theta_1(0)=\theta_2(0)=0$. However, when $t_1$ renormalizes to large values and $t_2$ 
renormalizes to zero, we expect a new boundary condition, from Eq. (\ref{phal}):
\be \cos \alpha \phi_\sigma (0)+\sin \alpha \phi_\rho (0)=0 \: {\rm or} \: r\sqrt{\pi}\label{bc}\ee
which couples the independent boson fields $\phi_{\rho}$ and $\phi_\sigma$ with different Luttinger parameters. 
This is only a partial specification of a conformally invariant boundary condition; we wish to deduce the 
complementary condition. To do this, it is convenient to first rewrite the Hamiltonian in terms of rescaled fields:
\be \bar \phi_\lambda \equiv \sqrt{K_\lambda}\phi_\lambda ,\ \  \bar \theta_\lambda \equiv \theta _\lambda  /\sqrt{K_\lambda}.\ee
Note that these obey canonical commutation relations:
\be [\bar \phi_\lambda (x),\bar \theta_{\lambda '}(y)]=-{i\over 2}\delta_{\lambda ,\lambda '}\epsilon (x-y).\ee
The Hamiltonian of Eq. (\ref{rg2.1}) now takes the simple form:
\be H={1\over 2}\sum_\lambda u_\lambda  \int   dx  \: 
[(\partial_x\bar \phi_\lambda )^2+(\partial_x\bar \theta_\lambda )^2].\label{hamcon}\ee
It is now convenient to rescale distance, $x$, differently for the $\rho$ and $\sigma$ fields:
\be \tilde \phi_\lambda (x/u_\lambda )\equiv \bar \phi (x),\ \  \tilde \theta_\lambda (x/u_\lambda )\equiv \bar \theta (x).
\ee
The Hamiltonian can then be written:
\be H={1\over 2}\sum_\lambda  \int dx [(\partial_x\tilde \phi_\lambda )^2+(\partial_x\tilde \theta_\lambda )^2]\label{hamcon2}\ee
where the $x$ integration variable now has dimensions of time. 
Next, we make a canonical  transformation, motivated by pinning of  
${\cos \alpha \over \sqrt{K_\sigma}}\tilde{\phi}_\sigma ( 0 )
+{\sin \alpha \over \sqrt{K_\rho}} \tilde{\phi}_\rho ( 0 ) $:
\be \left(\begin{array}{c}
\phi_1'\\
\phi_2'\end{array}\right)={\cal O}\left(\begin{array}{c}
\tilde \phi_\sigma \\
\tilde \phi_\rho \end{array}\right) ,\ \ 
 \left(\begin{array}{c}
\theta_1'\\
\theta_2'\end{array}\right)={\cal O}\left(\begin{array}{c}
\tilde \theta_\sigma  \\
\tilde \theta_\rho \end{array}\right) 
\ee
where ${\cal O}$ is the orthogonal matrix:
\beq
{\cal O}
= \frac{1}{\sqrt{\cos^2\alpha K_\rho+ \sin^2\alpha K_\sigma}} \left( \begin{array}{cc} \cos \alpha \sqrt{K_\rho} & 
                                          \sin \alpha   \sqrt{K_\sigma} \\
-\sin \alpha \sqrt{K_\sigma} & \cos \alpha \sqrt{K_\rho} 
                                            \end{array}  \right).  \label{phd.18}
\eneq
The Hamiltonian still takes the canonical form of Eq. (\ref{hamcon2}) in this basis and 
 our conformally invariant boundary conditions simply correspond to pinning  $\phi_1'(0)$ and
 $\theta_2'(0)$. 
To work out the scaling dimensions of the  various boundary operators at the $A\otimes N$ fixed point, we 
express the original fields $\phi_i$ and $\theta_i$ at the origin in terms of the transformed fields:
\bea \phi_{1}(0)&=& r^{-1}\sqrt{\cos^2\alpha K_\rho+ \sin^2\alpha K_\sigma \over K_\rho K_\sigma}\phi_1'(0)
\nonumber \\
\phi_2(0)&=&r\left[ {(K_\sigma -K_\rho )\sin 2\alpha \over \sqrt{4K_\rho K_\sigma (\cos^2\alpha K_\rho+ \sin^2\alpha K_\sigma )}}\phi_1'(0)
+{1\over \sqrt{\cos^2\alpha K_\rho+ \sin^2\alpha K_\sigma}}\phi_2'(0)\right] \nonumber \\
\theta_1(0)&=& r\left[ \sqrt{K_\rho K_\sigma \over \cos^2\alpha K_\rho+ \sin^2\alpha K_\sigma}\theta_1'(0)
+{\sin 2\alpha (K_\rho -K_\sigma )\over \sqrt{4(\cos^2\alpha K_\rho+ \sin^2\alpha K_\sigma )}}\theta_2'(0)\right] \nonumber \\
\theta_2(0)&=& r^{-1}\sqrt{\cos^2\alpha K_\rho+ \sin^2\alpha K_\sigma}\theta_2'(0)
.\label{'bc}
\eea
Using the conditions  that $\phi_1'(0)$ and
 $\theta_2'(0)$ are pinned, together with Eq. (\ref{'bc}),
 may now read off the dimensions of the various boundary operators. 
Normal back-scattering in channel 1 gives
\be \psi_{1L}^\dagger (0)\psi_{1R}(0)\propto e^{2i\sqrt{\pi}\theta_1}\propto \exp \left[ ir\sqrt{4\pi K_\rho K_\sigma \over \cos^2\alpha K_\rho+ \sin^2\alpha K_\sigma}\theta_1'(0)\right]
\ee
of dimension \cite{error2}:
\be d_{1n}={2r^2K_\rho K_\sigma \over \cos^2\alpha K_\rho+ \sin^2\alpha K_\sigma}.\ee
As above, from Eq. (\ref{di}), 
\be d_{1n}=1/d_1\ee
where $d_1$ is the dimension of $t_1$, the coupling of the Majorana mode to channel 1. Thus, whenever the coupling to the Majorana 
mode is relevant, so that a flow may occur to the $A\otimes N$ fixed point, normal backscattering in channel 1 is irrelevant. 

Now consider the coupling of channel 2 to the Majorana mode, from Eq. (\ref{Hbb}):
\be H_{b2}=2t_2 \tau_0^{-1+d_2} \gamma \Gamma_2\cos [\sqrt{\pi}\phi_2(0)]
.\ee
Making the Schrieffer-Wolff transformation discussed above
we obtain:
\be \delta H\propto {t_2^2\over t_1}\cos \left[r\sqrt{4\pi \over \cos^2\alpha K_\rho+ \sin^2\alpha K_\sigma}\phi_2'(0)\right]
\ee
which has scaling dimension:
\be d_{2,A\otimes N}={2r^2\over \cos^2\alpha K_\rho +\sin^2\alpha K_\sigma}.\ee
This has the value $d_{2,A\otimes N}=2$ for free fermions, $\alpha =0$, $K_\rho /r^2 =1$ and we generally expect it to increase 
with repulsive interactions, thus being strongly irrelevant. 

Finally, we can consider processes where the number of electrons in channel 1 and 2 simultaneously change 
by $\pm 1$. 
\bea H_{12} &\propto& \psi_1(0)\psi_2(0)\propto \cos  [\sqrt{\pi}\theta_1(0)]\exp [i\sqrt{\pi}\phi_2(0)]\nonumber \\
&\propto& \cos \left[ r\sqrt{\pi K_\rho K_\sigma \over \cos^2\alpha K_\rho+ \sin^2\alpha K_\sigma}\theta_1'(0)\right]
\nonumber \\ &\times&
\exp\left[ ir\sqrt{\pi \over \cos^2\alpha K_\rho+ \sin^2\alpha K_\sigma}\phi_2'(0)\right]\eea
of dimension
\be d_{12}=(1/4)(d_{1n}+d_{2,A\otimes N})={r^2(1+K_\rho K_\sigma )\over 2[\cos^2\alpha K_\rho+ \sin^2\alpha K_\sigma ]}.
\label{d12}\ee
This appears likely to be the most relevant operator at the $A\otimes N$ fixed point and may destabilize it 
is some cases as discussed above for the $T$-junction case. 
 It is  marginal in the SU(2) symmetric case, $r=1$, $\alpha =\pi /4$, $K_\sigma =1$. 
This is to be expected since, as discussed in Sec. II, in the SU(2) symmetric case there is a circle of 
rotated $A\otimes N$ fixed points in which the linear combination of the $\psi_i$'s defined in Eq. (\ref{SU2})  
experiences perfect Andreev reflection and the orthogonal linear combination perfect normal reflection. Thus 
the $A\otimes N$, fixed point, whose stability we are studying, is merely one point on this circle and the exactly 
marginal operator drives the system along the line.
Of course, setting $\alpha \to 0$, $K_\sigma \to K_1/r^2$, $K_\rho \to K_2r^2$
we recover the result of Eq. (\ref{d12d}).  As discussed below Eq. (\ref{d12d}) and in Fig. (\ref{fig_asym}), these processes
  can become relevant, destabilizing 
the $A\otimes N$ fixed point even when $K_1<K_2<1$. Similarly the $A\otimes N$ fixed point 
may be unstable for coupled channels when the self-interactions in channel 1 are more strongly 
repulsive than in channel 2.

\section{Stability of the non-trivial critical point}
\label{NTCP}
In Sec. II we showed that, for $0<\epsilon_i \ll 1$, there was a separatrix in the phase diagram, at 
$t_1=t_2$ in the case $\epsilon_1=\epsilon_2$, separating the $A\otimes N$ and $N\otimes A$ phases. 
The RG flow along this separatrix was found in Sec. II to be to a non-trivial critical point. 
In this Appendix wish to argue that this likely remains true for all Luttinger parameters such that 
both $A\otimes N$ and $N\otimes A$ fixed points are stable. This parameter range was calculated in 
\ref{AxN}.  For the $T$-junction (decoupled channels)  it is the region obeying the 4 inequalities:
\bea K_1&>&1/2\nonumber \\
K_2&>&1/2\nonumber \\
K_1+{1\over K_2}&>&2\nonumber \\
K_2+{1\over K_1}&>&2.
\eea
This region is labelled  number 1 in 
 Fig. (\ref{fig_asym}a). 

One way of arguing for this is based on  ``what else could happen''?  When 
both $A\otimes N$ and $N\otimes A$ fixed points are stable,  
something must separate these two phases. One logical possibility might seem to be an $A\otimes A$ 
fixed point. We  argue here that this is not possible. A quick way of seeing this is to consider, as in \ref{AxN}, 
processes that change the number of electrons in both channels by $\pm 1$. Imposing $A\otimes A$ boundary 
conditions, this perturbation bosonizes as:
\be \delta H\propto \cos [\sqrt{\pi}\theta_1(0)]\cos [\sqrt{\pi}\theta_2(0)].\label{rel}\ee
For decoupled channels this has dimensions $d=(K_1+K_2)/2$ and is thus relevant for repulsive interactions. 
In the general case, we can write:
\bea \delta H &\propto& \cos [\sqrt{\pi}(\theta_1+\theta_2)]+\cos [\sqrt{\pi}(\theta_1-\theta_2)] \nonumber \\
&=&\sum_{\pm}\cos [(r\cos \alpha \mp r^{-1}\sin \alpha )\theta_\sigma 
+ (r\sin \alpha \pm r^{-1}\cos \alpha )\theta_\rho ].
\eea
These two terms have dimensions:
\be d_{\pm}={1\over 2}[(r^2\cos^2\alpha + r^{-2}\sin^2\alpha \mp \sin 2\alpha )K_\sigma +
(r^2\sin^2\alpha +r^{-2}\cos^2\alpha \pm \sin 2\alpha )K_\rho ].\ee
For the case of equivalent channels, $r=1$, $\alpha =\pi /4$, these reduce to $d_+=K_\sigma$, $d_-=K_\rho$. 
In this case, this interaction is relevant  whenever $K_\rho$ or $K_\sigma <1$. 
This is consistent with our result in \ref{entropy}  that the impurity ground state degeneracy, $g$, is larger for the $A\otimes A$ 
fixed point than for the $N\otimes N$ fixed point at $t_i=0$. The ``$g$-theorem'' then implies 
that renormalization from $N\otimes N$ to $A\otimes A$ fixed points is impossible. 

The careful reader might wonder whether the interaction of Eq. (\ref{rel}) is really allowed at the $A\otimes A$ fixed point 
given the delicate entanglement of the Majorana mode with both channels at such a fixed point. In \ref{AxN} we found that 
a particular perturbation of another fixed point disrupted such an entangled state, driving the system into a 
high energy state and necessitating a Schrieffer-Wolff transformation resulting in a higher dimension operator. 
Does this also happen here? To check this point, it is useful to consider a rather contrived  
tight-binding model which really 
is described at low energies by an $A\otimes A$ fixed point.  We note that this {\it does not} correspond 
to simply taking large $t_1=t_2\equiv t$ in the Hamiltonian of Eq. (\ref{bione}). Setting $J_i=V_i=U=0$ in 
Eq. (\ref{bione}) gives the ``strong coupling Hamiltonian''
\be H_{sc}=-t(c^\dagger_0+c_0)(c_1^\dagger -c_1+c_{-1}^\dagger -c_{-1})/2.\ee
Expanding in Majorana fermions:
\bea c_0&\equiv& (\gamma +i\gamma ')/2\nonumber \\
 c_j&\equiv& (\gamma_j'+i\gamma_j)/2,\ \  (j=\pm 1),\label{gjdef} \eea
 $H_{sc}$ can be rewritten:
 \be H_{sc}=it\gamma (\gamma_1+\gamma_{-1})/2.\ee
The appropriate Dirac operator, annihilating the ground state of $H_{sc}$ is:
 \be \psi_0\equiv [\gamma +i(\gamma_1+\gamma_{-1})/\sqrt{2}]/2. \label{gjdef2}\ee
However, to check whether or not large $t$ really corresponds to an $A\otimes A$ fixed point we 
must consider the effect of turning on  the $J_i$, $V_i$ and $U$ interactions. To keep things as simple as 
possible, we consider only the non-interacting case with $V_i=U=0$. We also 
set $J_1=J_2$. The hopping terms 
from sites 1 to 2 and (-1) to (-2) are:
\bea H_{12}&=&-J(c_1^\dagger c_2+c_{-1}^\dagger c_{-2}+h.c.) 
=-J[\gamma_1'(c_2-c_2^\dagger )-i\gamma_1(c_2+c_2^\dagger )\nonumber \\ &+&
\gamma_{-1}'(c_{-2}-c_{-2}^\dagger )-i\gamma_{-1}(c_{-2}+c_{-2}^\dagger )]/2.
\label{H12}
\eea
The terms in $H_{12}$ involving the $\gamma_j'$ operators are harmless but the terms involving the 
$\gamma_j$ operators disrupt the localized state of the putative $A\otimes A$ fixed point. 
In order to study the stability of this fixed point we temporarily allow these couplings to be different, 
changing $H_{12}$ to:
\be H_{12}
=-J[\gamma_1'(c_2-c_2^\dagger )+\gamma_{-1}'(c_{-2}-c_{-2}^\dagger )]/2
+i\tilde J[\gamma_1(c_2+c_2^\dagger )+\gamma_{-1}(c_{-2}+c_{-2}^\dagger )]/2.
\ee
If we now set $\tilde J=0$, but include the full $J$ terms between all other pairs of sites, we 
indeed obtain a model which  we expect to renormalize, at low energies, to an $A\otimes A$ fixed point. 
The Majorana mode $\gamma_1'$ entangles with the Dirac fermions on sites $2,3,\ldots \infty$ to 
produce the analogue of the perfect Andreev scattering fixed point and likewise the Majorana 
mode $\gamma_{-1}'$ entangles with the Dirac fermions on sites $-2,-3,\ldots -\infty$. 
All of this can happen without disrupting the entanglement between the $\gamma$, $\gamma_1$ 
and $\gamma_{-1}$ Majorana modes that occurs in the ground state of $H_{sc}$. 
The resulting ground states are 4-fold degenerate corresponding to occupancy zero or one 
for the two local modes: one constructed from $\gamma_1'$ and the chain on the right and 
the other constructed from $\gamma_{-1}'$ and the chain on the left. As in Sec. II we may 
bosonize these two semi-infinite chains, introduce Klein factors $\Gamma_1$ and $\Gamma_2$, 
and construct the local 
Dirac operators  $(\gamma_1'+i\Gamma_1)/2$ and 
$(\gamma_{-1}'+i\Gamma_2)/2$. 
But now consider the effect of turning on a small $\tilde J$. Acting to first order in $\tilde J$ 
drives the system into a high energy state of $H_{sc}$. However, at second order in $\tilde J$ 
the system can return to the ground state of $H_{sc}$ and we develop, via a Schrieffer-Wolff 
type transformation, a perturbation of the $A\otimes A$ fixed point of the form:
\be \delta H\propto {\tilde J^2\over t}(c_2+c_2^\dagger )(c_{-2}+c_{-2}^\dagger ).\ee
This is precisely the relevant perturbation of Eq. (\ref{rel}), that changes 
the number of electrons in channels 1 and 2 by $\pm 1$. In this non-interacting example, we 
expect it to drive the system from the unstable $A\otimes A$ fixed point to the rotated 
$A\otimes N$ fixed point discussed in Sec. II. This argument goes through the 
same way for general $J_i$, $V_i$ and $U$. We could again artificially separate $V$ 
into terms that do and do not disrupt the Majorana entanglement in the $A\otimes A$ ground state. 
The relevant perturbation now also contains a term $\propto (\tilde V)^2/t$. In this case, 
we expect the relevant perturbation would drive the system to the nontrivial critical point 
(or else the $A\otimes N$ or $N\otimes A$ critical points off the separatrix).

The above arguments imply that, even along the separatrix, there is no stable $A\otimes A$ fixed point 
to complete with our non-trivial one. It is thus difficult for us to imagine how it could not exist 
in the phase diagram.    We provide further evidence for this in \ref{entropy}, where we calculate 
the impurity entropies at the various fixed points and invoke the ``g-theorem''.

\begin{figure}
 \center
\includegraphics*[width=0.44\linewidth]{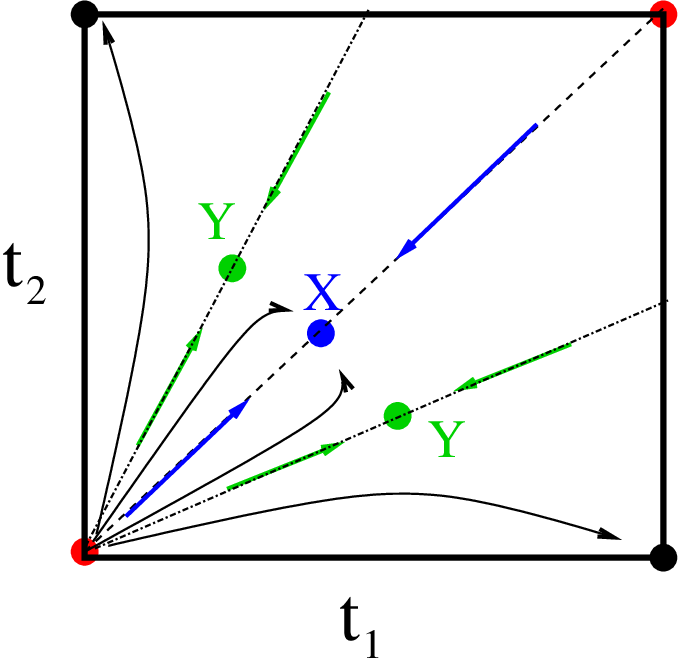}
\caption{Qualitative sketch of the putative phase diagram for a range of Luttinger parameters where 
the nontrival critical point, X,  might be stable.  In this case, two other fixed points, Y,  exist 
on the two separatrixes.} 
\label{sntcp}
\end{figure}

There is however, one interesting possibility that may warrant numerical investigation.
Could our nontrivial critical point become {\it completely} stable under arbitrary small variations of the $t_i$'s 
and other parameters 
for some range of bulk interaction parameters? In this case the $t_1$-$t_2$ plane would divide 
up into regions of finite area attracted to the $A\otimes N$, $N\otimes A$ and nontrivial critical points. Thus 
assuming continuous phase transitions between these phases, there would then need to be 
two other (equivalent) fixed points, unstable in one direction, on these separatrixes, as sketched 
in Fig. (\ref{sntcp}). 

\section{Uniform wire coupled to superconductor far from its end points}
\label{single}
\begin{figure}
\center
\includegraphics*[width=0.55\linewidth]{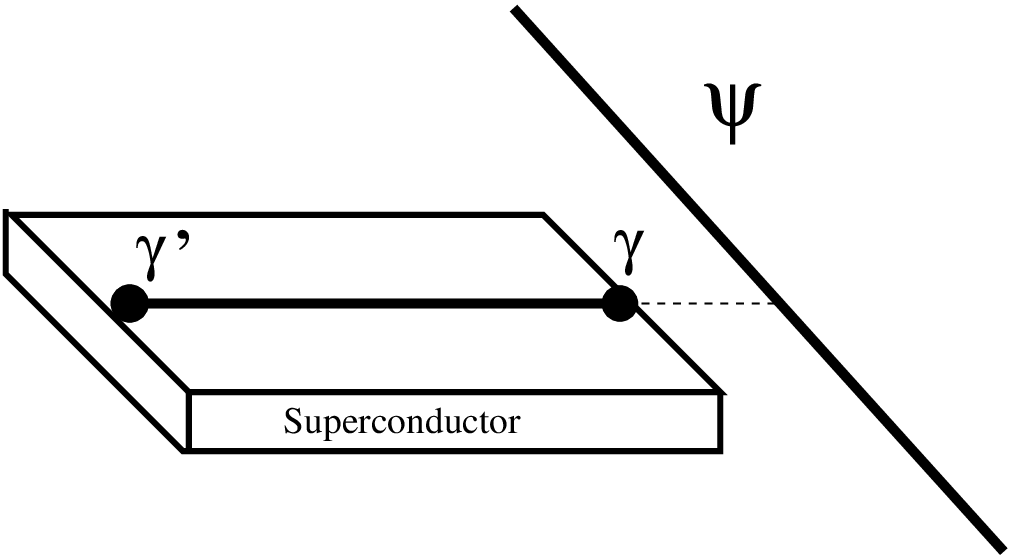}
\caption{A sketch of the uniform chain coupled to the topological superconductor far from its endpoints} 
\label{chiral}
\end{figure}
So far, the $T$-junction we have considered consists of two quantum wires, 
end-coupled to the topological superconductor, as sketched in Fig.\ref{fig_0}. 
In this Appendix we consider the opposite extreme of a single channel uniform quantum wire ``centre-coupled''  far from its endpoints to the 
topological superconductor, as sketched in Fig. (\ref{chiral}). Despite the extreme difference in the underlying model, we 
obtain the same phase diagram. In general, at low energies, the wire breaks up into two sections at the junction 
with one side coupling strongly to the Majorana mode and exhibiting perfect Andreev reflection while the other side decouples, 
exhibiting perfect normal reflection. Or, if a suitable parity 
symmetry is respected, the NTCP occurs. 

We begin by considering the low energy effective Hamiltonian:
\bea H_0&=&i\int_{-\infty}^\infty dx[\psi_R^\dagger \partial_x\psi_R-\psi_L^\dagger \partial_x\psi_L]\label{H0c}\nonumber \\
H_{\hbox{int}}&=&V\int_{-\infty}^\infty dx \psi_R^\dagger \psi_R\psi_L^\dagger \psi_L\nonumber \\
 H_b&=&\gamma [t_L(\psi_L^\dagger (0)-\psi_L(0))+t_R(\psi_R^\dagger (0)-\psi_R(0))].\label{Hbcc}\eea
Bosonizing we obtain the bulk Hamiltonian
\be H={1\over 2}u\int_0^\infty dx 
\left[ K\left( \frac{\partial \phi}{\partial x} 
\right)^2 + K^{-1}\left( \frac{\partial \theta}{\partial x} 
\right)^2\right] .\label{H0cb}\ee
The ``boundary'' Hamiltonian is:
\be H_b=i\gamma \{t_L
\cos [\sqrt{\pi}(\phi (0)+\theta (0)]+t_R\cos [\sqrt{\pi}(\phi (0)-\theta (0)]\}.\ee
Note that no boundary conditions are imposed on $\phi$ or $\theta$ in this case;  we start with 
a continuous translationally invariant chain at $t_L=t_R=0$. 
 The boundary interactions both have dimension:
\be d_b={1\over 4}\left(K+{1\over K}\right)\ee
1/2 at the free fermion point, $K=1$ and increasing when $K$ increases or decreases. They become 
marginal at $K_c=2\pm \sqrt{3} \approx 3.73$, $.268$. 

\begin{figure}
\center
\includegraphics*[width=0.3\linewidth]{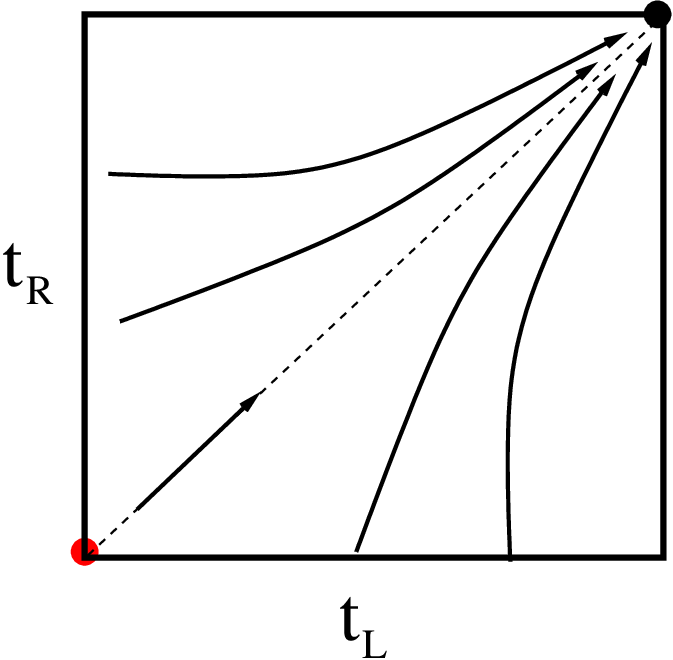}
\caption{RG flow of effective couplings for ${\cal F}<0$.} 
\label{RGF<0}
\end{figure}

We now calculate the cubic term in the $\beta$ functions at the marginal point. (Only the one 
at $K=.268\ldots$ is likely to be of physical interest.)
We first change variables to $\bar \phi \equiv \sqrt{K}\phi$, $\bar \theta \equiv \theta /\sqrt{K}$ as in \ref{AxN}. 
We then change variables to $\phi_{L/R}$ defined by:
\bea \bar \phi &\equiv& \phi_L+\phi_R \nonumber \\
\bar \theta &\equiv& \phi_L-\phi_R.
\eea
Then:
\bea &&\psi_L^\dagger (\tau_1)\psi_L(\tau_2)\psi_R(\tau_3)\propto  \nonumber \\
&& \exp \left\{i\sqrt{\pi} \left[  (1/\sqrt{K}+\sqrt{K})[\phi_L(\tau_2)-\phi_L(\tau_1)]+(1/\sqrt{K}-\sqrt{K})\phi_L(\tau_3)
\right] \right\}
\nonumber \\
&\cdot& \exp \left\{i\sqrt{\pi} \left[ 
(1/\sqrt{K}-\sqrt{K})[\phi_R(\tau_2)-\phi_R(\tau_1)]+(1/\sqrt{K}+\sqrt{K})\phi_R(\tau_3) \right] \right\}
\nonumber \\\eea
This gives an OPE:
\bea &&{\cal T}< \psi_L^\dagger (\tau_1)\psi_L(\tau_2)\psi_R(\tau_3)>\nonumber \\
&&\to \epsilon (\tau_1-\tau_2)
\left|{1\over \tau_{12}}\right|^{[(1/\sqrt{K}+\sqrt{K})^2+(1/\sqrt{K}-\sqrt{K})^2]/4}\cdot
\left|{\tau_{23}\over \tau_{13}}\right|^{(1/\sqrt{K}+\sqrt{K})(1/\sqrt{K}-\sqrt{K})/2}\psi_R(\tau_3)\nonumber \\
&=&\epsilon (\tau_1-\tau_2)\left({1\over \tau_{12}}\right)^2\left|{\tau_{23}\over \tau_{13}}\right|^{\sqrt{3}}\psi_R(\tau_3).
\eea
Here we used $K=2-\sqrt{3}$. 
Including the Majorana mode OPE of Eq. (\ref{gOPE}), 
this gives  Eq. (\ref{dS122})   with $\nu =\sqrt{3}>1$. As we see from Fig.  (\ref{fig_nu}), ${\cal F}(\nu )$ is 
negative, for $\nu >1$, with ${\cal F}(\sqrt{3})\approx -20$. So, in this case there is no nontrivial critical point for $\epsilon >0$; instead the 
flow is towards infinite coupling for any bare couplings which are both non-zero. Furthermore,  the negative ${\cal F}$  drives 
the couplings towards each other, rather than apart, as the floating cut-off, $D$, is reduced, as illustrated in Fig. (\ref{RGF<0}).
This might suggest a flow towards the NTCP, 
but this requires further substantiation. Note that, if we had instead found that $t_R\to 0$ and $t_L$ grew large 
under renormalization this would suggest very exotic behaviour indeed, with the left-moving chiral 
mode having perfect Andreev transmission and the right-moving chiral mode having perfect normal transmission.
 However, due to the fact that $\nu >1$, this {\it is not} what we are finding. 

\begin{figure}
\center
\includegraphics*[width=0.9\linewidth]{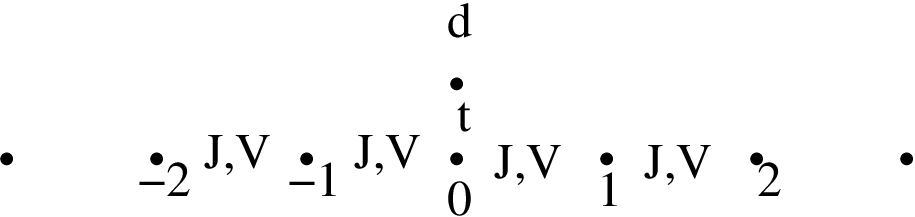}
\caption{Tight-binding model of topological superconductor coupled far from the ends of a single channel quantum wire.} 
\label{Ttb}
\end{figure}

To test our hypothesized flow to the NTCP, 
it is very convenient to consider a tight-binding model whose continuum limit 
gives the Hamiltonian of  Eq. (\ref{Hbcc}). This is represented in Fig. (\ref{Ttb}). 
The uniform chain has Hamiiltonian:
\be H_0+H_{\hbox{int}}=\sum_{j=-\infty}^\infty [-J(c^\dagger_jc_{j+1}+h.c.)+Vn_jn_{j+1}].\ee
The topological 
superconductor is represented by the impurity site, with annihilation operator $d=(\gamma +i\gamma ')/2$ and 
impurity coupling:
\be H_b=t\gamma (c_0-c_0^\dagger ).\ee
Using the low energy representation of the tight binding model operators
\be c_j\approx e^{ik_Fj}\psi_R(j)+e^{-ik_Fj}\psi_L(j)\label{cLR}\ee
establishes the low energy correspondence with the Hamiltonian of Eq. (\ref{Hbcc}) in the case $t_L=t_R=t$. Since the perturbative RG 
analysis suggests that $t$ renormalizes to large values, we consider the $t\to \infty$ limit of the tight-binding model. Writing:
\be c_0=(\gamma_0+i\eta_0)/2,\ee
we see that, in the $t\to \infty$  limit, $\eta_0$ combines with $\gamma$ to form a local Dirac operator:
\be \psi_0=(\gamma +i\eta_0)/2\ee
which is empty in the ground state. In this limit, to avoid driving the system into a high energy state, the hopping term 
between sites $0$ and $\pm 1$, is projected to:
\be -Jc^\dagger_0(c_1+c_{-1})+h.c.\to -J\gamma_0(c_1+c_{-1})+h.c.\ee
Up to a phase redefinition, this is precisely the tight-binding representation of our standard 2-channel model, with the 
sites at $j<0$ and $j>0$ corresponding to the two channels, introduced in \ref{spin}. Bosonizing all fermion operators except $\gamma$ and $c_0$, with open boundary conditions at $x=0$, gives
\be H_b=it\gamma \eta_0+iJ\gamma_0\{ \Gamma_1 \cos [\sqrt{\pi}\phi_1(0)]+\Gamma_2\cos [\sqrt{\pi}\phi_2(0)]\} .\ee
As established in Sec. II, \ref{RG} and \ref{NTCP}, this model 
renormalizes to the NTCP, confirming our conjecture based on naive extrapolation of the RG equations for 
the centre-coupled model. 
It is also interesting to analyze a tight-binding model which gives $t_L\neq t_R$. This is:
\be H_b=\gamma [tc_0-it'(c_1-c_{-1})]+h.c.]\ee
Using Eq. (\ref{cLR}) we see that this gives, at low energies, our continuum model of Eq. (\ref{Hbcc}) with
\be t_{L/R}=t\pm 2t'\sin k_F.\ee
Now the bosonized form is:
\bea H_b&=&it\gamma \eta_0+ iJ\gamma_0\{ \Gamma_1 \cos [\sqrt{\pi}\phi_1(0)]+\Gamma_2\cos [\sqrt{\pi}\phi_2(0)]\} \nonumber \\
&+&2 it'\eta_0 \{ \Gamma_1 \sin [\sqrt{\pi}\phi_1(0)]-\Gamma_2\sin [\sqrt{\pi}\phi_2(0)]\}.\eea
Again assuming $t$ renormalizes 
to large values, we see that the $t'$ term drives the system into a high energy state, due to the factor of $\eta_0$. Performing a Schrieffer-Wolff 
transformation, the perturbation due to $t'$ becomes:
\bea H_b' &\propto& {(t')^2\over t} \{ \Gamma_1 \sin [\sqrt{\pi}\phi_1(0)]-\Gamma_2\sin [\sqrt{\pi}\phi_2(0)]\}^2 \nonumber 
\\
&=&{(t')^2\over t} \{ \sin^2 [\sqrt{\pi}\phi_1(0)]+\sin^2 [\sqrt{\pi}\phi_2(0)]\}
\eea
where $\{\Gamma_i,\Gamma_j\}=\delta_{ij}$ was used in the last step. This has dimension 
$2/K$ and is  irrelevant for the range of physical interest, $K<1$.  This confirms our conjecture 
that the model renormalizes to the NTCP even when $t_L\neq t_R$. 

We might enquire as to whether there is a symmetry protecting the non-trivial critical point in this model. 
Basically, coupling more strongly to left or right {\it movers} does not correspond to coupling more strongly 
to left or right {\it sides} and so, does not lead to a flow from the NTCP to the $A\otimes N$ or $N\otimes A$ 
fixed point where the Majorana mode couples strongly to the left or right side of the system.  The operative 
symmetry is parity $\times$ time-reversal, PT. This is an anti-unitary operator which complex conjugates 
c-numbers and takes:
\be \psi_L(x)\to \psi_L(-x),\ \  \psi_R(x)\to \psi_R(-x), \ \  \gamma \to \gamma \ee
in the continuum model and:
\be c_j\to c_{-j}\ee
in the lattice model.  This is readily seen to be a symmetry of the Hamiltonian, in continuum and tight-binding forms, 
for all $t$ and $t'$.   Time reversal acts on the components of the conductance tensor, defined in Sec. III as $G_{ij}\to G_{ji}$. 
On the other hand, parity takes $G_{01}\to G_{02}$ and $G_{10}\to G_{20}$. Therefore PT takes $G_{01}\to G_{20}$. 
This is a symmetry of $G_{ij}$ at the NTCP, Eq. (\ref{GNTCP}), in the parity symmetric case where the $G_{jc}$ defined 
in Eq. (\ref{Gjc}) are equal. However, it is {\it not} a symmetry of $G_{ij}$ at the $A\otimes N$ critical point, Eq. (\ref{CAN}). 
Thus, PT symmetry prevents a flow to $A\otimes N$, stabilizing the NTCP.  When  PT symmetry is broken, 
a flow {\it does} occur to the $A\otimes N$ or $N\otimes A$ critical point.  In the continuum model, PT symmetry is broken by:
\be \delta H=iV(\psi^\dagger_L\psi_R-h.c.)\ee
which is relevant at the $t_L=t_R=0$ fixed point, but less relevant than $t_{L/R}$. 
This interaction arises in the continuum limit from the perturbation in the lattice model:
\be \delta H=J'c^\dagger_0(c_1-c_{-1})+h.c.\label{xPT}\ee
giving $V=2J'\sin k_F$. Bosonizing Eq. (\ref{xPT}) and projecting $c_0$ gives:
\be \delta H=iJ'\eta_0\{ \Gamma_1 \cos [\sqrt{\pi}\phi_1(0)]-\Gamma_2\cos [\sqrt{\pi}\phi_2(0)]\}\ee
which, as we know from Sec. II, leads to an RG flow to $A\otimes N$ or $N\otimes A$.

To conclude this Appendix, even the very different centre-coupled model exhibits the same phase diagram with stable 
$A\otimes N$ and $N\otimes A$ critical points and a NTCP which is stabilized by an appropriate parity symmetry, 
providing further evidence for the universality of our proposed phase diagram. 

\section{Impurity Entropy}
\label{entropy}
Critical points of quantum impurity models  with boundary conformal invariance can be characterized by a 
universal zero temperature impurity entropy \cite{g,aflud}, whose exponential is denoted by $g$, the ``ground state 
degeneracy''. This thermodynamic impurity entropy is experimentally measurable for 
some systems such as dilute magnetic impurities in metals. Furthermore, the same universal quantity, $\ln g$, 
appears \cite{pasqjohn} as an impurity contribution to the ground state entanglement entropy, a convenient 
quantity for characterizing phases of one dimensional models via DMRG methods.  $g$ is known 
to always decrease under RG flows \cite{g,aflud,Friedan}.  Thus determining $g$ at various fixed points can 
put constraints on possible phase diagrams.  In this section we calculate $g$ for the various stable 
and unstable fixed points discussed in this paper.

Conformally invariant critical points of quantum impurity models are characterized by conformally invariant 
boundary conditions (CIBC's).  We label these by an integer, $A$ with the corresponding ground state degeneracy 
$g_A$. Imposing boundary conditions $A$ and $B$ at the two ends of a strip of length $\ell$ determines a finite 
size spectrum of energies $E^n=x_{AB}^nu/\ell$ for some dimensionless universal real numbers $x_{AB}^n$. 
($u$ is the velocity.) The corresponding partition function, 
at temperature $T$, is:
\be Z_{AB}[u/(\ell T)]=\sum_n\exp [-x_{AB}^n(u/\ell T)].\ee
(Non-universal terms in the ground state energy, of $O(\ell)$ and $O(1)$ are dropped from $Z$.) 
It is important to note that $Z_{AB}[u/(\ell T)]$ is a universal function of the dimensionless 
ratio $u/(lT)$ only. In the limit $u/(\ell T)\to 0$, it has the asymptotic form:
\be Z_{AB}[u/(\ell T)]\to g_Ag_Be^{\pi \ell Tc/(6u)}.\label{Z}\ee
Here $c$ is the ``conformal anomaly'' parameter which characterizes the bulk conformal field theory and is 
independent of the boundary conditions $A$ and $B$. For the model we are considering, $c=2$. The exponential 
factor in Eq. (\ref{Z}) gives a term in the free energy quadratic in temperature and 
hence the universal low temperature {\it bulk} entropy \cite{c}:
\be S_{\hbox{bulk}}=\pi \ell Tc/(3u)\ee
independent of the boundary conditions. In addition there is an {\it impurity} entropy:
\be S_{\hbox{imp}}=\ln g_A+\ln g_B\ee
which is independent of $l$ and $T$ and is a sum of contributions from both boundaries of the system. 

Note that 
the order of limits is important here. We take $\ell \to \infty$ first, then take $T\to 0$. In this limit, many non-universal 
contributions to the partition function, associated with irrelevant operators, become negligible and the 
asymptotic form of Eq. (\ref{Z}) applies. An additional ground state energy factor:
\be Z_{AB}\to Z_{AB}\exp [-e_0\ell /T-e_1/T]\ee
will generally  be present.  However, the boundary-dependent term in the ground state energy, $e_1$, can easily 
be distinguished from the impurity entropy by its temperature-dependence. The ground state degeneracy can 
also be defined by this method for a system with 2 or more channels of decoupled gapless bulk excitations
 with different velocities, $u_j$. Since 
we are only concerned with critical phenomena at the boundary here, we can formally rescale the lengths differently 
for each channel to make the $u_j$ equal. Without rescaling, Eq. (\ref{Z}) still applies in the more general form:
\be Z_{AB}[ u/(\ell T)]\to g_Ag_B\exp [{\pi \ell T\sum_jc_j/(6u_j)}].\label{Zg}\ee
where $c_j$ is the conformal anomaly for the bulk sector with velocity $u_j$. In our case we have 2 channels 
of free bosons, so $c_1=c_2=1$. 

\begin{figure}
 \center
\includegraphics*[width=0.7\linewidth]{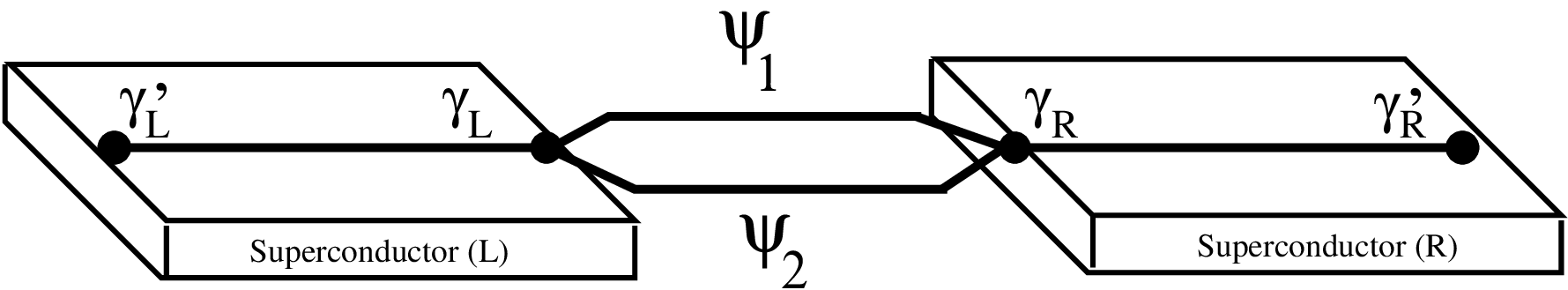}
\caption{A finite system of 2 channels coupled to topological superconductors at both ends. } 
\label{fss}
\end{figure}

Thus a straightforward method to determine $g_A$ for some 
CIBC $A$, is to first calculate the finite size spectrum with $A$-type boundary conditions at both 
ends of a system of length $l$. The corresponding partition function, $Z_{AA}\propto g_A^2$ in the appropriate limit.
We will follows this procedure, restricting our Luttinger liquid channels to have length $\ell$ and coupling them 
to topological superconductors at both ends with identical couplings, as sketched in Fig. (\ref{fss}). To get a 
well-defined value for $g_A$, we are careful to include the other Majorana mode, $\gamma '$ localized 
at the other end of the superconductor, far from the junction. 

\subsection{$N\otimes N$ fixed point}
We begin with the limit $t_i=0$ where the superconductor is decoupled from the two Luttinger liquid channels 
which obey ``open'', that is ``perfect normal reflection'' boundary conditions at both ends: $\psi_L=\psi_R$. 
From Eq. (\ref{bos}) this 
corresponds to: 
\be \theta_i(0)=\theta_i(\ell )=0,\ \  (\hbox{mod}\sqrt{\pi}).\label{modeN}\ee
We may also derive these boundary conditions by considering a large normal scattering boundary interaction:
\be H_{bN}=-t_N\sum_j[\psi_{Li}^\dagger \psi_{Ri}+h.c.]=-t_N\sum_j\cos [2\sqrt{\pi}\theta_j ].\ee
Requiring $\theta_j$ to be at the minimum of $H_{bN}$ at $x=0$ and $\ell$ gives Eq. (\ref{modeN}). 

Let us begin with the simplest case of decoupled channels, the T-junction. Then the fields 
$\theta_i$ and $\phi_i$ have the mode expansions:
\bea  \theta_j ( x   ) &=&  \frac{ \sqrt{\pi} x}{\ell}p_j+ i  \sum_{n = 1}^\infty 
\sqrt{ \frac{K_j}{  \pi n } } 
\: \sin \left[ \frac{ \pi n x}{\ell} \right] [ \alpha_{n ,j} 
-   \alpha_{n ,j}^\dagger    ] \nonumber \\
\phi_j( x   ) &=& \phi_j^0 + \sum_{n = 1}^\infty \frac{1}{ \sqrt{K_j\pi n } } 
\: \cos \left[ \frac{ \pi n x}{\ell} \right] [ \alpha_{n ,j} 
+   \alpha_{n ,j}^\dagger   ].\label{moex}
\eea
Here the $p_j$ are integers, the $\phi_j^0$'s are constants and the $\alpha_{n, j}$ are harmonic 
oscillator annihilation operators. In deriving Eq. (\ref{moex}) we have used the fact that when 
$\theta_j(x)$ obeys Dirichlet boundary conditions, $\phi_j(x)$ must obey Neumann boundary conditions, 
$d\phi_j/dx (0)=d\phi_j/dx (\ell)=0$. Letting $m_{n,j}=0,1,2,\ldots \infty$ label the eigenvalues of 
$\alpha_{n,j}^\dagger \alpha_{n,j}$, the finite size spectrum becomes:
\be E[p_j,m_{n,j}]=\sum_{j=1}^2{\pi u_j\over \ell}\left[{p_j^2\over 2K_j}+\sum_{n=1}^\infty m_{n,j}n\right].\ee
This gives the partition function:
\be Z_{N\otimes N, N\otimes N}=4\prod_{j=1}^2\left\{ \left[ \sum_{p=-\infty}^\infty \exp [-\pi u_jp^2/(2\ell TK_j)]\right] \cdot 
\prod_{n=1}^\infty \{ 1-\exp [-\pi u_jn/(\ell T)]\}^{-1}
\right\} \label{fssN}\ee
The prefactor of 4 was inserted to account for the zero modes in the superconductor on the left and right side 
of the wires. 
To extract $g_N$, we must take the limit $u_j/(\ell T)\to 0$. The second factor in Eq. (\ref{fssN}) is proportional to 
the Dedekind $\eta$-function and has the asymptote:
\be \prod_{n=1}^\infty \{ 1-\exp [-\pi u_jn/(\ell T)]\}^{-1}\to \sqrt{u_j\over 2\ell T}\exp [\pi \ell T/(6u_j)].\ee
The sum in Eq. ({\ref{fssN}) can be approximated, when $u_j/\ell T \ll 1$ by an integral:
\be  \sum_{p=-\infty}^\infty \exp [-\pi u_jp^2/(2\ell TK_j)]\approx \int_{-\infty}^\infty dp \exp [-\pi u_jp^2/(2\ell TK_j)]
=\sqrt{2\ell TK_j\over u_j}.\ee
Thus:
\be Z_{N\otimes N,N\otimes N}\to 4\prod_{j=1}^2\sqrt{K_j}\exp [\pi T/(6u_j)].\ee
This has the expected form of Eq. (\ref{Zg}) and allows us to extract the ground state degeneracy with $N\otimes N$
boundary conditions, in this simple case \cite{gN}:
\be g_{N\otimes N}=2(K_1K_2)^{1/4}.\label{gN2}\ee

Now consider the general $N\otimes N$ case, with arbitrary bulk parameters, as defined in Sec. II. We then have:
\bea \cos \alpha \theta_\sigma (\ell)+\sin \alpha \theta_\rho (\ell)&=&\sqrt{\pi}p_1/r \nonumber \\
-\sin \alpha \theta_\sigma  (\ell) +\cos \alpha \theta_\rho (\ell)&=&\sqrt{\pi}p_2r\eea
for arbitrary integers $p_1$ and $p_2$. Thus the mode expansions are modified to:
\bea \theta_\sigma (x)&=&\frac{ \sqrt{\pi} x}{\ell}  (\cos \alpha \ p_1/r-\sin \alpha \ p_2r)+ i  \sum_{n = 1}^\infty 
\sqrt{ \frac{K_\sigma}{  \pi n } } 
\: \sin \left[ \frac{ \pi n x}{\ell} \right] [ \alpha_{n ,\sigma} 
-   \alpha_{n ,\sigma}^\dagger    ]\nonumber \\
\theta_\rho (x)&=&\frac{ \sqrt{\pi} x}{\ell}  (\sin \alpha p_1/r+\cos \alpha p_2r)+ i  \sum_{n = 1}^\infty 
\sqrt{ \frac{K_\rho}{  \pi n } } 
\: \sin \left[ \frac{ \pi n x}{\ell} \right] [ \alpha_{n ,\rho} 
-   \alpha_{n ,\rho}^\dagger    ] \nonumber \\
 \eea
 and the finite size spectrum becomes:
\bea E[p_j,m_{n,\lambda }]&=&{\pi \over \ell}\biggl[ {u_\sigma\over 2K_\sigma}(\cos \alpha \ p_1/r-\sin \alpha \ p_2r)^2
+ {u_\rho\over 2K_\rho}(\sin \alpha \ p_1/r+\cos \alpha \ p_2r)^2 \nonumber \\
&+&\sum_\lambda u_\lambda \sum_{n=1}^\infty m_{n,\lambda}n\biggr].\eea
We may evaluate $Z_{N\otimes N,N\otimes N}$ at $u_i/(\ell T)\to 0$ as before. The needed Gaussian integral
\bea && \int_{-\infty}^\infty dp_1dp_2\exp \left\{ -{\pi\over 2\ell}\left[{u_\sigma\over K_\sigma}(\cos \alpha \ p_1/r-\sin \alpha \ p_2r)^2
+{u_\rho\over K_\rho}(\sin \alpha \ p_1/r+\cos \alpha \ p_2r)^2\right]\right\}\nonumber \\
&=&2lT\sqrt{K_\sigma K_\rho \over u_\sigma u_\rho},
\eea
is independent of $\alpha$ and $r$, as can be seen by making the rotation of the vector of integration co-ordinates 
$(p_1,p_2)$, $\vec p'=M^T\vec p$, where the matrix $M$ is defined in Eq. (\ref{M}),  before performing the integration. 
(We use the fact that Det $M=1$.) 
Thus, our general result is:
\be g_{N\otimes N}=2(K_\rho K_\sigma )^{1/4},\label{gNNg}\ee
independent of $\alpha$.

\subsection{$A\otimes N$ fixed point}
Now suppose we have a relevant boundary interaction Eq. (\ref{Hbb}) in channel 1 but channel 2 
still obeys normal boundary conditions. 

Let us begin with the simpler T-junction case of decoupled channels. Then $Z_{A\otimes N,A\otimes N}$ factorizes 
into contributions from each channel. The normal channel 2 contributes a factor of $K_2^{1/2}$, as above. 
To calculate the partition function for channel 1 we write the sum of boundary Hamiltonians at each boundary as:
\be H_b=2ti\{ \gamma_L \Gamma_1 \cos [\sqrt{\pi}\phi (0)]+ \gamma_R \Gamma_1 \cos [\sqrt{\pi}\phi (\ell)]\}
\ee
Here $\gamma_{L/R}$ are the Majorana modes at the SN junctions at $x=0$ and $\ell$ respectively 
and $\Gamma_{1}$ is the  Klein factors corresponding to channel-1. $t$ is assumed to be large.
As discussed above, the $A$ boundary conditions imply pinning of both $\phi ( 0 )$ and $\phi ( \ell )$ to 
integer multiples of $\sqrt{\pi}$. Then we have the mode expansion:
\be \phi_1(x)=\sqrt{\pi}{mx\over \ell}+\ldots \label{Amode}\ee
for for $m$ an arbitrary integer.
Accordingly,   the 2 possible 
states of the superconductors correspond to 
$ \frac{1}{\sqrt{2}} \{ \gamma_L \Gamma_1 \cos [\sqrt{\pi}\phi (0)]+ \gamma_R \Gamma_1 
\cos [\sqrt{\pi}\phi (\ell)]\} =\pm 1$. As the system lies within the minimum 
energy state, there is a real fermion ''left-over'', corresponding to the combination of 
$\gamma_L$ and $\gamma_R$ ``orthogonal'' to $ \frac{1}{\sqrt{2}} \{ \gamma_L \Gamma_1 \cos [\sqrt{\pi}\phi (0)]+ \gamma_R \Gamma_1 
\cos [\sqrt{\pi}\phi (\ell)]\}$. Together with the Klein factor corresponding to 
channel-2, $\Gamma_2$, this yields a degeneracy factor of 2, which, when calculating the 
partition function as above, for $\ell T\gg u_i$ now gives:
\be Z_{A\otimes N,A\otimes N}=2\sqrt{K_2/K_1}\prod_je^{\pi \ell T/(6u_j)}.\ee
 Thus we obtain, for a single channel 
with Andreev boundary conditions:
\be g_A=\sqrt{2}/K_1^{1/4}\label{gA}\ee
and for the 2 channel case:
\be g_{A\otimes N}=\sqrt{2}(K_2/K_1)^{1/4}.\label{gANd}\ee
Note that $g_{N\otimes N}/g_{A\otimes N}=\sqrt{2K_1}$, so the $g$-theorem implies the RG flow is from $N \otimes N$ to 
$A \otimes N$ fixed points 
for $K_1>1/2$, consistent with the RG scaling dimension of the boundary interactions discussed 
in Sec. II and \ref{AxN}. 

Now consider the general case of coupled channels. 
As discussed in \ref{bosonize}, the corresponding boundary conditions on the boson fields are 
most conveniently written in terms of the rotated and rescaled fields $\phi_i'$, $\theta_i'$. 
Using Eq. (\ref{bos}) and Eq. (\ref{'bc}) these conditions require:
\bea \phi_1'(0)&=&\phi_1'(\ell),\ \  [\hbox{mod}\  r\sqrt{\pi K_\rho K_\sigma/(\cos ^2\alpha K_\rho +\sin^2\alpha K_\sigma )}]
\nonumber \\
\theta_2'(0)&=&\theta_2'(\ell),\ \  [\hbox{mod}\  r\sqrt{\pi /(\cos ^2\alpha K_\rho +\sin^2\alpha K_\sigma )}].
\eea
Thus the mode expansions for the primed fields are:
\bea \phi_1'(x)&=& \sqrt{\pi K_\rho K_\sigma \over \cos ^2\alpha K_\rho +\sin^2\alpha K_\sigma}{x\over l}p_1r+\ldots 
\nonumber \\
\theta_2'(x) &=& \sqrt{\pi  \over \cos ^2\alpha K_\rho +\sin^2\alpha K_\sigma}{x\over l}p_2r+\ldots 
\eea
Repeating the above calculation gives, in general:
\be g_{A\otimes N}=g_{A \otimes N} =\sqrt{2}\left\{  \frac{ 4 K_\rho K_\sigma  }{r^4  \left[ \cos^2 ( \alpha )   K_\sigma^2 + 
\sin^2 ( \alpha )   K_\rho^2  \right] } \right\}^\frac{1}{4}.\ee
Combining this with Eq. (\ref{gNNg}) we have:
\be
\left( \frac{g_{A \otimes N}}{g_{N \otimes N}} \right)^2 =
\frac{1}{\sqrt{2 K_1} } \: \left[ 1 -  \frac{\tilde{U}^2}{u_1 u_2  } \right]^\frac{1}{4} 
=1/\sqrt{d_1}
\ee
where $d_1$, given in Eq. (\ref{di}), is the dimension of $t_1$ at the normal fixed point. Again we 
obtain consistency with the $g$-theorem: the flow from $N\otimes N$ to $A\otimes N$ fixed points 
only occurs when $d_1<1$ so that $g$ decreases. 

\subsection{$A\otimes A$ fixed point}
As discussed in \ref{NTCP}, by artificially tuning parameters we can reach an $A\otimes A$ fixed point, formally 
corresponding to $t_1=t_2\to \infty$.  We argued in \ref{NTCP} that this is an unstable fixed point, renormalizing 
to the NTCP if the couplings are tuned to lie on the separatrix, and otherwise renormalizing to $A\otimes N$ or $N\otimes A$. 
It is interesting to calculate $g$ at this unstable fixed point since this will help us to confirm its instability, 
once we invoke the $g$-theorem. 

We first consider the T-junction (decoupled channels) and follow the approach of \ref{entropy}.2. From \ref{NTCP},  we see that at the $A\otimes A$ fixed point each channel is coupled 
to a separate Majorana mode, $\gamma_1'$ and $\gamma_{-1}'$. The original Majorana mode $\gamma$ is eliminated 
from the low energy effective Hamiltonian since it combines to make a gapped local Dirac mode defined in  
Eq. (\ref{gjdef2}).
 Multiplying 
together the two $g_A$ 
factors, from Eq. (\ref{gA})   would give $2/(K_1K_2)^{1/4}$. However, we get an extra factor of $2$ due to two other Majorana modes 
which both commute with the low energy effective Hamiltonian. These are $(\gamma_1-\gamma_{-1})/\sqrt{2}$ and $\gamma '$. 
Here $\gamma_{\pm 1}$ are defined in (\ref{gjdef2}) and $\gamma '$ is the ever-present Majorana mode at the other end
of the topological 
superconductor. We may construct a Dirac zero mode operator out of these and the corresponding state 
can be filled or empty, giving the extra factor of 2. Thus, we obtain:
\be  g_{A \otimes A} = \frac{2 \sqrt{2}}{ [ K_1 K_2 ]^\frac{1}{4}}.\label{gAAd}\ee

Now consider coupled chains. The low energy Hamiltonian is:
\bea H_b&=&2i\gamma_{-1}'\Gamma_1\cos \{ \sqrt{\pi} r^{-1} [\cos \alpha \phi_\sigma (0)+\sin \alpha \phi_\rho (0)]\}
\nonumber \\ &+& 2i\gamma_{1}'\Gamma_2\cos \{ \sqrt{\pi} r[-\sin \alpha \phi_\sigma (0)+\cos \alpha \phi_\rho (0)]\}.
\eea
Following the same logic as in \ref{entropy}.2, the mode expansions are:
\bea \cos \alpha \phi_\sigma (x)+\sin \alpha \phi_\rho (x)&=&r \sqrt{\pi}{m_1x\over l}+\ldots \nonumber \\
-\sin \alpha \phi_\sigma (x)+\cos \alpha \phi_\rho (x)&=&r^{-1} \sqrt{\pi}{m_2x\over l}+\ldots
\eea
Here $m_1$ and $m_2$ are arbitrary integers.  Solving:
\bea \phi_\sigma (x)&=&\sqrt{\pi}{x\over l}(r \cos \alpha \ m_1 -\sin \alpha \ m_2 / r)+\ldots \nonumber \\
 \phi_\rho (x)&=&\sqrt{\pi}{x\over l}(r \sin \alpha \ m_1+\cos \alpha \ m_2 / r )+\ldots 
 \eea
 The corresponding terms in the energy are:
 \be E={\pi \over 2l}[u_\sigma K_\sigma (r \cos \alpha \ m_1-\sin \alpha \ m_2 / r)^2+
 u_\rho K_\rho (r \sin \alpha \ m_1+\cos \alpha \ m_2 / r)^2]
 +\ldots \ee
 The corresponding factor in the partition function is:
 \bea Z&\propto &\int dm_1dm_2\exp \biggl\{ -{\pi \over 2lT}\biggl[ u_\sigma K_\sigma 
 (r \cos \alpha \ m_1-\sin \alpha \ m_2 / r)^2\nonumber \\
 &&+u_\rho K_\rho (r \sin \alpha \ m_1+\cos \alpha \ m_2 / r)^2\biggr] \biggr\}\nonumber \\
 &=&{2lT\over \sqrt{u_\sigma u_\rho K_\sigma K_\rho}}.
 \eea
(As in \ref{entropy}.1 the above integrals are easily done using the transformation $\vec m'=M^T\vec m$.) 
 This determines:
 \be g_{A \otimes A} = \frac{2 \sqrt{2}}{ [ K_\rho K_\sigma ]^\frac{1}{4}}.\label{gAA}\ee
 
   Now consider the implications of the $g$-theorem. From Eq. (\ref{gAA}) and Eq. (\ref{gNNg}), we see that 
   \be g_{A \otimes A}/g_{N \otimes N} = \sqrt{\frac{2}{K_\rho K_\sigma}}
   \ee  
This is larger than 1 for the range of Luttinger parameters likely to be of physical relevance, $K_\rho$, $K_\sigma <1$. 
This implies that an RG flow from the $N\otimes N$ fixed point to the $A\otimes A$ fixed point is impossible, providing 
further evidence for the instability of the $A\otimes A$ fixed point.   It can also be checked that quite generally 
$g_{A\otimes A}>g_{A\otimes N}$. For instance, in the case of decoupled channels, using Eqs. (\ref{gANd}) and (\ref{gAAd}), 
\be g_{A \otimes A}/g_{A \otimes N} = 2 \sqrt{\frac{1}{K_2}}\ee
which is larger than 1 for the physically relevant range, $K_2<1$. In the next sub-section, we calculate $g$ at the 
non-trivial critical point, using the $\epsilon$-expansion. Since, for small $\epsilon$ this critical point is close to 
the unstable $N\otimes N$  point, we obtain a value of $g_{\rm NTCP}$ which is only slightly 
less than $g_{N\otimes N}$.  This implies that $g_{A\otimes A}> g_{\rm NTCP}$, so that the flow 
from the unstable $A\otimes A$ point to the non-trivial critical point respects the $g$-theorem.

\subsection{Non-trivial critical point}
In general, the value of $g$ at the NTCP cannot be calculated analytically. It corresponds to a non-trivial 
universal property of the critical point, like its conductance, considered in Sec. III.  
Here we 
calculate it in the $\epsilon$-expansion, introduced in Sec. II and used in Sec. III to obtain the conductance. 
The method we use is similar to that introduced in \cite{aflud} with one important difference. 
In \cite{aflud} a barely relevant ($0<\epsilon \ll 1$) boundary perturbation was considered 
that had a cubic term in its $\beta$-function. An expression for $g$ was obtained, $\propto \epsilon^3$, 
in terms of the coefficient of the cubic term. For the models cosidered here the cubic term in the 
$\beta$-function vanishes and it is necessary to analyse the quartic term. We find it convenient 
to develop a perturbative expansion for the impurity entropy, $\ln g$, rather than $g$ itself:
\be S_{\hbox{imp}}\equiv \ln g = \ln g_{N\otimes N}-a\cdot \epsilon^2 \ee
where $g_{N\otimes n}$ is given in Eq. (\ref{gNNg}) and $a$ is a number of order one which we will calculate. Of course, to 
second order in $\epsilon$, 
\be g=g_{N\otimes N}[1-a\cdot \epsilon^2 ].\ee
In principle, the method is straightforward; we simply expand the log of the partition function in powers of the $t_i$ and 
eventually evaluate the $t_i$ at their critical values. What makes the calculation a bit tricky is that, 
in addition to corrections to the impurity entropy, we will also obtain non-universal corrections, $e_1$, to 
the ground state energy:
\be \delta \ln Z=-e_1\beta + \delta S_{\hbox{imp}}.\ee
These are distinguished by their dependence on the inverse temperature, $\beta$. $e_1$ is temperature 
independent. $\delta S_{\hbox{imp}}$ has a weak temperature dependence $\propto \beta ^{2\epsilon}$, 
associated with the RG flow of $t(\beta )$. 
To simplify the calculation, we first consider the symmetric case $\epsilon_1=\epsilon_2$, along the 
separatrix, $t_1=t_2=t$. 

We begin by expanding $\ln Z$ to second order in $t$:
\be \ln Z\approx \ln Z_0+{1\over 2!}\int d\tau_1d\tau_2{\cal T}<H_b(\tau_1)H_b(\tau_2)>.\ee
At zero temperature, summing over both channels, we have:
\be {\cal T}<H_b(\tau_1)H_b(\tau_2)>=4\tau_0^{-2\epsilon}t^2{1\over |\tau_{12}|^{2(1-\epsilon )}}.\ee
At finite temperature we make the replacement:
\be \tau_{12}\to (\beta /\pi )\sin (\pi \tau_{12}/\beta )\label{finT}\ee
which follows from a conformal transformation. Assuming $\tau_0\ll \beta$, this gives:
\be \delta \ln Z \approx 4t^2\tau_0^{-2\epsilon}\left({\pi\over \beta}\right)^{2(1-\epsilon )}\beta \int_{\tau_0}^{\beta /2}{d\tau \over 
\sin^{2(1-\epsilon )} (\pi \tau /\beta )}.
\ee
Changing variables to
\be u\equiv \tan (\pi \tau /\beta )\label{u}\ee
this becomes:
\be \delta \ln Z \approx 4t^2\pi \left({\beta \over \pi \tau_0}\right)^{2\epsilon}\int_{u_0}^\infty {du\over (1+u^2)^\epsilon 
u^{2(1-\epsilon )}}.\ee
Here, the ultra-violet cut-off has become:
\be u_0\equiv \pi \tau_0/\beta \ee
for $\tau_0\ll \beta$. In order to separate the ground state energy correction from the entropy correction, it is 
convenient to integrate by parts:
\be  \delta \ln Z \approx -4t^2\pi \left({\beta \over \pi \tau_0}\right)^{2\epsilon}{1\over 1-2\epsilon}
\left[ {u_0^{2\epsilon -1}\over (1+u_0^2)^{\epsilon}}+2\epsilon \int_{u_0}^\infty du {u^{2\epsilon}\over (1+u^2)^{1+\epsilon}}\right] .\ee
The remaining integral is finite at $u_0\to 0$.  Since it already has a prefactor of $\epsilon$ we may evaluate it 
at $\epsilon =0$, giving a simple integral. Thus, to lowest order in $\epsilon$:
\be  \delta \ln Z \approx -4t^2\pi \left[ {\beta \over \pi \tau_0}+\epsilon \pi \left({\beta \over \tau_o}\right)^{2\epsilon}
\right].\ee
We have succeeded in separating this into a non-universal correction to the ground state energy, $e_1=4t^2\pi /\tau_0$, 
of no interest, together with a correction to the impurity entropy:
\be \delta S_{\hbox{imp}}=-4t^2\pi^2\epsilon \left({\beta \over \tau_o}\right)^{2\epsilon}.\label{ds2}\ee
We recognize:
\be t \left({\beta \over \tau_o}\right)^{\epsilon}=t(\beta ),\ee
the renormalized coupling constant at scale $\beta$.  To order $t^2$ only the lowest order renormalization 
of $t(\beta )$ appears but we expect that higher order terms will continue to give an expression 
for $\delta S_{\hbox{imp}}(\beta )$ which can be expressed in terms of the renormalized coupling 
constant $t(\beta )$ only. (This was shown explicitly in \cite{aflud}.) Thus, we write:
\be  \delta S_{\hbox{imp}}(\beta )=-4\pi^2\epsilon t(\beta )^2+\ldots \ee
where the $\ldots$ represents higher orders in the expansion in $t(\beta )$. Thus, in this approximation, 
the correction to the zero temperature impurity entropy is obtained by setting:
\be t(\beta )\to t_c=\sqrt{\epsilon /{\cal F}(\nu )}.\ee
($t_c$ was calculated in Sec. II.) Thus:
\be  \delta S_{\hbox{imp}}=-{4\pi^2\epsilon^2\over {\cal F}(\nu )}.\label{ds2f}\ee
It is important to note that the impurity entropy {\it decreases} under renormalization, as required 
by the g-theorem. 

Note that the correction to $S_{\hbox{imp}}$ obtained so far, Eq. (\ref{ds2}), is second order in $t$ 
and contains an explict factor of $\epsilon$, making it $O(\epsilon^2)$. We must now go to fourth 
order in $t$ looking for a term in $\delta S_{\hbox{imp}}$ of order $\epsilon^0t^4$ which is also 
$O(\epsilon^2)$. We will now show that no such term exists and therefore Eq. (\ref{ds2f}) contains 
the entire correction to $S_{\hbox{imp}}$ of $O(\epsilon^2)$. We begin with:
\bea \delta \ln Z&=&{1\over 4!} \prod_{i=1}^4\int^{'} d\tau_i
 {\cal T}<H_b(\tau_1)H_b(\tau_2)H_b(\tau_3)H_b(\tau_4)>\nonumber \\
 &-&{1\over 8}\left[\int^{'} d\tau_1d\tau_2{\cal T}<H_b(\tau_1)H_b(\tau_2)>\right]^2
.\label{dz4}\eea
(Because we are calculating $\delta \ln Z$ rather than $\delta Z$ itself, we subtract off the second order iteration 
of the term of $O(t^2)$ calculated above. This leads to very convenient cancellations. The $'$ on the 
integral signs signifies that we apply our ultra-violet cut-off, $|\tau_{ij}|>\tau_0$.) 
The fourth order matrix element is readily evaluated following the methods of \ref{RG} 
There are two distinct contributions, $\propto t_j^4$ (with equal contributions for $j=1$ or $2$), and $\propto t_1^2t_2^2$. 
We first consider the $t_j^4$ term.
 The needed fermionic factors are simply:
\be {\cal T}<\gamma (\tau_1)\gamma (\tau_2)\gamma (\tau_3)\gamma (\tau_4)>=
 {\cal T}<\Gamma_1 (\tau_1)\Gamma_1 (\tau_2)\Gamma_1 (\tau_3)\Gamma_1 (\tau_4)>=\epsilon (\tau_1,\tau_2,\tau_3,\tau_4)\ee
and square to one. Thus, at zero temperature we have:
\bea \delta \ln Z&=&\tau_0^{-4\epsilon}{t^4\over 6} \prod_{i=1}^4\int^{'} d\tau_i\Biggl[ 
\left| {\tau_{12}\tau_{34}\over \tau_{13}\tau_{14}\tau_{23}\tau_{24}}\right|^{2(1-\epsilon )}
+\left| {\tau_{13}\tau_{24}\over \tau_{12}\tau_{14}\tau_{23}\tau_{34}}\right|^{2(1-\epsilon )}
+\left| {\tau_{14}\tau_{23}\over \tau_{12}\tau_{13}\tau_{24}\tau_{34}}\right|^{2(1-\epsilon )}
\nonumber \\ &&
-2\left|{1\over \tau_{12}\tau_{34}}\right|^{2(1-\epsilon )}
-2\left|{1\over \tau_{13}\tau_{24}}\right|^{2(1-\epsilon )}
-2\left|{1\over \tau_{14}\tau_{23}}\right|^{2(1-\epsilon )}
\Biggr] .\eea
As for the $O(t^2)$ term calculated above, we actually do the calculation at a low but finite temperature, $\beta^{-1}$, 
resulting in the substitution: $\tau_{i}\to (\beta /\pi )\sin (\pi \tau_{i}/\beta )$. It is convenient 
to use time-translation invariance to do the integral over $\tau_1$, giving a factor of $\beta$. We then change 
variables to:
\bea u&=&\tan (\pi \tau_2/\beta )\nonumber \\
v&=&\tan (\pi \tau_3/\beta )\nonumber \\
w&=&\tan (\pi \tau_4/\beta ).\label{uvw}\eea
After a little algebra, the integral then becomes:
\bea \delta \ln Z&=&{\pi t^4\over 6}\left({\beta \over \pi \tau_0}\right)^{4\epsilon}
\int^{'}{dudvdw\over (1+u^2)^\epsilon (1+v^2)^\epsilon (1+w^2)^\epsilon} \nonumber \\
& &\Biggl[
\left|{u(v-w)\over vw(u-v)(u-w)}\right|^{2(1-\epsilon)}+\left|{v(u-w)\over uw(v-u)(v-w)}\right|^{2(1-\epsilon)}
+
\left|{w(u-v)\over uv(w-u)(w-v)}\right|^{2(1-\epsilon)} \nonumber \\&-&2\left|{1\over u(v-w)}\right|^{2(1-\epsilon )}
-2\left|{1\over v(u-w)}\right|^{2(1-\epsilon )}-2\left|{1\over w(u-v)}\right|^{2(1-\epsilon )}
\Biggr] .\label{lZ4}
\eea
The $'$ on the integral sign now indicates that the $u$, $v$ and $w$ integrals run from $-\infty$ to $\infty$, 
subject to the ultra-violet cut-off $|u|$, $|v|$, $|w|$, $|u-v|$, $|u-w|$, $|v-w|>u_0\equiv \pi \tau_0/\beta$. 
Once again, we must disentangle terms in $\delta \ln Z \propto \beta /\tau_0$ which correspond to 
non-universal ground state energy corrections from terms $\propto t^4(\beta /\tau_0)^4\propto t(\beta )^4$, 
corresponding to corrections to the impurity entropy. 
Note that the integrand in Eq. (\ref{lZ4}) diverges as $u^{-4(1-\epsilon)}$ when $u$, $v$ and $w$ all go to zero 
proportional to each other. 
Since the integral is 3-dimensional, this implies $(\beta /\tau_0)^{1-4\epsilon}$ behavior, which when combined 
with the prefactor gives $\beta /\tau_0$, corresponding to a ground state energy term. 
In order to separate this ground state energy correction from the impurity entropy term, 
it is convenient to first change variables to $v'\equiv v/u$, $w'\equiv w/u$, then 
integrate by parts with respect to $u$.
The $u$ integral is now:
\bea I&& (v',w')\equiv \int_{u_0}^\infty du u^{4\epsilon -2}\left\{ {1\over (1+u^2)[1+(uv')^2][1+uw')^2]}\right\}^\epsilon
\nonumber \\
&=&{u_0^{4\epsilon -1}\over 1-4\epsilon }\left\{ {1\over (1+u_0^2)[1+(u_0v')^2[1+(u_0w')^2]}\right\}^\epsilon
\nonumber \\ &&
-{2\epsilon \over 1-4\epsilon}\int_{u_0}^\infty duu^{4\epsilon -1}
\left\{ {1\over (1+u^2)[1+(uv')^2][1+(uw')^2]}\right\}^\epsilon \nonumber \\
&& \left[{u^2\over 1+u^2}+{(uv')^2\over 1+(uv')^2}+{(uw')^2\over 1+(uw')^2}\right].\label{parts}
 \eea
Inserting the first term in Eq. (\ref{parts}) back into Eq. (\ref{lZ4}) gives a term in $\delta \ln Z \propto \beta /\tau_0$ 
corresponding to a ground state energy correction, while the second term in Eq. (\ref{parts}) gives 
a correction to $S_{\hbox{imp}}$ which can be seen to be $\propto t^4\epsilon \propto \epsilon^3$ and 
therefore negligible to the order we are working.  To see this, we can set some 
unessential factor of $\epsilon \to 0$ in Eq. (\ref{parts}). 
Then the first term in Eq. (\ref{parts}) gives:
\bea \delta \ln Z&=& {\pi t^4\over 3}{\beta \over \pi \tau_0}
\int^{'}dv'dw'\Biggl[ \left( {v'-w'\over v'w'(1-v')(1-w')}\right)^2+\left( {v'(1-w')\over w'(1-v')(w'-v')}\right)^2 \nonumber 
\\
&& +\left( {w'(1-v')\over v'(1-w')(w'-v')}\right)^2 \nonumber \\ &&
-2\left({1\over v'-w'}\right)^2-2\left({1\over v'(1-w')}\right)^2-2\left({1\over w'(1-v')}\right)^2
\Biggr]\label{eint}
 .\eea
This integral can be seen to be ultraviolet finite when the cut-off goes to zero, due to cancellations between 
the six terms.  For instance, the singular part of the integrand when $v'\to 0$ is
\be I\to \left({1\over v'(1-w')}\right)^2\left[\left({1-v'/w'\over 1-v'}\right)^2+\left({1-v'\over 1-v'/w'}\right)^2-2\right]
\to {4\over (1-w')^2}\left[{1\over (w')^2}+1\right] .
\ee
The sum of the three  terms inside the brackets scale as $(v')^2$ as $v'\to 0$, leaving a finite limiting value for the integrand.
The integral in Eq. (\ref{eint}) can also be seen to converge at $|v'|$, $|w'|\to \infty$. (At 
large  $v'$, $w'$ the second, third and fourth terms almot cancel each other.) 
Therefore 
Eq. (\ref{eint}) gives a term in $\delta \ln Z$ which is $\beta /\tau_0$ times a finite number, 
corresponding to a ground state energy correction.
Of greater interest is the result of inserting the second term in Eq. (\ref{parts}) back into Eq. (\ref{lZ4}). 
This gives the correction to $S_{\hbox{imp}}$. Converting back to the original $v$ and $w$ integration 
variables for convenience, and setting inessential factors of $\epsilon$ to zero, this gives:
\bea \delta \ln Z &=& -{2\pi t^4\over 6}\left({\beta \over \pi \tau_0}\right)^{4\epsilon}\epsilon
\int^{'}dudvdw\left[{u^2\over 1+u^2}+{v^2\over 1+v^2}+{w^2\over 1+w^2}\right]\nonumber \\ &&
\Biggl\{  \left( {u(v-w)\over vw(u-v)(u-w)}\right)^2+\left( {v(u-w)\over uw(u-v)(v-w)}\right)^2
+\left( {w(u-v)\over vu(u-w)(v-w)}\right)^2\nonumber \\ &&
-2\left({1\over u(v-w)}\right)^2-2\left({1\over v(u-w)}\right)^2-2\left({1\over w(v-u)}\right)^2\Biggr\}
\eea
Again, Taylor expanding the expression in curly brackets around the singular points, $u\to 0$, $u\to v$ et cetera shows that 
this integral is finite when the ultra-violet cut-off is taken to zero. It also converges at $|u|$, $|v|$, $|w| \to \infty$.
(This is fairly obvious since the integrand of this 3-dimensional integral 
scales as $1/u^4$ when $|u|$, $|v|$ and $|w|$ all go to $\infty$ 
proportional to each other.)   
Therefore it gives a correction  $\delta \ln Z\propto \epsilon t^4(\beta /\tau_0)^{4\epsilon}=\epsilon t(\beta )^4$. 
In the zero temperature limit, we can replace the running coupling constant, $t(\beta )$ by its 
fixed point value $t_c$, giving a correction $\delta \ln Z\propto \epsilon^3$, negligible 
compared to Eq. (\ref{ds2f}).

Finally, we consider the term in $\delta \ln Z \propto t_1^2t_2^2$, showing that it is again $\propto \epsilon t_c^4\propto \epsilon^3$. 
The fermion factors now give:
\bea && {\cal T}<\Gamma_1(\tau_1)\Gamma_1(\tau_2)\Gamma_2(\tau_3)\Gamma_2(\tau_4)>
{\cal T}<\gamma (\tau_1)\gamma (\tau_2)\gamma (\tau_3)\gamma (\tau_4)>\nonumber \\ &&
=\epsilon (\tau_1-\tau_2)\epsilon (\tau_3-\tau_4)
\epsilon (\tau_1,\tau_2,\tau_3,\tau_4)=\epsilon (\tau_1-\tau_3)\epsilon (\tau_1-\tau_4)\epsilon (\tau_2-\tau_3)\epsilon (\tau_2-\tau_4).
\nonumber \\ \eea
$\epsilon (\tau )$ and $\epsilon (\tau_1,\tau_2,\tau_3,\tau_4)$ are the 
anti-symmetric step functions defined in \ref{RG}, not to be confused with $\epsilon =d_b-1$, the scaling 
dimension.
The bosonic factor is:
\bea &&16<\cos[\sqrt{\pi}\phi_1(\tau_1)]\cos[\sqrt{\pi}\phi_1(\tau_2)]\cos[\sqrt{\pi}\phi_2(\tau_3)]
\cos[\sqrt{\pi}\phi_1(\tau_1)]>\nonumber \\ &&
=2\left({1\over (\tau_1-\tau_2)(\tau_3-\tau_4)}\right)^{2(1-\epsilon )}
\left[ \left|{(\tau_1-\tau_3)(\tau_2-\tau_4)\over (\tau_1-\tau_4)(\tau_2-\tau_3)}\right|^\nu
+ \left|{(\tau_1-\tau_4)(\tau_2-\tau_3)\over (\tau_1-\tau_3)(\tau_2-\tau_4)}\right|^\nu \right]
\nonumber \\
\eea
where $\nu$ is defined in Eq. (\ref{nu}). Thus, at zero temperature:
\bea &&\delta \ln Z=t^4\prod_{i=1}^4\int^{'}d\tau_i\left({1\over (\tau_1-\tau_2)(\tau_3-\tau_4)}\right)^{2(1-\epsilon )}
\biggl\{ - 2 
\nonumber \\ &+&
\biggl\{\left[ \left|{(\tau_1-\tau_3)(\tau_2-\tau_4)\over (\tau_1-\tau_4)(\tau_2-\tau_3)}\right|^\nu
+ \left|{(\tau_1-\tau_4)(\tau_2-\tau_3)\over (\tau_1-\tau_3)(\tau_2-\tau_4)}\right|^\nu\right] 
\epsilon (\tau_1-\tau_3)\epsilon (\tau_1-\tau_4)\epsilon (\tau_2-\tau_3)\epsilon (\tau_2-\tau_4)\biggr\}
\nonumber \\
\eea
Passing to finite temperature via Eq. (\ref{finT}), integrating over $\tau_1$, and then 
changing variables as in Eq. (\ref{uvw}) turns this into:
\bea \delta \ln Z&=&\pi t^4\left({\beta \over \pi \tau_0}\right)^{4\epsilon}
\int^{'}{dudvdw\over (1+u^2)^\epsilon  (1+v^2)^\epsilon (1+w^2)^\epsilon}\left|{1\over u(v-w)}\right|^{2(1-\epsilon)}
\nonumber \\ &&
\cdot \left\{\left[ \left|{v(u-w)\over w(u-v)}\right|^\nu +\left|{w(u-v)\over v(u-w)}\right|^\nu
 \epsilon (v)\epsilon (w)\epsilon (u-v)\epsilon (u-w)\right]-2
\right\}.
\nonumber \\
\eea
To separate the ground state energy correction from the impurity entropy correction, we again rescale 
and integrate by parts, as above. The needed $u$-integral is the same one as in Eq. (\ref{parts}). 
The first term in Eq. (\ref{parts}) then gives, after dropping inessential factors of $\epsilon$:
\bea \delta \ln Z &=& 2\pi t^4{\beta \over \pi \tau_0}\int^{'}{dv'dw'\over (v'-w')^2}\nonumber \\ &\times&
\left\{\left[ \left|{v'(1-w')\over w'(1-v')}\right|^\nu +\left|{w'(1-v')\over v'(1-w')}\right|^\nu\right]
\epsilon (v')\epsilon (w')\epsilon (1-v')\epsilon (1-w')-2\right\}. \nonumber \\
\eea
Taylor expanding the quantity in curly brackets around the point $v'=w'$ it can be seen that 
it is $O[(v'-w')^2]$, and so there is no ultra-violet divergence in that limit. For the range 
of Luttinger parameters and anisotropy parameter $\alpha$ that we are considering, $0<\nu \leq 1$, so 
there is no ultraviolet divergence at $v',w'\to 0,1$. To see this in the limiting $SU(2)$ invariant 
case, $\nu =1$, we may use:
\bea && \left[ \left|{v'(1-w')\over w'(1-v')}\right| +\left|{w'(1-v')\over v'(1-w')}\right|\right]
\epsilon (v')\epsilon (w')\epsilon (1-v')\epsilon (1-w') \nonumber \\
&=&{(v')^2(1-w')^2+(w')^2(1-v')^2\over v'w'(1-v')(1-w')}.
\label{E042}\eea
Then integrating symmetrically around the singularities at $v',w'=0,1$ gives a finite result. 
Thus we take the ultraviolet cut off to zero in evalulating this integral. The integral can 
also be seen to converge at $|v'|,|w'|\to \infty$ and thus Eq. (\ref{E042}}) gives $\beta /\tau_0$ 
times a finite number, corresponding to a ground state energy correction. The second term in Eq. (\ref{parts}), 
after talking inessential factors of $\epsilon \to 0$, 
gives:
\bea \delta \ln Z&=&2\pi t^4\left({\beta \over \pi \tau_0}\right)^{4\epsilon}\epsilon 
\int^{'}dudvdw\left({1\over u(v-w)}\right)^2
\left[{u^2\over 1+u^2}+{v^2\over 1+v^2}+{w^2\over 1+w^2}\right]
\nonumber \\ &&
\cdot \left\{\left[ \left|{v(u-w)\over w(u-v)}\right|^\nu +\left|{w(u-v)\over v(u-w)}\right|^\nu\right]
 \epsilon (v)\epsilon (w)\epsilon (u-v)\epsilon (u-w)-2
\right\}.
\eea
Again by Taylor expanding the quantity in curly brackets we can see that there is no divergence at $v\to w$ or $u\to 0$. 
Nor are there any divergences at $v,w\to 0,u$ for $0<\nu \leq 1$. The integral can also be seen to converge 
at $|u|, |v|, |w|\to \infty$. Thus we obtain $t(\beta )^4\epsilon$ times a finite constant. This again 
gives a correction to $S_{\hbox{imp}}$ of $O(\epsilon^3)$, completing our proof that Eq. (\ref{ds2f}) 
gives the entire correction of $O(\epsilon^2)$.

In the above calculation, we assumed $\epsilon_1=\epsilon_2$, for simplicity. But the quadratic term 
giving Eq. (\ref{ds2f}) just consisted of a sum of contributions from each channel, so it follows that in general:
\be \delta S_{\hbox{imp}}=-{2\pi^2(\epsilon_1^2+\epsilon_2^2)\over {\cal F}}+O(\epsilon^3)\ee
and $g$ at the non-trivial critical point is given by:
\be g_{\rm NTCP}\approx g_{N\otimes N}\left[ 1-{2\pi^2\over {\cal F}}(\epsilon_1^2+\epsilon_2^2)\right]
=2(K_\sigma K_\rho)^{1/4}\left[ 1-{2\pi^2\over {\cal F}}(\epsilon_1^2+\epsilon_2^2)\right]
\ee
where the function ${\cal F}(\nu )$ is given by the integral in Eq. (\ref{fnu}), plotted in Fig. (\ref{fig_nu}). 
[$\nu$ is given in terms of Luttinger parameters and anisotropy parameter $\alpha$, in Eq. (\ref{nu}).]


\Bibliography{50}

\bibitem{Mourik} V. Mourik, K. Zuo, S.M. Frolov, S.R. Plissard, E.P.A.M. Bakkers and L.P. Kowenhoven, 
Science, {\bf 336}, 1003 (2012).

\bibitem{Lutchyn} R. M. Lutchyn, J. D. Sau, and S. Das Sarma, Phys. Rev. Lett. {\bf 105},
077001 (2010).

\bibitem{Oreg} Y. Oreg, G. Refael, and F. von Oppen, Phys. Rev. Lett. {\bf 105}, 177002
(2010).

\bibitem{Law} K. T. Law, P. A. Lee, and T. K. Ng, Phys. Rev. Lett. {\bf 103}, 237001
(2009).

\bibitem{Sau} J. D. Sau, R. M. Lutchyn, S. Tewari, and S. Das Sarma, Phys. Rev. Lett. {\bf 104}, 040502  (2010).

\bibitem{Alicea0} J. Alicea, Phys. Rev. {\bf B 81}, 125318 (2010).

\bibitem{Wimmer} M. Wimmer, A. R. Akhmerov, J. P. Dahlhaus, and C. W . J. Beenakker, New J. Phys. {\bf 13}, 053016 (2011).

\bibitem{Cook} A. Cook and M. Franz, Phys. Rev. {\bf B84}, 201105 (2011).

\bibitem{Klinovaja} J. Klinovaja and D. Loss, Phys. Rev. {\bf B86}, 085408 (2012).

\bibitem{Sticlet} D. Sticlet, C. Bena, and P. Simon, Phys. Rev. Lett. {\bf 108}, 096802 (2012).

\bibitem{Chevalier} D. Chevallier, D. Sticlet, P. Simon, and C. Bena, Phys. Rev. {\bf B85}, 235307 (2012).

\bibitem{Beri0} B. B\'eri, N.R. Cooper, Phys. Rev. Lett. {\bf 109}, 156803 (2012).

\bibitem{lut_majo} L. Fidkowski, J. Alicea, N. H. Lindner, R. M. Lutchyn,
 and M. P. A. Fisher, Phys. Rev. {\bf B 85}, 245121 (2012).

\bibitem{ACZ} I. Affleck, J.-S. Caux and A. Zagoskin, Phys, Rev. {\bf B62}, 1433 (2000).

\bibitem{Alicea} J. Alicea, Y. Oreg, G. Refael, F. von Oppen and M.P.A. Fisher, Nat. Phys. {\bf 7}, 412 (2011).

\bibitem{giu_af1} D. Giuliano and I. Affleck,   in preparation.

\bibitem{giu_af2} D. Giuliano and I. Affleck,   J. Stat. Mech.P02034 (2013).

\bibitem{Beri} B. B\'eri,  Phys. Rev. Lett. {\bf 110}, 216803 (2013).

\bibitem{affl_1} A. H. Castro Neto, E. Novais, L. Borda, G. Zar\'and, and I. Affleck, 
Phys. Rev. Lett 91, 096401 (2003).

\bibitem{affl_2} E. Novais, A. H. Castro Neto, L. Borda, I. Affleck, and G. Zar\'and,
Phys. Rev. {\bf B 72}, 014417 (2005).

\bibitem{Novais} E. Novais, F. Guinea, and A. H. Castro Neto, Phys. Rev. Lett. {\bf 94}, 170401 (2005).

\bibitem{Callan} C.G. Callan and D. Freed, Nucl. Phys. {\bf B374}, 543 (1992).

\bibitem{Zarand}  G. Zar\'and and E. Demler, Phys. Rev. {\bf B66}, 024427 (2002).

\bibitem{lukyanov} S. M. Lukyanov, A. M. Tsvelik, and A. B. Zamolodchikov, 
Nucl.Phys. {\bf B719}, 103 (2005). 

\bibitem{KF} C.L. Kane and M.P.A. Fisher, Phys. Rev. {\bf B46}, 15233 (1992).

 \bibitem{typos} Eq. (3.12) of \cite{KF} appears to be missing a factor of $(e^2/h)\tau_c^{2n^2/g}$.  We can ignore the 
powers of $\tau_c$, or $\tau_0$ in our notation, since we have effectively adsorbed them into our definition of the $t_j$ as follows from Eq. (\ref{norm}).  
We included the factor of $e^2/h$ in extracting the conductance from  \cite{KF}.

\bibitem{oca} M. Oshikawa, C. Chamon and I. Affleck,  J.Stat.Mech.0602:P02008 (2006). 

\bibitem{Cardy}  See, for example, the textbook: J.L. Cardy, {\it Scaling and Renormalization in Statistic Physics}, Cambridge 
University Press, 1996.

\bibitem{AL}The third order term in the $\beta$-function of the Kondo model was calculated by a similar 
technique in I. Affleck and A.W.W. Ludwig, Nucl. Phys. {\bf B360}, 641 (1991). 

\bibitem{Ian_1} I. Affleck, J. Phys. A: Math. Gen. {\bf 31}, 2761 (1998).

\bibitem{SW} J.R. Schrieffer and P.A. Wolff, Phys. Rev. {\bf 149}, 491–492 (1966).

\bibitem{error2} \cite{lut_majo} gives this dimension at the $A\otimes N$ fixed point, in the symmetric 
case, $\alpha =\pi /4$, but seems to contain 
a ``typo'' about which operator it corresponds to.

\bibitem{g} I. Affleck and A.W.W. Ludwig, Phys. Rev. Lett. {67}, 161 (1991).

\bibitem{aflud} I. Affleck and A.W.W. Ludwig, PRB {\bf 48}, 7279 (1993).

\bibitem{Friedan} D. Friedan and A. Konechny, Phys. Rev. Lett. {\bf 93}, 030402, (2004).

\bibitem{pasqjohn} P. Calabrese, J. Cardy,  J. Stat. Mech. 0406:P06002 (2004).

\bibitem{c} H.W.J. Bl\"ote, J.L. Cardy and M.P. Nightingale, Phys. Rev. Lett. {\bf 56}, 742 (1986);
I. Affleck, ibid 746 (1986).

\bibitem{gN} This formula, for the case of a single channel and no superconductor, 
was derived  by this method in S. Eggert and I. Affleck, Phys. Rev. {\bf B46}, 10866 (1992) but the square root was omitted
 from the final answer. 
It was rederived  correctly by a different method in M. Oshikawa and I. Affleck, Nucl. Phys. {\bf B495}, 533 (1997).

\end{thebibliography}
\end{document}